%% file: usenixsecurity2026.tex
\documentclass[letterpaper,twocolumn,10pt]{article}
\usepackage{usenix}

\usepackage{cite}
\usepackage{multirow}
\usepackage{amsmath,amssymb,amsfonts}
\usepackage{graphicx}
\usepackage{textcomp}
\usepackage{longtable}
\usepackage{tabularray}
\usepackage{xcolor}
\usepackage{cite}
\usepackage{subcaption}
\usepackage{caption}
\usepackage{booktabs}
\usepackage{xspace}
\usepackage{array}
\usepackage{bbm}
\usepackage{makecell}
\usepackage{tikz}
\usepackage{pgfplots}
\usepackage{pgfplotstable}
\usepackage{filecontents}
\usepackage{adjustbox}
\usepackage{fixltx2e}
\usepackage{stfloats}
\usepackage{url}
\usepackage{algorithm}
\usepackage{algpseudocode}
\usepackage{amsmath}
\usepackage{amssymb}
\usepackage{setspace}
\usepackage{hyperref}
\usepackage{multirow}
\usepackage{xurl}
\usepackage{tikz}
\usepackage{xcolor}

\newcommand{\circlednum}[1]{%
  \tikz[baseline=(char.base)]{
    \node[shape=circle, draw=black, fill=black, text=white, inner sep=1pt] (char) {\small\bfseries #1};}}

\usepackage{enumitem}
\newenvironment{smitemize}{
  \begin{itemize}[topsep=1pt, partopsep=0pt, itemsep=1pt, parsep=0pt, leftmargin=10pt, itemindent=1pt]
}{\end{itemize}}

\usepackage[color=green]{todonotes}

\newcommand{\msr}[1]{\todo[inline,color=green!40]{SR: #1}}

\newcommand{\meanstd}[2]{#1{\tiny\,$\pm$\,#2}}

\newcommand{\paragraphB}[1]{\vskip 5pt \noindent {\textbf{#1}.}\xspace}

\newcommand{\system}{{CITADEL}\xspace}

\begin{document}

\date{}


\title{\system: A Semi-Supervised Active Learning Framework for Malware Detection Under Continuous Distribution Drift}

\author{
Md~Ahsanul~Haque$^{1}$, Md~Mahmuduzzaman~Kamol$^{1}$, Suresh~Kumar~Amalapuram$^{2}$\\
Vladik~Kreinovich$^{1}$, Mohammad~Saidur~Rahman$^{1}$\\[0.5ex]
\small
$^{1}$University of Texas at El Paso \quad
$^{2}$Indian Institute of Technology Hyderabad\\
\texttt{\{mhaque3,mkamol,vladik,msrahman3\}@utep.edu}, \texttt{apskumarkrc@gmail.com}
}


\maketitle


\input{0_abstract}

\input{1_introduction}

\input{2_related_work}

\input{3_methodology}

\input{4_experimental_setup}

\input{5_experimental_results_and_evaluation}

\input{6_discussion}
\input{7_conclusion}
\cleardoublepage
\bibliographystyle{plainurl}
\bibliography{main}

\appendix

\input{appendix}

\end{document}

%% file: 0_abstract.tex
\begin{abstract}
Android malware detection systems suffer severe performance degradation over time due to concept drift caused by evolving malicious and benign app behaviors. Although recent methods leverage active learning and hierarchical contrastive loss to address drift, they remain fully supervised, computationally expensive, and ineffective on long-term real-world benchmark. Moreover, expert labeling does not scale to the monthly emergence of nearly 300K new Android malware samples, leaving most data unlabeled and underutilized.


To address these challenges, we propose CITADEL, a semi-supervised active learning framework for Android malware detection. 
Existing semi-supervised methods assume continuous and semantically meaningful input transformations, and fail to generalize well to high-dimensional binary malware features.
We bridge this gap with malware-specific augmentations, Bernoulli bit flips and feature masking, that stochastically perturb feature to regularize learning under evolving malware distributions.
\system further incorporates supervised contrastive loss to improve boundary sample discrimination and combines it with a multi-criteria active learning strategy based on prediction confidence, $L_p$-norm distance, and boundary uncertainty, enabling effective adaptation under constrained labeling budgets. Extensive evaluation on four large-scale Android malware benchmarks---APIGraph, Chen-AZ, MaMaDroid, and LAMDA, demonstrates that \system outperforms prior work, achieving F1 score of over 1\%, 3\%, 7\%, and 14\% respectively, using only 40\% labeled samples. Furthermore, \system shows significant efficiency over prior work incurring $24\times$ faster training and $13\times$ fewer operations.
\end{abstract} 

\paragraphB{Availability}
The code is available at \url{https://github.com/IQSeC-Lab/CITADEL.git}.

%% file: 1_introduction.tex
\section{Introduction}


The Android ecosystem, with over 3 billion active devices, has become a primary target for malware developers~\cite{abusnaina2025exposing}. The AV‑TEST Institute reports around 5 million new malware and potentially unwanted applications (PUAs) annually~\cite{avtest}, posing a growing challenge to reliable threat detection. Traditional methods such as sandbox analysis and signature matching are labor-intensive, brittle to obfuscation, and struggle to scale~\cite{rastogi2013droidchameleon,zhou2012dissecting}. As a scalable alternative, machine learning (ML)-based detection systems have gained widespread adoption in both academia and industry~\cite{arp2014drebin,tesseract}, including integration by vendors such as Avast~\cite{avast2021ai}. These systems learn from structural and behavioral features, and have demonstrated strong performance and generalization~\cite{transcendent,chen2023continuous}.


\begin{figure}
    \centering
    \includegraphics[width=\linewidth]{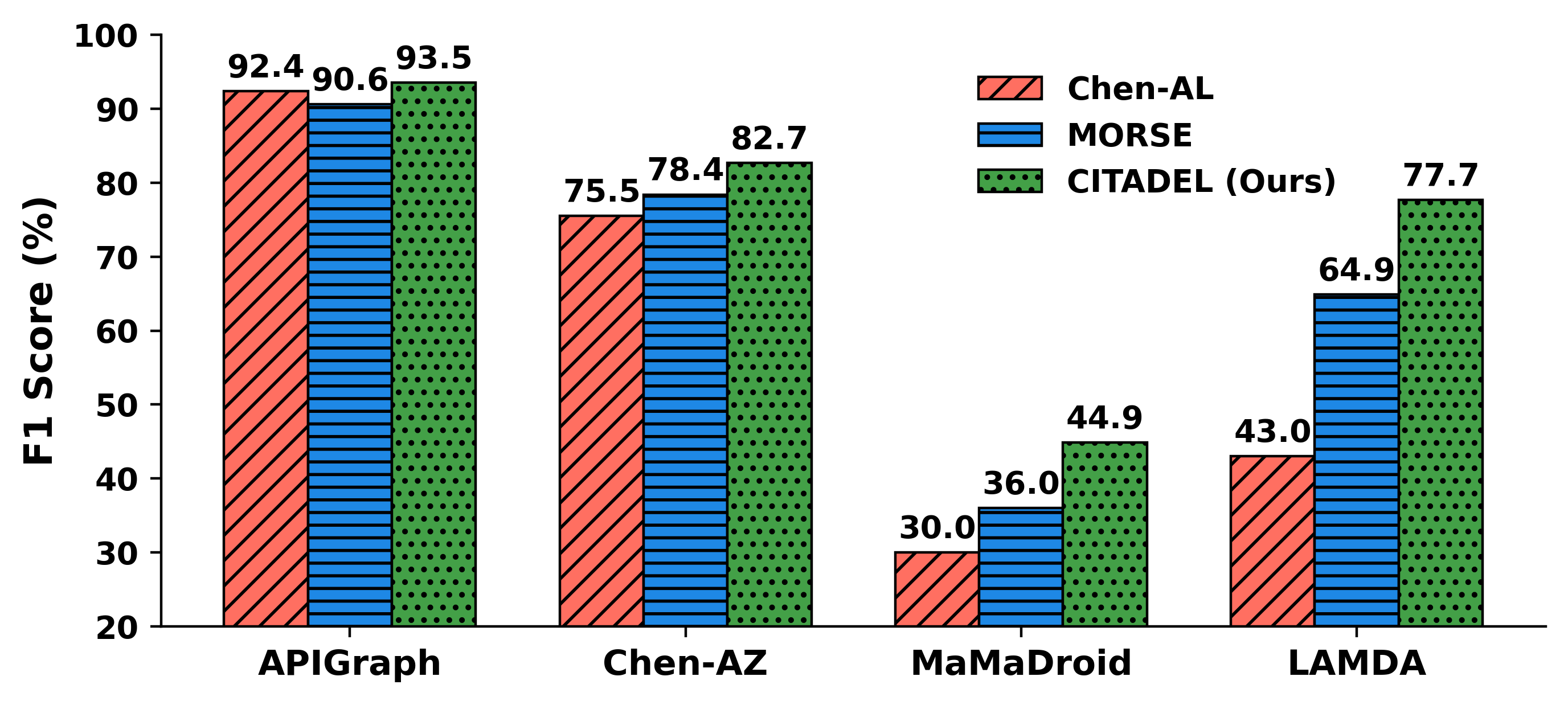}
    \caption{
    Concept drift adaptation performance comparison among Chen-AL~\cite{chen2023continuous}, MORSE~\cite{wu2023grim}, and our proposed semi-supervised active learning system, \system, on four Android malware benchmark datasets---APIGraph~\cite{api_graph_dataset}, Chen-AZ~\cite{chen2023continuous}, MaMaDroid~\cite{mamadroid}, and LAMDA~\cite{haque2025lamda}.
    }
    \label{fig:citadel-sota-comparison}
\end{figure}

A core challenge in ML-based malware detection is {\em concept drift}--the change in the statistical properties of malware and benign samples over time, which degrades detector performance. Prior work has addressed concept drift using contrastive learning, with active learning (AL) incorporated through uncertainty-based sample selection~\cite{chen2023continuous}. However, these approaches are evaluated under limited temporal settings. Although Chen-AL, as the state of the art, performs effectively on short temporal spans—achieving a 92.4\% F1-score on the 7-year APIGraph dataset~\cite{api_graph_dataset}—its performance degrades substantially as the temporal span increases, dropping to 43\% on the 12-year LAMDA benchmark~\cite{haque2025lamda} and further to 30\% on a 12-year curated dataset derived from MaMaDroid~\cite{mamadroid} features (Figure~\ref{fig:citadel-sota-comparison}).

Furthermore, the scale of Android malware makes sustained labeling impractical. With over 300,000 new android malware samples emerging each month~\cite{avtest}, producing high-quality labels is costly and slow; human analysts can review only around 80 samples per day~\cite{miller2016reviewer}, and vendors may take weeks to agree on newly discovered variants~\cite{zhu2020measuring, kantchelian2015better}. 
Despite these constraints, existing concept drift adaptation methods~\cite{yang2021cade, chen2023continuous} rely heavily on supervised learning, implicitly assuming continuous access to labeled data and focusing on short-term evaluations. 
Such assumptions {\em do not hold in realistic, long-term drift settings}, where labels are scarce and the malware distribution evolves unpredictably. These challenges highlight the need for semi-supervised learning (SSL), which can leverage abundant unlabeled data to sustain detection performance under long-term distribution shifts~\cite{apruzzese2022sok}.

To address long-term malware concept drift adaptation under limited labeling budgets, we propose~\system, an SSL framework adapted from FixMatch~\cite{sohn2020fixmatch} and tailored for Android malware detection under distribution shift. 
While prior work applies FixMatch to malware~\cite{wu2023grim}, it primarily tackles label noise and data imbalance rather than concept drift. Moreover, FixMatch is originally designed for computer vision and performs poorly when directly applied to malware feature representations (see Table~\ref{tab:baseline_results}).
To address this gap, \system adapts FixMatch with domain-specific feature transformations---Bernoulli bit flip and Bernoulli feature mask, designed for structured binary malware features. 

In practice, malware evolution introduces adversary-driven changes in the feature space that are difficult to predict. Malware families may stop using certain API calls, permissions, or intent filters, while adopting others to realize the same behavior~\cite{api_graph_dataset}. Although such drift is often structured, to the best of our knowledge, prior works do not identify generalizable patterns that characterizes how static feature distributions evolve over time~\cite{mcfadden2025drmd, kaya2025ml,transcend,tesseract,yang2021cade}. As a result, we adopt a stochastic approach to modeling feature variation.



In particular, \system employs Bernoulli-based~\cite{ross2014introduction} feature augmentations to inject controlled noise into the feature space. 
In Bernoulli bit flip, a randomly selected subset of binary features is flipped, simulating behavioral changes where malware starts or stops using specific functionalities. In Bernoulli feature mask, a randomly selected subset of features is deactivated, modeling the disappearance of observable patterns such as deprecated API calls or obsolete permissions in newer malware variants. When applied with low perturbation probabilities, these transformations introduce mild variation without significantly altering class semantics, whereas excessive perturbation can distort feature representations and degrade detection performance (see Appendix~\ref{app:hyperparameter}). 
While static features cannot capture entirely new APIs introduced in the future~\cite{tesseract}, these feature-space augmentations act as noise injection that regularizes the model against feature instability, encouraging robustness to unpredictable distribution shift.

While feature-space augmentations improve robustness to evolving malware behavior, effective adaptation also depends on which samples are selected for labeling over time. 
Adapting to such evolving patterns therefore requires active learning (AL) to selectively query informative samples~\cite{chen2023continuous}. 
However, malware drift is often abrupt and unpredictable, limiting the effectiveness of selection strategies that assume gradual or ordered change. 
Under these conditions, many drifted samples lie near the decision boundary, where the model is most uncertain and prone to misclassification~\cite{ducoffe2018adversarial, vzliobaite2011active}. 
In addition, outlier samples often appear far from existing cluster in the latent space, further challenging accurate classification~\cite{yang2021cade}. 

To address these challenges, we propose a multi-criteria sample selection strategy that combines: (i) low-margin samples with small confidence gaps, (ii) samples with large $L_p$-norm distances indicating distributional deviation, and (iii) low-confidence predictions. These criteria focus labeling effort on samples that are both ambiguous and influential for refining the decision boundary under distribution shift.

Moreover, our analysis shows that samples near the decision boundary where benign and malware features overlap are particularly difficult to classify. 
To better handle such drifted and ambiguous cases, \system 
incorporates an auxiliary objective that promotes improved class separation, complementing the semi-supervised loss. Figure~\ref{fig:ssl-malware-system-design} provides an overview of the architecture of \system.
Our extensive evaluation shows that \system achieves strong performance across multiple datasets, with F1 scores of 93.5\% on APIGraph, 82.7\% on Chen-AZ, 44.9\% on MaMaDroid, and 77.7\% on LAMDA. In addition, \system improves training efficiency, achieving up to 24$\times$ faster training and requiring 13$\times$ fewer operations compared to prior methods.




\noindent In summary, the contributions of this work are as follows:

\begin{smitemize}
    \item We propose \system, a semi-supervised active learning (SSAL) framework for Android malware detection under long-term concept drift. \system introduces domain-specific feature augmentations, {\em Bernoulli bit flip} and {\em Bernoulli feature mask}, for structured malware features.
    
    \item To better discriminate ambiguous boundary samples, we integrate a supervised contrastive loss~\cite{khosla2020supervised} with the standard SSL objective, improving class separation of low-margin samples in the latent space. 

    \item We design a Multi-Criteria AL strategy that selects informative samples by jointly considering boundary margin, $L_p$-norm distance, and low-confidence predictions---enabling efficient adaptation under labeling budget constraints.
    
    \item We evaluate \system on four Android malware benchmarks---APIGraph~\cite{api_graph_dataset}, Chen-AZ~\cite{chen2023continuous}, MaMaDroid~\cite{mamadroid}, and LAMDA~\cite{haque2025lamda}---under both static and AL settings. \system consistently outperforms prior methods, achieving F1 gains of $+1.1\%$ on APIGraph, $+4.3\%$ on Chen-AZ, $+8.9\%$ on MaMaDroid, and $+13.2\%$ on LAMDA, while providing up to $24\times$ faster training and requiring $13\times$ fewer operations.

    \item We construct a new dataset based on MaMaDroid~\cite{mamadroid} features spanning 12 years, mirroring the temporal characteristics of LAMDA~\cite{haque2025lamda}. 

\end{smitemize}

%% file: 2_related_work.tex
\section{Related Work}

\paragraphB{Concept Drift Detection and Adaptation}
Concept drift is commonly detected with respect to the performance degradation of a trained model on future unseen data~\cite{tesseract, anoshift}. In malware detection, identifying concept drift is critical, as both malware and benign applications continuously evolve, leading to shifts in feature distributions. Prior work explored various strategies for concept drift adaptation. Several approaches leverage contrastive learning to identify distributional changes or out-of-distribution samples~\cite{chen2023continuous, yang2021cade, he2025combating}. For instance, CADE~\cite{yang2021cade} proposes a contrastive framework for detecting out-of-distribution malware samples, while Chen-AL~\cite{chen2023continuous} utilizes active learning with uncertainty-based sample selection to efficiently query labels from malware analysts under a limited budget. TRANSCENDENT~\cite{transcendent} employs conformal rejection to identify unreliable predictions and retrain the model to mitigate performance degradation. However, {\em most of these methods lack evaluation under long-term, large-scale, and highly pronounced drift scenarios, limiting their applicability to real-world malware evolution}.




\paragraphB{Semi-Supervised Learning (SSL)}
SSL techniques are significantly valuable where labeled data are scarce. Recent SSL techniques are widely recognized for their outstanding performance, such as FixMatch~\cite{sohn2020fixmatch}, FlexMatch~\cite{zhang2021flexmatch}, FreeMatch~\cite{wang2022freematch}, SoftMatch~\cite{chen2023softmatch}, and Dash~\cite{xu2023dash}, in computer vision. \cite{wu2023grim} is the first to utilize SSL technique like FixMatch for malware detection on noisy label dataset {\em but they did not address concept drift}. DeepReflect~\cite{downing2021deepreflect} integrates analyst feedback to minimize reverse engineering, while Kan et al.~\cite{kan2021investigating} demonstrate that pseudo-labeling helps classifiers adapt to new threats without frequent retraining. Apruzzese et al.~\cite{apruzzese2022sok} show that leveraging unlabeled data improves detection across diverse malware families. 



\paragraphB{Active Learning (AL) in Malware Detection}
AL reduces labeling costs while enabling malware detectors to adapt to evolving threats by querying only the most uncertain samples for expert annotation. This is particularly effective in malware detection, where labeling is expensive and malware behavior changes rapidly. Prior work, such as Chen-AL~\cite{chen2023continuous}, shows that uncertainty sampling improves detection performance, while APIGraph~\cite{api_graph_dataset} enhances robustness by merging semantically similar API calls. Other paradigms have also been explored such as Casandra~\cite{demontis2017casandra} applies online learning for continual adaptation but incurs high computational overhead, while DroidEvolver~\cite{nguyen2019droidevolver} uses self-training to reduce labeling effort at the risk of self-poisoning~\cite{demetrio2021explaining}.



\paragraphB{Curriculum Learning (CurL)}
CurL primarily focuses on structuring training from easy to hard examples to improve model stability and convergence. Zhou et al.~\cite{zhou2020curriculum} adjusts sample difficulty based on model uncertainty, allowing the curriculum to evolve during training. Prior work proposed CurL for malware analysis~\cite{shashidhar2016teaching}, combining theoretical foundations with practical exercises such as reverse engineering and dynamic analysis which is out of scope in this paper.

\paragraphB{Continual Learning (CL)}
Unlike AL, which reduces annotation cost by selecting informative samples, continual learning (CL) aims to maintain performance under evolving data distributions by mitigating catastrophic forgetting~\cite{parisi2019continual}. Rahman et al.~\cite{continual-learning-malware} showed that CL methods developed for computer vision perform poorly in the malware domain, motivating domain-specific approaches, while MADAR~\cite{madar} explores replay-based strategies for malware evolution. However, CL addresses a different problem setting and is out of the scope of this work.

%% file: 3_methodology.tex
\section{Preliminaries}

\subsection{Problem Setup and Notations}

\begin{figure*}[!t]
    \centering
    \includegraphics[width=\textwidth]{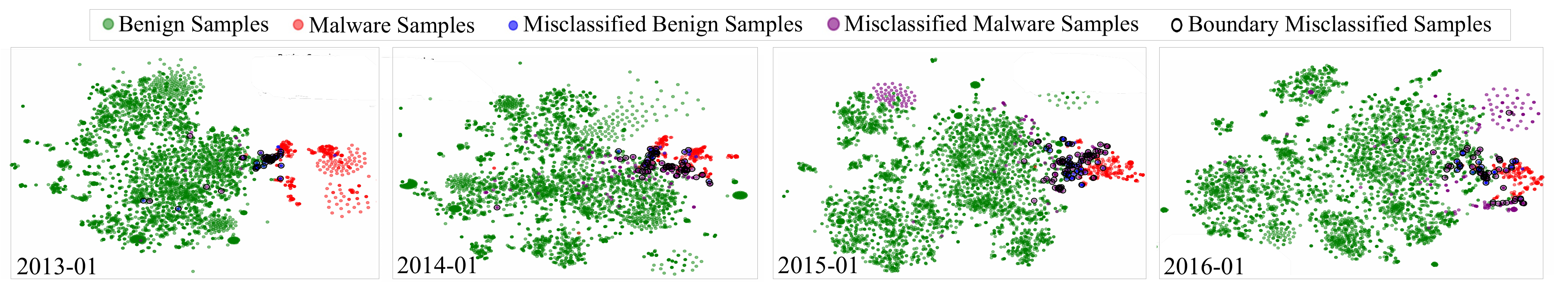}
    \caption{t-SNE projection of APIGraph latent representations from the MLP in \system across multiple time periods, illustrating the distribution of benign and malware samples along with boundary and misclassified samples.}
    \label{fig:samples-distribution-example}
\end{figure*}


We formulate our task as a \textit{semi-supervised malware detection problem under concept drift}, where the distribution of the malware and benign samples evolves over time. Let $\mathcal{X} \subset \{0, 1\}^d$ denote the input space of feature representations extracted from Android applications; 
and let $\mathcal{Y} = \{0, 1\}$ denote the label space, where 0 corresponds to benign software and 1 to malware. 
At each time step $t \in \{1, \dots, T\}$, the detector receives a small labeled dataset $\mathcal{D}_l^{(t)} = \{(x_i, y_i)\}_{i=1}^{n_t}$ and a large pool of unlabeled data $\mathcal{D}_u^{(t)} = \{x_j\}_{j=1}^{m_t}$, where $n_t \ll m_t$. The goal is to learn a malware detector $f_{\theta_t}: \mathcal{X} \rightarrow [0, 1]$ with parameters $\theta_t$ that estimates a score corresponding to 
$P(y = 1\mid x)$ under the current data distribution $\mathcal{P}_t(X, Y)$, while maintaining strong detection performance as $\mathcal{P}_t$ changes over time.

Concept drift arises from the emergence of new malware variants, evolving evasion techniques, and benign software mimicking malicious behavior. Such drift can affect the input distribution $\mathcal{P}_t(X)$ (covariate drift), the conditional distribution $\mathcal{P}_t(Y | X)$ (real concept drift), or both. We model concept drift as a temporal changes in the 
joint distribution $\mathcal{P}_t(X, Y)$ such that:
\begin{equation}
    \exists\ t_1, t_2 \in \{1, \dots, T\},\ \text{where}\ \mathcal{P}_{t_1}(X, Y) \neq \mathcal{P}_{t_2}(X, Y).
\end{equation}

Figure~\ref{fig:samples-distribution-example} illustrates this phenomena using 2d projections of latent space representations from APIGraph samples, spanned across several months. The visualization depicts increased overlap between malware (red) and benign (green) samples, along with our preference for outliers and boundary samples that are more sensitive to distribution shifts. In particular, boundary samples that are misclassified by the detector are captured using the softmax margin and are highlighted with black circles. The visualization is obtained by projecting penultimate-layer MLP embeddings into two dimensions using t-SNE with a perplexity of 30 and a fixed random seed (random\_state = 42) for reproducibility.





We argue that addressing evolving distributions requires a learning framework that supports:
(1) updating the model $f_{\theta_t}$ without full retraining,
(2) effective use of unlabeled data as domains shift,
and (3) efficient selection of informative samples when labeling budgets are constrained.


\subsection{Motivating Example}

A core challenge in malware detection is the continual emergence of new variants that introduce unseen and rare semantic features. 
Models trained on static, labeled datasets often struggle to generalize to these changes, resulting in degraded performance over time. 
Compounding this issue, producing high-quality labels for malware is both time-consuming and resource-intensive. Prior studies show that vendors may take weeks to reach consensus on newly discovered variants~\cite{zhu2020measuring, kantchelian2015better}. Moreover, malware analysts can manually review only about 80 samples per day~\cite{miller2016reviewer}, leaving the vast majority of malwares unlabeled. 

This imbalance creates creates a large pool of unlabeled samples that is unused by traditional detection pipelines~\cite{apruzzese2022sok}. Motivated by this observation, we explore semi-supervised learning (SSL) as a means to improve malware detection under evolving data distributions by jointly leveraging labeled data and abundant unlabeled samples. In particular, we investigate the FixMatch framework~\cite{sohn2020fixmatch}, which has demonstrated strong performance in settings where labeled data is scarce.



FixMatch combines pseudo-labeling with consistency regularization through weak and strong data augmentation techniques. For an unlabeled sample \( x \), the model first produces a pseudo-label \( \hat{y} \) by applying a weak augmentation \( \alpha(x) \).
This pseudo-label then supervises
the corresponding strongly augmented version \( A(x) \). The learning objective enforces consistency between the model's predictions for \( \alpha(x) \) and \( A(x) \).
Weak augmentations (e.g., horizontal flips and rotations) preserve the semantic content of the input and facilitate reliable pseudo-labeling. Strong augmentations (e.g., color jitter and cutout) introduce stochastic feature perturbations that can mask or distort key input features~\cite{li2024towards}. 
Despite this, the model is trained to maintain prediction consistency, thereby forcing it to generalize beyond specific feature dependencies.


In the context of malware detection, this mechanism is well suited to learning from features introduced by distribution shifts. 
Let \( x_i \) be a malware sample containing both a known semantic feature \( v_{i,1} \) and a novel, unseen feature \( v_{i,2} \). The weak augmentation \( \alpha(x_i) \) retains both features, enabling the model to generate a high-confidence pseudo-label \( \hat{y}_i \) based on \( v_{i,1} \). If the strong augmentation \( A(x_i) \) masks \( v_{i,1} \) via stochastic feature perturbation, the detector must rely on \( v_{i,2} \) to maintain consistency with \( \hat{y}_i \), thereby incorporating \( v_{i,2} \) into the learned decision boundary. 
This implicit adaptation supports drift learning by progressively integrating novel features without explicit annotation. 
Prior work shows that such training supports better alignment between labeled and unlabeled samples under distribution drift~\cite{li2024towards}.


In our setting, FixMatch leverages this behavior by combining a supervised loss on labeled samples with an unsupervised consistency loss on confidently pseudo-labeled data.
The training dataset \(\mathcal{D}\) at step \circlednum{1} in the Figure~\ref{fig:ssl-malware-system-design} is divided into two sets labeled, \(\mathcal{D}_l\) at step \circlednum{2} and unlabeled \(\mathcal{D}_u\) at step \circlednum{3}. Let \( \mathcal{D}_l = \{(x_i, y_i)\}_{i=1}^{N_l} \) denote the labeled dataset and \( \mathcal{D}_u = \{x_j\}_{j=1}^{N_u} \) the unlabeled dataset. The learning objective is:
\begin{equation}
\mathcal{L} = \mathcal{L}_{\text{sup}} + \lambda_u \cdot \mathcal{L}_{\text{unsup}}
\label{eq:ssl_loss}
\end{equation}
where \( \mathcal{L}_{\text{sup}} \) is the standard cross-entropy (CE) loss on labeled samples.



\begin{equation}
\mathcal{L}_{\text{sup}} = \frac{1}{N_l} \sum_{i=1}^{N_l} \text{CE}(f_\theta(x_i), y_i)
\end{equation}

and \( \mathcal{L}_{\text{unsup}} \) is the unsupervised loss enforcing consistency on strongly augmented unlabeled inputs, conditioned on high-confidence pseudo-labels:

\begin{equation}
\mathcal{L}_{\text{unsup}} = \frac{1}{N_u} \sum_{j=1}^{N_u} \mathbf{1}_{[\max(p_j) \geq \tau]} \cdot \text{CE}(f_\theta(A(x_j)), \hat{y}_j)
\end{equation}


Here, \( f_\theta \) denotes the malware detector parameterized by \( \theta \), \( A(x_j) \) is the strongly augmented view of the unlabeled input \( x_j \), and \( p_j \) is the predicted softmax probability from the weakly augmented version used to obtain pseudo-label \( \hat{y}_j = \arg\max(p_j) \). The indicator function \( \mathbf{1}_{[\max(p_j) \geq \tau]} \) filters pseudo-labels based on a confidence threshold \( \tau \). The hyperparameter \( \lambda_u \) controls the influence of the unsupervised loss. In our experiments, we set \( \tau = 0.95 \) and \( \lambda_u = 1 \), following best practices from prior work~\cite{sohn2020fixmatch}. This high confidence threshold filters out noisy samples, preventing performance degradation even under significant drift.

\section{Overview of \system}



\begin{figure}[!t]
    \centering
    \includegraphics[width=0.475\textwidth]{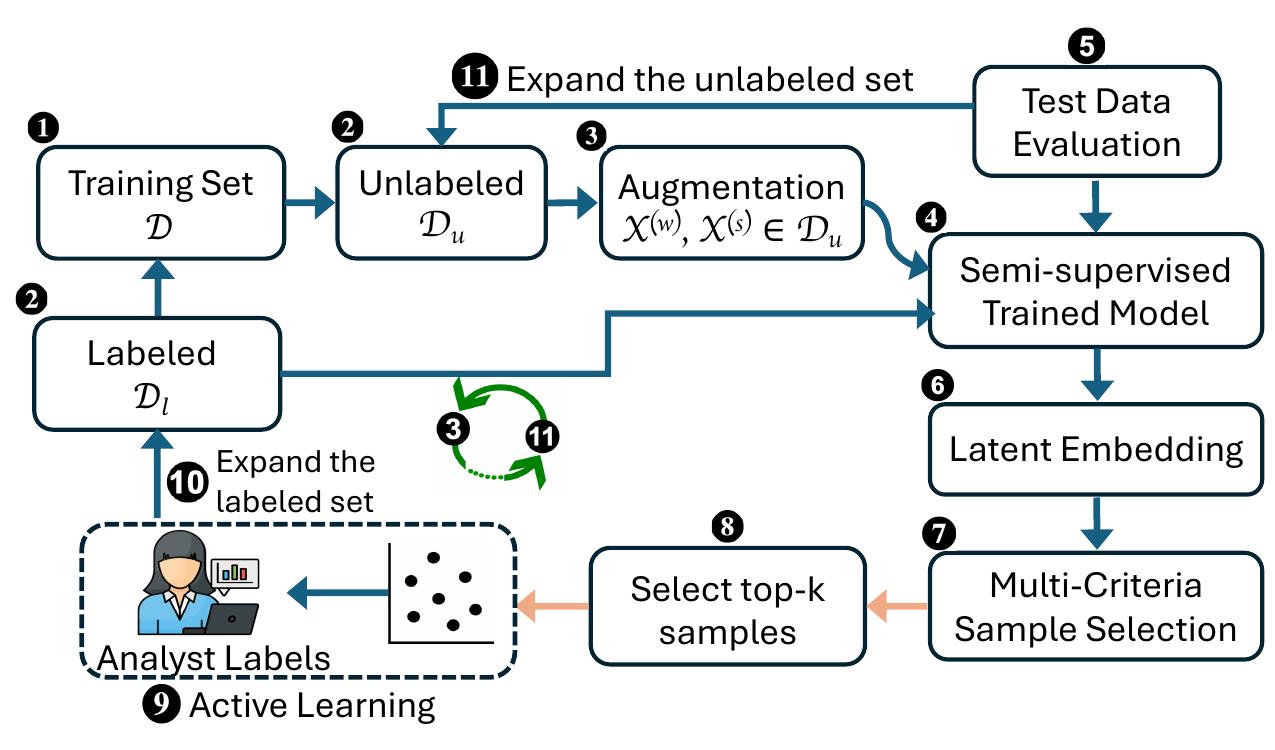}
    \caption{Overview of CITADEL for concept drift adaption with semi-supervised + active learning.}
    \label{fig:ssl-malware-system-design}
\end{figure}

In this section, we introduce, \system, a semi-supervised active learning framework for adapting to concept drift in malware detection. The overall workflow is outlined in Algorithm~\ref{algo:ssl} 
, Algorithm~\ref{algo:mc-selector}, and Algorithm~\ref{algo:active-learning}, and illustrated in Figure~\ref{fig:ssl-malware-system-design}.

To support robust learning under evolving feature distributions, \system applies two domain-specific augmentations---Bernoulli bit flip and feature masking--- to unlabeled data. As shown in Figure~\ref{fig:ssl-malware-system-design}, the unlabeled dataset \(\mathcal{D}_u\) is divided into a weekly perturbed set \(\mathcal{X}^{(w)}\) and a strongly perturbed set \(\mathcal{X}^{(s)}\). These perturbed variants are incorporated into a unified semi-supervised training objective (in step~\circlednum{4}) that combines supervised contrastive loss with a standard semi-supervised loss, improving class separation for samples affected by distribution shift. To support adaptation under limited labeling budgets, \system incorporates a multi-criteria sample selection strategy in step~\circlednum{7} that guides the active learning process in step~\circlednum{9}, based on the budget defined in step~\circlednum{8}. This process repeats as new data becomes available, enabling the detector to progressively adapt to concept drift over time.

\subsection{Augmentations for Malware Features}

\system is built on FixMatch~\cite{sohn2020fixmatch}, which is originally developed for the computer vision tasks where augmentations such as rotation and color distortion preserve semantic meaning. Malware detection, however, operates over structured feature representations with strict semantic constraints~\cite{continual-learning-malware}. 
In the case of Android malware detection, features are commonly represented as binary vectors derived from static analysis, where each dimension indicates the presence or absence of specific behavioral attributes, permissions, or API calls~\cite{arp2014drebin}. These features encode meaningful structural and behavioral properties of malware samples, making image-based augmentations inappropriate in this domain. As a result, effective SSL in this domain requires features perturbations that respect the underlying semantics of malware samples.

\input{algo/1-fix-match-andours}

\begin{figure}[t]
    \centering
    \begin{subfigure}[b]{0.485\linewidth}
        \centering
        \includegraphics[width=\linewidth]{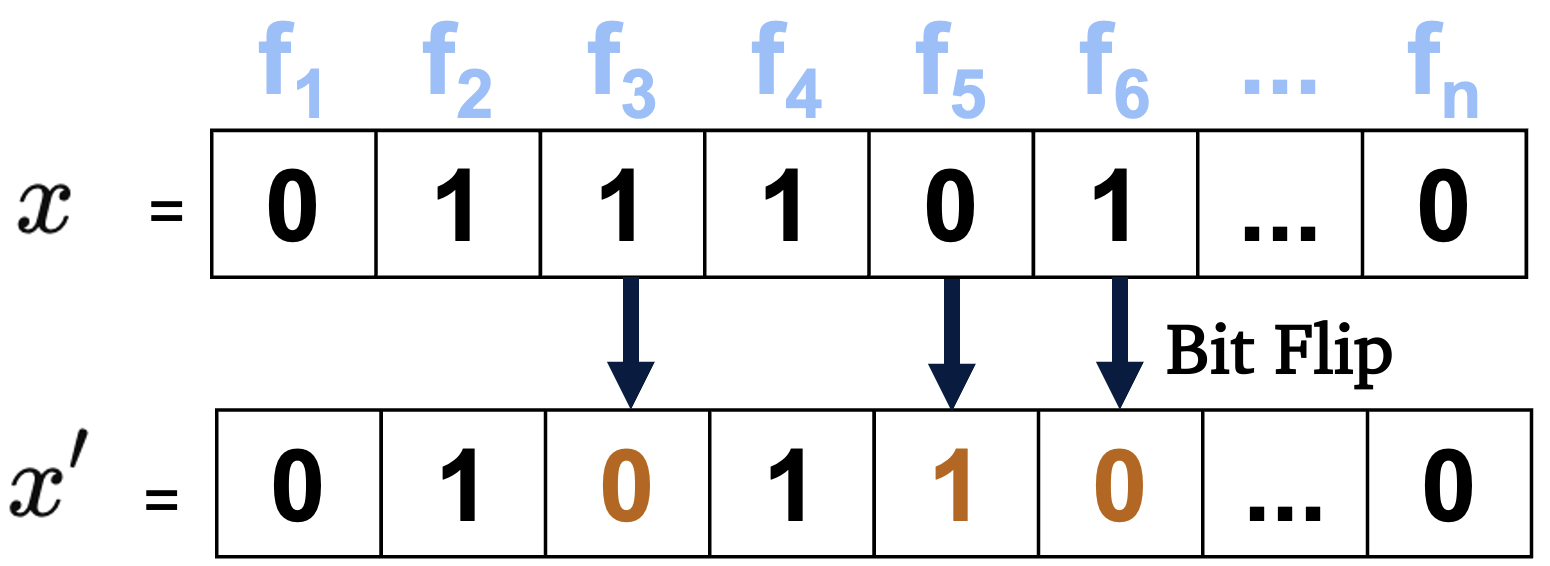}
        \caption{Bernoulli Bit Flip.}
        \label{fig:bitflip}
    \end{subfigure}
    \hfill
    \begin{subfigure}[b]{0.485\linewidth}
        \centering
        \includegraphics[width=\linewidth]{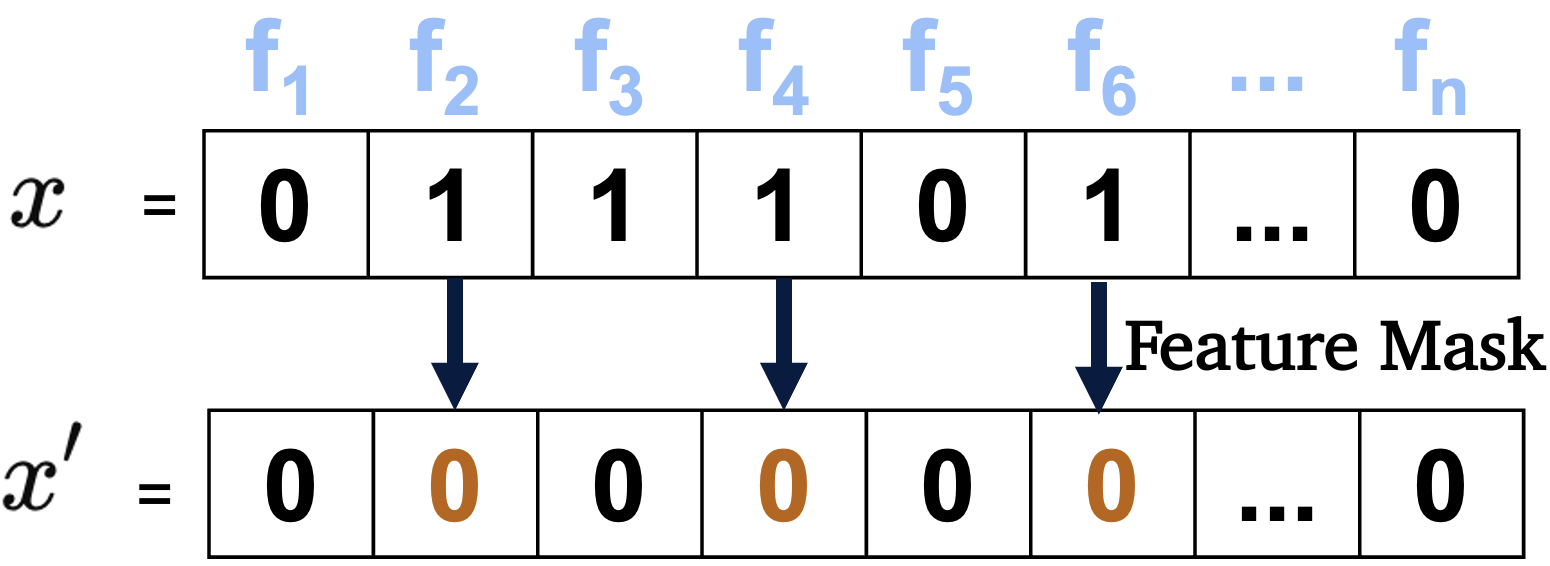}
        \caption{Bernoulli Feature Mask.}
        \label{fig:featuremask}
    \end{subfigure}
    \caption{Augmentation strategies used in \system.
    }
    \label{fig:bernoulli_augmentations}
\end{figure}

To address this challenge, we propose two augmentation strategies tailored for static binary feature vectors—\textit{Bernoulli bit flip} and \textit{Bernoulli feature mask}, as shown in Algorithm~\ref{algo:ssl} lines 6–7 and Figure~\ref{fig:bernoulli_augmentations}. These augmentations are designed to model feature-level variation induced by concept drift. In Android malware, concept drift often manifests as changes in attack strategies rather than objectives. For example, an early variant may abuse SMS permissions to exfiltrate sensitive data via text messages, while a later variant achieves the same objective by encrypted HTTPS requests, activating different static features. Since such drifts do not follow a deterministic pattern~\cite{tesseract}, we adopt a randomized perturbation strategy to inject controlled noise into the feature space.

In particular, we use Bernoulli distribution to perturb features since it enables controlled but probabilistic perturbation. Lower probabilities result in mild perturbations that preserve most features, while high probabilities induce stronger variation. This design enables consistency-based regularization within the feature space, encouraging the classifier to rely on features that remain predictive under variation and supporting adaptation to evolving malware behavior.


Prior work such as MORSE~\cite{wu2023grim} also applies FixMatch to malware analysis but focuses on handling label noise rather than concept drift. In contrast, our augmentations are explicitly designed to model feature changes arising from evolving attack strategies.

\paragraphB{Bernoulli Bit Flip} 
This augmentation models the feature-level variation across malware variants by probabilistically flipping bits in the binary feature vector. In this representation, a value of 1 indicates the presence of a specific API call or permission, while 0 indicates its absence. Given an input feature vector \( x \in \{0, 1\}^d \), we apply independent Bernoulli noise to each feature to control the extent of perturbation. Specifically, we sample a binary mask \(\mathbf{m} \in \{0,1\}^d\) such that each entry is drawn independently from a Bernoulli distribution with parameter \(p\). The perturbed input is obtained via an element-wise \texttt{XOR} operation: 

\begin{equation}
x' = x \oplus m, \quad \text{where} \quad m \sim \text{Bernoulli}(p)
\end{equation}
where \( \oplus \) denotes the element-wise \texttt{XOR}. Small values of \(p\) introduce subtle perturbations that preserve most features, while larger values of \(p\) causes  substantial perturbations that may degrade semantic coherence.

\paragraphB{Bernoulli Feature Mask} 
In this augmentation, we intend to model scenarios in which observable behaviors are suppressed due to obfuscation or changes in implementation by randomly zeroing out a subset of the features, effectively removing features from the sample. 
Given an input vector \(x\), we sample a binary mask \(m \sim \text{Bernoulli}(q) \) and compute the masked input as:
\begin{equation}
x' = x \odot m, \quad m \sim \text{Bernoulli}(q)
\end{equation}
where $\odot$ denotes element-wise multiplication.

\paragraphB{Weak and Strong Augmentation} Within the \system framework, weak and strong augmentation serve distinct roles. 
Weak augmentation applies a low perturbation probability preserving class semantics of a sample and is used to generate confident pseudo-labels \( \hat{y}_j\) (shown in Algorithm~\ref{algo:ssl} line: 8). Strong augmentation apples a higher perturbation probability to introduce greater variation during training.

To evaluate the impact of each strategy, we explore the effect of \textit{Bernoulli bit flip} and \textit{Bernoulli feature mask} independently on model generalization and concept drift adaptation. 
For \textit{Bernoulli bit flip}, we apply weak (\(p_w\)) and strong (\(p_s\)) perturbations using flip probabilities \(p_w = 0.01\) and \(p_s = 0.05\), respectively. For \textit{Bernoulli feature mask}, we use the same weak and strong masking probabilities, \(q_w = 0.01\) and \(q_s = 0.05\), respectively. These values are selected based on empirical evaluation (see Appendix~\ref{app:hyperparameter}), as they provide the best performance while minimizing label distortion and preventing model performance degradation.

\input{algo/AL-SSL}

\subsection{\system's Objective Function}

We propose an improved objective function incorporating supervised contrastive loss, $\mathcal{L}_{\text{con}}$, computed using the labeled data $\mathcal{D}_l$ with $N_l$ labeled samples along with EQ~\ref{eq:ssl_loss} to better distinguish boundary samples and realize effective adaptation to concept drift. This enforces the model to pull together feature representations of samples belonging to the same class while pushing apart those from different classes.

The total loss, \(\mathcal{L}\) (shown in Algorithm~\ref{algo:ssl} line: 15), is defined as a combination of supervised loss $\mathcal{L}_{\text{sup}}$ and unsupervised consistency loss $\mathcal{L}_{\text{unsup}}$ based on FixMatch, along with the supervised contrastive loss $\mathcal{L}_{\text{con}}$:

\begin{equation}
\mathcal{L} = \mathcal{L}_{\text{sup}} + \lambda_u \cdot \mathcal{L}_{\text{unsup}} + \lambda_{\text{con}} \cdot \mathcal{L}_{\text{con}}
\label{eq:total_loss}
\end{equation}

where $\lambda_u$ and $\lambda_{\text{con}}$ are weighting hyperparameters for the unsupervised and contrastive terms, respectively.
All hyperparameter exploration to select the optimal values of $\lambda_u$, $\lambda_{\text{con}}$ are provided in Appendix~\ref{app:hyperparameter}.

The supervised contrastive loss $\mathcal{L}_{\text{con}}$ follows the formulation on~\cite{khosla2020supervised}. Given a batch of encoded feature vectors $\mathbf{z}_i$ and their corresponding labels $y_i$, the loss is defined as:

{\small 
\begin{equation}
\mathcal{L}_{\text{con}} = -\frac{1}{N_l} \sum_{i=1}^{N} \frac{1}{|P(i)|} \sum_{p \in P(i)} \log \left( 
\frac{\exp\left(\mathbf{z}_i \cdot \mathbf{z}_p / \tau\right)}
{\sum\limits_{a=1}^{N} 1_{[a \neq i]} \exp\left(\mathbf{z}_i \cdot \mathbf{z}_a / \tau\right)}
\right)
\end{equation}}

where $\tau$ is the temperature scaling parameter that regulates the concentration of the distribution, $P(i)$ is the set of indices of positive samples (same label as $i$) excluding $i$ itself. The dot product $\mathbf{z}_i \cdot \mathbf{z}_p$ measures similarity between the features of anchor and positive samples in the representation space. This loss formulation facilitates intra-class compactness (i.e., pulling together samples of the sample class) and inter-class separability (i.e., pushing apart samples of different class) in the representation space. These properties are particularly advantageous in the presence of concept drift where samples near the decision boundary may become ambiguous due to evolving semantic features.  

While our semi-supervised framework effectively leverages unlabeled data through pseudo-labeling, it remains vulnerable to real-world drift, where certain samples may be uncertain or misclassified, or poorly represented in the feature space. To further improve adaptability and sample efficiency in such conditions, we integrate an active learning mechanism that selectively queries labels for the most informative and uncertain samples, thereby guiding the model toward better generalization with minimal labeling overhead.

\subsection{Multi-Criteria Sample Selection}

Active learning offers a promising approach to reducing labeling costs in malware detection~\cite{chen2023continuous}, particularly in the presence of concept drift. 
Prior work utilizes Mean Absolute Deviation (MAD) to find the distance from its class center using contrastive classifier and applying threshold to find the out-of-distribution (OOD) samples~\cite{yang2021cade}, Chen-AL~\cite{chen2023continuous} uses complex local pseudo loss for selecting uncertain samples; however, both fail to adapt longitudinal and pronounced drift scenario such as in LAMDA~\cite{haque2025lamda}. We empirically observe that no single sampling criterion reliably captures all informative instances under such dynamics. To address this, we introduce a multi-criteria sampling strategy (Algorithm~\ref{algo:mc-selector}) based on three complementary indicators: (1) \textit{boundary proximity}, via the softmax margin; (2) \textit{feature displacement},via the $L_p$-norm in embedding space; and (3) \textit{prediction uncertainty}, via low-confidence outputs. We hypothesize that their combination enables more robust and stable adaptation in the face of continual malware evolution. Algorithm~\ref{algo:active-learning} presents the detailed procedure of active learning method.


\paragraphB{Softmax Margin for Boundary Samples} 
Through softmax prediction margin of the classifier reflects the proximity of a sample to the decision boundary. For an unlabeled input \( x \), let \( p(x) \) denote the classifier’s softmax output vector. The softmax margin is defined as the difference between the top two predicted class probabilities:
\begin{align}
\mathcal{M}(x) &= p_{1}(x) - p_{2}(x)
\label{eq:margin_function}
\end{align}

where \( p_{1}(x) \) and \( p_{2}(x) \) are the highest and second-highest values in \( p(x) \), respectively. A smaller margin indicates greater uncertainty and typically corresponds to samples near the model's decision boundary. Such samples are more likely to be informative when labeled. Margin-based uncertainty sampling is a well-established strategy in the active learning literature~\cite{gal2017deep}, and has demonstrated robustness even under distributional shift~\cite{ash2019deep}.



\input{algo/Multi-Criteria}

\paragraphB{$L_p$-Norm for Feature-Space Drift} While the softmax margin captures local uncertainty near the classifier's decision boundary, it may fail to identify outliers or samples that are semantically distant from the training distribution—a common scenario under concept drift. To address this, we compute the \( L_p \)-norm distance between each unlabeled sample and its nearest labeled neighbor in the classifier’s learned feature space.

Formally, let \( f(x) \) denote the embedding of a sample \( x \) obtained from the model's penultimate layer. For each unlabeled sample \( x_i \in \mathcal{U} \) and each labeled sample \( x_j \in \mathcal{L} \), we compute:

\begin{equation}
\mathcal{D}_{\text{lp}}(x_i) = \min_{x_j \in \mathcal{L}} \left\| f(x_i) - f(x_j) \right\|_p
\label{eq:lp_distance}
\end{equation}

This minimum \( L_p \)-distance quantifies the proximity of an unlabeled sample to the closest known (labeled) point in the embedding space. Samples with larger distances are more likely to be unfamiliar or semantically novel, making them strong candidates for labeling under drift.

To select an appropriate \( p \) value, we empirically evaluated the active learning performance for \( p \in [1.0, 9.0] \) in increments of 0.1. Our results showed only minor performance variation across values. We ultimately select \( p = 2 \) due to its consistent performance across multiple settings and its theoretical grounding in high-dimensional vector spaces~\cite{ busemann2005geometry}.

\paragraphB{Low-Confidence Predictions} The third sampling criterion relies on the model's raw predictive confidence, defined as the maximum probability assigned to any class:
\begin{equation}
\mathcal{C}(x) = \max \left( p(x) \right)
\label{eq:confidence_score}
\end{equation}
Samples with low confidence often correspond to ambiguous instances or previously unseen malware classes. Prior work has demonstrated that such samples are strongly correlated with misclassification~\cite{hendrycks2017baseline} and can be effectively utilized for active learning under distributional shift~\cite{wang2022freematch, zhang2021flexmatch}. In the context of malware detection, this criterion enables the model to prioritize uncertain or suspicious samples that may not be captured by margin-based or distance-based strategies alone.








\paragraphB{Sample Selection Strategy} 
To select the uncertain samples for labeling, we combine three complementary uncertainty criteria: softmax margin (\(\mathcal{M}\)), feature-space distance (\(\mathcal{D}_{\text{lp}}\)), and prediction confidence (\(\mathcal{C}\)). For each unlabeled sample, we first compute all three scores. Since these scores have different ranges and interpretations, we apply min--max normalization to each criterion so that all values lie in the range \([0,1]\).

We then align the direction of the scores such that higher values consistently indicate higher uncertainty. Specifically, samples with smaller softmax margins and lower prediction confidence are more uncertain, so we invert their normalized values using \(1-\hat{\mathcal{M}}(x)\) and \(1-\hat{\mathcal{C}}(x)\), respectively. After this alignment, we combine the three uncertainty signals by summation to obtain a unified uncertainty score for each unlabeled sample and select the top-\(k\) samples with the highest scores for labeling. The unified uncertainty score for an unlabeled sample \(x\) is computed as:
\begin{equation}
\text{score}(x) = \left(1 - \hat{\mathcal{M}}(x)\right)
+ \hat{\mathcal{D}}_{\text{lp}}(x)
+ \left(1 - \hat{\mathcal{C}}(x)\right).
\label{eq:hybrid_score}
\end{equation}

Here, \( \hat{\cdot} \) denotes the normalized value. This multi-criteria strategy does not rely on a single uncertainty signal. Instead, it jointly considers boundary ambiguity, feature-space novelty, and low-confidence predictions, which is particularly important under concept drift where predictive confidence alone can be misleading due to shifts in the decision boundary~\cite{transcend}.


This formulation assigns higher selection scores to samples with low prediction margins, high feature-space distance, and low model confidence—each of which signals potential informativeness under concept drift. The top-\( k \) samples with the highest scores are selected for labeling.

The sample selection and retraining procedures are detailed in Algorithm~\ref{algo:active-learning}, lines 10–13, which invoke the semi-supervised training routine in Algorithm~\ref{algo:ssl}. Compared to prior active learning approaches in malware detection that rely on a single uncertainty metric~\cite{yang2021cade, chen2023continuous}, our multi-criteria strategy enables more robust adaptation to drift by jointly targeting decision boundary ambiguity and semantic outliers in the learned feature space.

%% file: algo/1-fix-match-andours.tex
\begin{algorithm}[!t]
\caption{Semi-Supervised Initial Training without Concept Drift Adaptation}
\label{algo:ssl}
\begin{algorithmic}[1]
\Require Labeled set $\mathcal{D}_l$, unlabeled set $\mathcal{D}_u$, model $f_\theta$, threshold $\tau$, weights $\lambda_u$, $\lambda_{\text{con}}$
\For{$e = 1$ to $E$}
    \State Sample minibatches $\mathcal{B}_l \subset \mathcal{D}_l$, $\mathcal{B}_u \subset \mathcal{D}_u$
    \State $\mathcal{L}_{\text{sup}} \gets \frac{1}{|\mathcal{B}_l|} \sum_{(x_i, y_i) \in \mathcal{B}_l} \text{CE}(f_\theta(\text{WeakFlip}(x_i, p=0.01)), y_i)$
    \State $\mathcal{L}_{\text{unsup}} \gets 0$
    \ForAll{$x_j \in \mathcal{B}_u$}
        \State $x_j^{(w)} \gets \text{WeakFlip}(x_j, p=0.01)$
        \State $x_j^{(s)} \gets \text{StrongFlip}(x_j, p=0.05)$
        \State $p_j \gets f_\theta(x_j^{(w)}), \quad \hat{y}_j \gets \arg\max(p_j)$
        \If{$\max(p_j) \geq \tau$}
            \State $\mathcal{L}_{\text{unsup}} \gets \mathcal{L}_{\text{unsup}} + \text{CE}(f_\theta(x_j^{(s)}), \hat{y}_j)$
        \EndIf
    \EndFor
    \State Encode $\mathbf{z}_i \gets \text{Enc}(x_i)$ for all $x_i \in \mathcal{B}_l$
    \State Compute $\mathcal{L}_{\text{con}}$ using supervised contrastive loss
    \State $\mathcal{L} \gets \mathcal{L}_{\text{sup}} + \lambda_u \cdot \mathcal{L}_{\text{unsup}} + \lambda_{\text{con}} \cdot \mathcal{L}_{\text{con}}$
    \State Update model parameters using $\nabla_\theta \mathcal{L}$
\EndFor
\State \Return $f_\theta$
\end{algorithmic}
\end{algorithm}

%% file: algo/AL-SSL.tex
\begin{algorithm}[!t]
\caption{Model Retraining with Active Learning Using Multi-Criteria Sample Selection}
\label{algo:active-learning}
\begin{algorithmic}[1]
\Require Initial labeled set $\mathcal{D}_l$, unlabeled pool $\mathcal{D}_u$, data stream $\{\mathcal{D}_{\text{stream}}^{(m)}\}_{m=1}^M$ over $M$ months, initially trained model $f_\theta$, selection budget $k$, norm $lp$
\For{$m = 1$ to $M$}
    \State $\mathcal{D}_u^{(m)} \gets \mathcal{D}_{\text{stream}}^{(m)}$
    \State $(\mathcal{Q}_u, \mathcal{Q}_l) \gets \textsc{MultiCriteria}(\mathcal{D}_l, \mathcal{D}_u^{(m)}, f_\theta, k, p)$
    \State $\mathcal{D}_l \gets \mathcal{D}_l \cup \mathcal{Q}_l$; \quad $\mathcal{D}_u \gets \mathcal{D}_u^{(m)} \setminus \mathcal{Q}_u$

    \State Retrain model $f_\theta$ using Algorithm~\ref{algo:ssl} on $(\mathcal{D}_l, \mathcal{D}_u)$
\EndFor
\State \Return $f_\theta$
\end{algorithmic}
\end{algorithm}


%% file: algo/Multi-Criteria.tex
\begin{algorithm}[!t]
\caption{Multi-Criteria Sample Selector}
\label{algo:mc-selector}
\begin{algorithmic}[1]
\Require Labeled set $\mathcal{D}_l$, unlabeled data $\mathcal{D}_u$, model $f_\theta$, budget $k$, norm $l_p$
\Ensure Selected unlabeled set $\mathcal{Q}_u$ and labeled set $\mathcal{Q}_l$
\ForAll{$x_j \in \mathcal{D}_u$}
    \State Compute softmax output $p(x_j)$ using $f_\theta$
    \State $\mathcal{M}(x_j) \gets p_1(x_j) - p_2(x_j)$
    \State $\mathcal{D}_{\text{lp}}(x_j) \gets \min\limits_{(x_i,y_i)\in \mathcal{D}_l} \|\text{Enc}(x_j) - \text{Enc}(x_i)\|_p$
    \State $\mathcal{C}(x_j) \gets \max(p(x_j))$
\EndFor
\State Min--max normalize over $\mathcal{D}_u$: $\hat{\mathcal{M}}, \hat{\mathcal{D}}_{\text{lp}}, \hat{\mathcal{C}}$
\ForAll{$x_j \in \mathcal{D}_u$}
    \State $\text{score}(x_j) \gets (1-\hat{\mathcal{M}}(x_j)) + \hat{\mathcal{D}}_{\text{lp}}(x_j) + (1-\hat{\mathcal{C}}(x_j))$
\EndFor
\State $\mathcal{Q} \gets$ top-$k$ samples in $\mathcal{D}_u$ by $\text{score}(\cdot)$
\State Label $\mathcal{Q}$ via human analysts to obtain $\mathcal{Q}_l$
\State $\mathcal{Q}_u \gets \mathcal{D}_u \setminus \mathcal{Q}_l$
\Return $\mathcal{Q}_u, \mathcal{Q}_l$    
\end{algorithmic}
\end{algorithm}

%% file: 4_experimental_setup.tex
\input{tables/dataset}

%% file: tables/dataset.tex
\begin{table}[!t]
\scriptsize
\setlength{\tabcolsep}{3pt}
\centering
\caption{Year-wise distribution of Benign (B) and Malware (M) samples for APIGraph, Chen-AZ, LAMDA, and MaMaDroid datasets.}
\begin{tabular}{c|cc|cc|cc|cc}
\toprule
\multirow{2}{*}{\textbf{Year}} 
 & \multicolumn{2}{c|}{\textbf{APIGraph}~\cite{api_graph_dataset}} 
 & \multicolumn{2}{c|}{\textbf{Chen-AZ}~\cite{chen2023continuous}} 
 & \multicolumn{2}{c|}{\textbf{LAMDA}~\cite{haque2025lamda}} 
 & \multicolumn{2}{c}{\textbf{MaMaDroid}~\cite{mamadroid}} \\ 
\cline{2-9}
 & \textbf{B} & \textbf{M} 
 & \textbf{B} & \textbf{M} 
 & \textbf{B} & \textbf{M}
 & \textbf{B} & \textbf{M} \\ 
\midrule
2012 & 27,613 & 3,066 & - & - & - & - & - & - \\
2013 & 43,873 & 4,871 & - & - & 42,048 & 44,383 & 9,000 & 1,000 \\
2014 & 52,843 & 5,871 & - & - & 55,427 & 45,756 & 9,000 & 1,000 \\
2015 & 52,173 & 5,797 & - & - & - & - & - & - \\
2016 & 50,859 & 5,651 & - & - & 64,059 & 45,134 & 9,000 & 1,000 \\
2017 & 24,930 & 2,620 & - & - & 77,785 & 21,359 & 9,000 & 1,000 \\
2018 & 38,214 & 4,213 & - & - & 64,942 & 39,350 & 9,000 & 1,000 \\
2019 & - & - & 40,947 & 4,542 & 49,465 & 41,585 & 9,000 & 1,000 \\
2020 & - & - & 34,921 & 3,982 & 55,718 & 46,355 & 9,000 & 1,000 \\
2021 & - & - & 13,985 & 1,676 & 45,528 & 35,627 & 9,000 & 1,000 \\
2022 & - & - & - & - & 44,768 & 41,648 & 9,000 & 1,000 \\
2023 & - & - & - & - & 46,462 & 7,892 & 9,000 & 1,000 \\
2024 & - & - & - & - & 47,633 & 794 & 9,000 & 674 \\
2025 & - & - & - & - & 44,640 & 23 & 9,000 & 18 \\
\midrule
\textbf{Total} 
 & 290,505 & 32,089 
 & 89,853 & 10,200 
 & 638,475 & 369,906 
 & 108,000 & 10,692 \\ 
\bottomrule
\end{tabular}
\label{tab:yearly-distribution-all}
\end{table}

%% file: 5_experimental_results_and_evaluation.tex
\section{Evaluation}


In this section, we evaluate \system across multiple settings to assess its effectiveness under concept drift and limited labeling budgets. We first describe the datasets in Section~\ref{sec:datasets}. then analyzes the behavior of the \system without active learning under varying labeled ratios and feature augmentations in Section~\ref{sec:common-setup}. Section~\ref{sec:baselineAL} compares \system against baseline methods without active learning (AL), while Section~\ref{sec:ALResults} extends this comparison to the AL framework. Section~\ref{sec:compcomplex} reports the computational complexity. 


Additional analyses are provided in the Appendix, including dataset composition statistics (Appendix~\ref{app:singleton}), visual analysis comparing CITADEL and Chen-AL performance (Appendix~\ref{app:CITADEL_Chen_visual_comp}), hyperparameter sensitivity and augmentation analysis (Appendix~\ref{app:hyperparameter}), robustness to label noise (Appendix~\ref{app:label_noise}), curriculum learning integration (Appendix~\ref{app:curriculum_learning}), results on additional feature sets such as EMBER (Appendix~\ref{app:exp_other_feature}), and details of computational resources (Appendix~\ref{app:computational_resources}).


\input{tables/ssl_baselines}

\subsection{Datasets}
\label{sec:datasets}


We evaluate \system on four Android malware datasets that span different time periods and features: (i) APIGraph~\cite{api_graph_dataset}, (ii) Chen-AZ~\cite{chen2023continuous}, (iii) LAMDA~\cite{haque2025lamda}, and (iv) MaMaDroid~\cite{mamadroid}. \textbf{APIGraph} spans 2012 to 2018 and contains 32,089 malware samples from 1,120 malware families and  290,505 benign samples~\cite{api_graph_dataset}. \textbf{Chen-AZ} derived from AndroZoo~\cite{androzoo} and is used by Chen-AL~\cite{chen2023continuous}, spans 2019 to 2021 and includes 10,200 malware samples across 254 families and 89,853 benign samples. \textbf{LAMDA} is the large-scale longitudinal dataset explicitly designed for studying concept drift. It spans 2013–2025 (excluding 2015) and contains 1,008,381 APKs, of which 369,906 are malware and 638,475 are benign, covering 1,380 malware families~\cite{haque2025lamda}. We further construct a \textbf{MaMaDroid} dataset by processing APKs from LAMDA using  MaMaDroid’s feature representation~\cite{mamadroid}. The resulting dataset contains approximately 10,000 samples per year (9000 benign and 1000 malware), except for 2024 and 2025 where fewer samples are available. Table~\ref{tab:yearly-distribution-all} summarizes the yearly counts of benign (B) malware (M) samples across all datasets. Appendix~\ref{app:singleton} reports detailed statistics. 

\input{tables/baseline_comparison_table}

\subsection{Label Ratios and Feature Augmentations}
\label{sec:common-setup}

We evaluate \system under varying supervision levels using multiple label ratios (10\%–90\%) with our proposed augmentations: random bit flip, random bit flip using Bernoulli, random feature masking, and a combined bit flip and masking strategy. 

\paragraphB{Experimental Setup}
Following the protocol from prior work~\cite{chen2023continuous}, we adopt the similar multilayer perceptron (MLP) architecture across all experiments to ensure a fair, model-agnostic comparison. 
For APIGraph, we use 2012 for training, Jan–Jun 2013 for validation, and Jul 2013–Dec 2018 for testing. For Chen-AZ, we use 2019 for training, Jan–Jun 2020 for validation, and Jul 2020–Dec 2021 for testing. LAMDA provides a predefined split--2013 for training, Jan–Jun 2014 for validation, and Jul 2014–Jan 2025 for testing. For MaMaDroid, we adopt the same temporal split as LAMDA for training, validation, and testing. 
Model performance is reported as the average over all test months. 
For each label ratio, the training data is randomly divided into labeled and unlabeled subsets, simulating different levels of supervision. 

\paragraphB{Results} Our results show that the proposed augmentation strategies yield promising performance across varying label ratios. As expected, F1 scores generally improve with increased supervision; however, concept drift still constrains performance, particularly in datasets with severe distributional shifts (see Table~\ref{tab:ssl_results}). Among the four datasets, LAMDA exhibits the most significant drift. 
For instance, with 40\% labeled data and Bernoulli bit flip augmentation, it achieves around 30\% F1 with a false negative rate (FNR) of about 75\%. 
For MaMaDroid under the same labeled ratio with Bernoulli masking, it achieves 26\% F1 and FNR of 75\%. 
In comparison, APIGraph with the same setting reaches around 70\% F1 with a reduced FNR of around 43\%. Chen-AZ lies in between, with around 50\% F1 and around 50\% FNR, reflecting moderate drift. 
These results indicate that our augmentations, while effective, must operate within a continuously updating learning system. 
Due to the evolving nature of malware, one-time training is insufficient, models must adapt incrementally to temporal concept drift~\cite{chen2023continuous}. 


Across most settings, the Bernoulli bit flip augmentation outperforms others except for MaMaDroid. 
This suggests that stochastic noise yields more meaningful transformations in binary feature space, enhancing generalization under drift. 
In addition, we also analyzed the impact of the label ratio and found that performance does not improve significantly beyond 40\%. After this point, any performance gains are subtle across all datasets. This finding aligns with established research showing that the benefit of SSL diminishes as the proportion of labeled data grows~\cite{sohn2020fixmatch}. 
Based on this analysis, we use Bernoulli bit flip augmentation with a 40\% label ratio in subsequent experiments.

\input{tables/active_learning_table}

\subsection{Baseline without Active Learning}
\label{sec:baselineAL}

We first compare \system with prior methods in a static temporal setting, where neither active learning nor model retraining is performed. Models are trained once on an initial labeled dataset and evaluated on future, temporally shifted data without incremental updates. This setting highlights each model's ability to generalize under distributional shift and provides a direct measure of concept drift severity.



\paragraphB{Experimental Setup}
For the baseline comparison, we follow the training and evaluation protocol described in Section~\ref{sec:common-setup}. 
In this setting, no active learning or model retraining are performed; models are trained once on the initial labeled data and evaluated on future, temporally shifted test sets without retraining.
We compare \system against representative baseline methods. TRANSCENDENT~\cite{transcendent} uses conformal prediction to reject uncertain samples without adaptation. CADE~\cite{yang2021cade} detects out-of-distribution samples using a contrastive autoencoder but does not update the model. Chen-AL~\cite{chen2023continuous} applies hierarchical contrastive learning for drift adaptation. We also include a FixMatch and FlexMatch baseline using image-based augmentations, which are not designed for binary malware features. Furthermore, we evaluate MORSE~\cite{wu2023grim}, which applies FixMatch-based SSL to noisy malware datasets. In contrast, \system employs Bernoulli bit-flip augmentation and is trained with a 40\% label ratio. 

\paragraphB{Results}
Table~\ref{tab:baseline_results} shows that all baseline methods suffer substantial performance degradation over time due to concept drift. On APIGraph, \system achieves 67.2\% F1 using only 40\% labeled data, outperforming Chen-AL by nearly 7\% and reducing the FNR by over 1.8\%. Similar trends are observed on Chen-AZ, where \system attains 59.4\% F1, approximately 12\% higher than Chen-AL with a 10\% reduction in FNR.
On LAMDA, \system achieves 30.9\% F1, significantly outperforming Chen-AL by 22.6\%, highlighting the limitations of static retraining and supervised contrastive learning. Across all datasets, baselines consistently underperform without active learning, and FixMatch with image-based augmentations performs particularly poorly, underscoring the need for malware-specific feature transformations.
These findings emphasize that semi-supervised learning with tailored augmentations is essential for maintaining performance under drift, especially in low-label regimes.

\subsection{Comparison with Active Learning}
\label{sec:ALResults}

Building on the baseline evaluation, we incorporate active learning (AL) into \system to enable continual adaptation under concept drift. This section presents CL setup and compares \system against prior approaches.

\paragraphB{Experimental Setup}
We follow the temporal data partitioning protocol described in Section~\ref{sec:common-setup}. Consistent with prior work~\cite{chen2023continuous}, we adopt a \textit{warm-start} strategy, where the model trained on the initial labeled dataset serves as the starting point for iterative adaptation, providing a more stable alternative to cold-start methods. At each test month, we first evaluate the model on the current data. We then apply the proposed multi-criteria sample selection strategy to identify the most informative unlabeled samples under various labeling budgets (i.e., 50, 100, 200, and 400). The selected samples are annotated and added to the labeled pool, while the remaining samples are added to the unlabeled pool. The model is then retrained on the updated data, enabling progressively adaptation to evolving malware distributions.

\paragraphB{Comparison with Existing Methods}
We compare \system with five representative techniques: Binary SVM, Binary GBDT, TRANSCENDENT~\cite{transcendent}, CADE~\cite{yang2021cade}, MORSE~\cite{wu2023grim}, and Chen-AL~\cite{chen2023continuous}. 

Chen-AL is the first to explicitly propose an AL framework for Android malware detection. Earlier approaches such as CADE~\cite{yang2021cade} focuses on out-of-distribution (OOD) detection using contrastive autoencoders, while TRANSCENDENT~\cite{transcendent} employs non-conformity-based criteria to reject uncertain samples in order to mitigate performance degradation. However, neither incorporates AL with explicit sample selection. Although MORSE~\cite{wu2023grim} does not address concept drift, its semi-supervised approach for handling noisy malware data makes it a relevant baseline. To ensure fair comparison, we evaluate MORSE within the same AL framework by integrating an uncertainty-based sample selection strategy.



\paragraphB{Results}
Table~\ref{tab:main_results} presents the summary of the results.
The reported results are in F1 score, false positive rate (FPR), and false negative rate (FNR) for all four datasets.


Across all datasets and budgets, \system consistently outperforms existing methods. On APIGraph, \system achieves an F1 score of 90.17\% with a budget of 50, increasing to 93.52\% at a budget of 400. Compared to Chen-AL, our method shows consistent F1 gains (e.g., $+$1.13\% at budgets 100 and 200), while simultaneously reducing FNR and FPR, demonstrating improved handling of uncertain and evolving malware samples. In contrast, CADE and TRANSCENDENT yield lower F1 scores and higher FNRs, particularly under lower labeling budgets.

The performance gap is even more pronounced on the Chen-AZ dataset. At a budget of 400, \system achieves an F1 score of 82.69\%, outperforming Chen-AL by 7.12\% and reducing the FNR from 31.43\% to 23.95\%. Even with only 50 labeled samples per month, \system maintains a strong F1 of 72.76\%, demonstrating higher label efficiency under moderate concept drift and unlabeled data. LAMDA and the curated MaMaDroid datasets, spanning over a decade, presents the most challenging evaluation due to their severe and long-term distributional drift. At a budget of 400, \system attains F1 scores of 77.6\% and 44.9\% on LAMDA and MaMaDroid, respectively, improving over Chen-AL by more than 34\% and 14.9\%. Although Chen-AL reports a low FNR of 2.5\%, it exhibits an extremely high FPR of 90\%, indicating poor generalization (See Appendix~\ref{app:CITADEL_Chen_visual_comp}). In contrast, \system achieves balanced performance with an FNR of 24.05\% and a remarkably low FPR of 2.35\%, demonstrating robustness and adaptability under realistic, evolving malware scenarios. 

On the LAMDA dataset, MORSE achieves an F1 score of 64.5\%, with an FNR of 33.8\%. On MaMaDroid, its performance drops further to an F1 of 36.0\%, FNR of 68.1\%. In contrast, \system attains substantially higher F1 scores on LAMDA and on MaMaDroid, while also maintaining lower FNR and FPR values. This comparison highlights that \system demonstrates superior performance over MORSE, primarily due to its more effective augmentations, objective function and sample selection techniques.

\begin{figure}[!t]
\centering
\begin{subfigure}[t]{0.48\columnwidth}
    \centering
    \includegraphics[width=\linewidth]{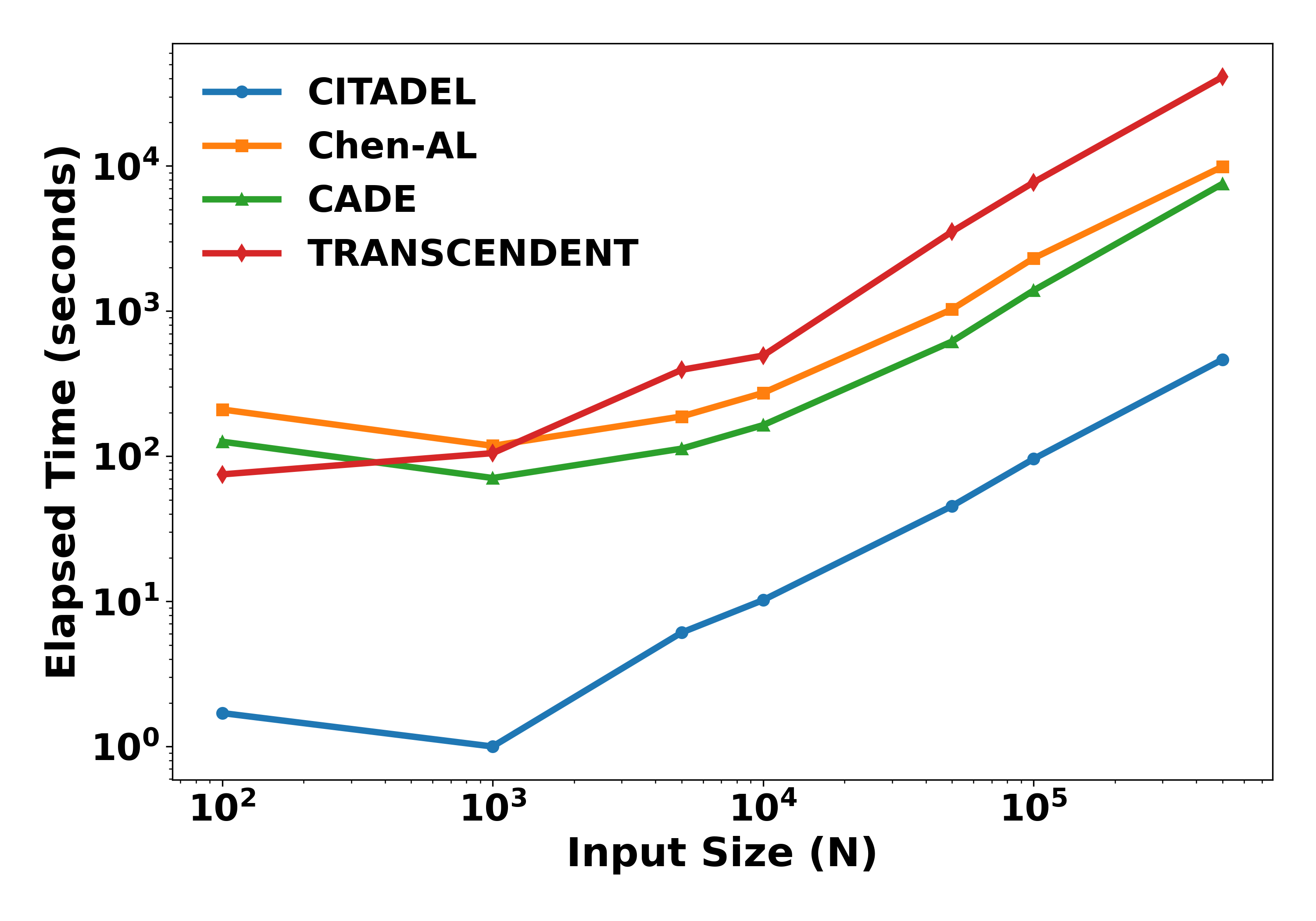}
    \subcaption{Elapsed Time vs Sample Size}
    \label{fig:runtime_time}
\end{subfigure}
\hfill
\begin{subfigure}[t]{0.48\columnwidth}
    \centering
    \includegraphics[width=\linewidth]{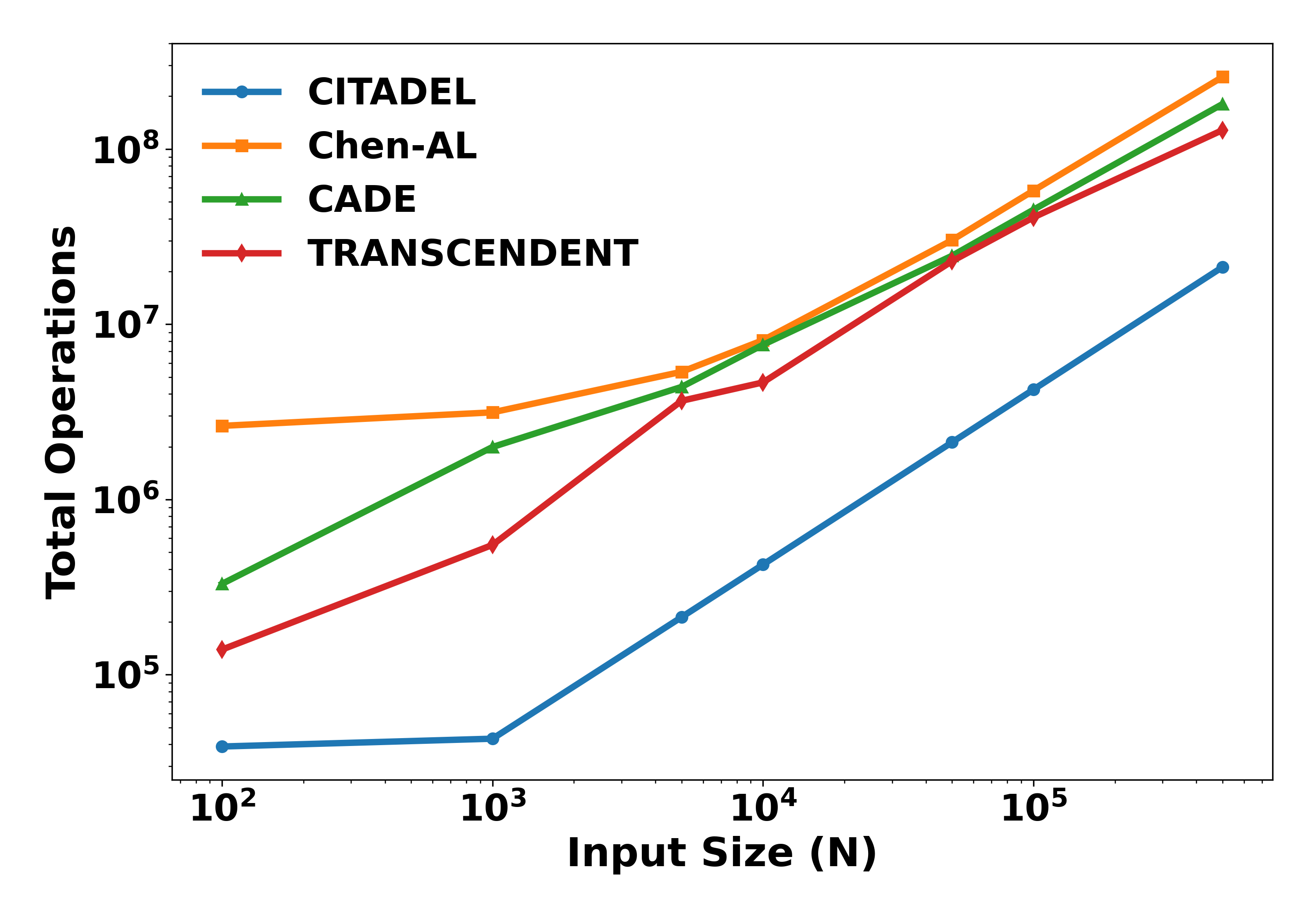}
    \subcaption{Operations vs Sample Size}
    \label{fig:runtime_ops}
\end{subfigure}
\caption{Runtime complexity comparison between Chen-AL and CITADEL using random APIGraph subsamples.}
\label{fig:runtime_comparison}
\end{figure}

\input{tables/complexity_comparison_table}

\input{tables/ablation-study-APIGraph-LAMDA}

To ensure a fair and credible comparison, we re-evaluated Chen-AL under strict temporal partitioning, with all validation samples excluded from the active learning retraining process. These experiments are conducted independently by multiple researchers, each repeated over five runs. Final results are reported as the mean $\pm$ standard deviation. Our re-evaluation yields average F1 scores of 75.57\% on the Chen-AZ dataset and 92.39\% on the APIGraph dataset.

\subsection{Complexity Analysis}
\label{sec:compcomplex}

We evaluate the computational scalability of \system in comparison to prior methods. Using a fixed batch size of 10, input dimensionality, and training budget of 400 samples, we measure runtime and operation count over one epoch of initial training and one epoch of active learning retraining. Experiments are conducted on random subsamples from the APIGraph dataset with increasing sizes (\(n \in \{100, 1000, 5000, 10000, 50000, 100000, 500000\}\)).

To quantify computational cost, we use \texttt{PyTorch Profiler} to record total runtime and number of operations across all methods. As shown in Figure~\ref{fig:runtime_time}, \system consistently maintains significantly lower runtime across increasing dataset sizes. Figure~\ref{fig:runtime_ops} further illustrates that \system requires substantially fewer operations. At \(n = 100{,}000\), \system completes both training and retraining epochs in just 95.8 seconds and requires approximately 4.2 million operations. In comparison, Chen-AL~\cite{chen2023continuous} takes 2,312 seconds and performs 57.9 million operations for the same task. This results in a 24.1$\times$ reduction in runtime and a 13.7$\times$ reduction in operation count, demonstrating the computational efficiency of \system. CADE~\cite{yang2021cade} incurs comparable overhead to Chen-AL due to clustering and latent-space projections, while TRANSCENDENT~\cite{transcendent} is even slower, particularly at scale and in high-dimensional input spaces. 




All detailed comparisons of runtime and operation counts between \system and Chen-AL are summarized in Table~\ref{tab:complexity_comparison}. This level of efficiency allows our system to adapt more quickly to emerging threats and deliver more timely updates to end users in realistic deployment environments. Experimental computational resources are detailed in Appendix~\ref{app:computational_resources}.

\section{Ablation Study}

We conduct an ablation study to assess the contribution of individual components within \system. First, we evaluate the FixMatch~\cite{sohn2020fixmatch} baseline without our feature augmentations. Then, we incorporate our proposed augmentation strategy (Bernoulli bit flip) and evaluate the model without any active learning using 40\% labeled ratio. To analyze the effectiveness of our multi-criteria sample selection method, we compare it against several alternative active learning strategies for 400 budgets, including selection based solely on decision boundary sample selector~\cite{ash2019deep, gal2017deep} (softmax margin), $L_p$-norm uncertainty sample selector, low-confidence samples, and random sample selector. Finally, we evaluate the contribution of \system objective function, which explicitly helps the separation of boundary samples in the latent space. The results demonstrate that each component of \system contributes to improved adaptation under concept drift. 

We conducted the ablation study on three datasets--APIGraph~\cite{api_graph_dataset}, MaMaDroid~\cite{mamadroid} and LAMDA~\cite{haque2025lamda}. LAMDA and Chen-AZ are both DREBIN~\cite{arp2014drebin} feature-based and collected from AndroZoo repository~\cite{androzoo}. In contrast, APIGraph contains structured API call graph features and MaMaDroid utilizes abstraction. Therefore, we choose these three dataset because of their difference in feature representation. 

Table~\ref{tab:ablation_apigraph_mamadroid_lamda} show each component contributes to performance improvements on the three datasets APIGraph, MaMaDroid and LAMDA. By applying our tailored augmentation techniques, the model achieves improved performance of 67.2\% on APIGraph and 30.9\% on LAMDA. Building on this, the addition of our muti-criteria sample selection strategy with \system objective function further enhances the model's ability to adapt to concept drift. Overall, \system achieves final F1-scores of 93.5\%, 44.9\%, and 77.7\% on APIGraph, MaMaDroid, and LAMDA respectively.

%% file: tables/ssl_baselines.tex
\begin{table*}[!t]

\centering
\caption{CITADEL's performance analysis on F1 Score, FNR, and FPR with varying labeled ratios and augmentation strategies across four malware datasets. All values are reported in percentage for multiple runs with mean{\tiny$\pm$std}.
}

\label{tab:ssl_results}
\resizebox{\textwidth}{!}{
\begin{tabular}{l|l|ccc|ccc|ccc|ccc}
\toprule
\multirow{2}{*}{\textbf{LR}} & \multirow{2}{*}{\textbf{Augmentation}} 
& \multicolumn{3}{c|}{\textbf{APIGraph}} 
& \multicolumn{3}{c|}{\textbf{Chen-AZ}} 
& \multicolumn{3}{c|}{\textbf{LAMDA}} 
& \multicolumn{3}{c}{\textbf{MaMaDroid}} \\ \cline{3-14}
 & & F1 & FNR & FPR & F1 & FNR & FPR & F1 & FNR & FPR & F1 & FNR & FPR \\
\midrule

\multirow{4}{*}{10\%} 
& Random Bit Flip            
& 59.8{\tiny$\pm$0.7} & 52.9{\tiny$\pm$0.2} & 7.3{\tiny$\pm$0.9}
& 39.0{\tiny$\pm$0.6} & 72.0{\tiny$\pm$1.0} & 0.7{\tiny$\pm$0.0}
& 23.1{\tiny$\pm$1.3} & 82.5{\tiny$\pm$0.6} & 3.4{\tiny$\pm$0.2}
& \meanstd{12.5}{0.5} & \meanstd{92.0}{0.8} & \meanstd{2.8}{0.3}
\\

& Bernoulli Bit Flip  
& \textbf{61.8{\tiny$\pm$1.4}} & 51.4{\tiny$\pm$1.4} & 5.9{\tiny$\pm$0.2}
& \textbf{50.0{\tiny$\pm$0.5}} & 59.7{\tiny$\pm$0.7} & 1.1{\tiny$\pm$0.0}
& 23.0{\tiny$\pm$0.4} & 82.5{\tiny$\pm$0.1} & 3.2{\tiny$\pm$0.7}
& \meanstd{14.3}{0.4} & \meanstd{90.1}{0.6} & \meanstd{2.6}{0.2}
\\

& Random Flip + Masking   
& 60.3{\tiny$\pm$1.2} & 52.8{\tiny$\pm$1.4} & 0.6{\tiny$\pm$0.1}
& 37.4{\tiny$\pm$1.7} & 73.5{\tiny$\pm$1.2} & 0.6{\tiny$\pm$0.0}
& 22.5{\tiny$\pm$0.3} & 83.1{\tiny$\pm$0.0} & 3.0{\tiny$\pm$0.6}
& \meanstd{13.8}{0.4} & \meanstd{90.7}{0.7} & \meanstd{2.7}{0.2}
\\

& Bernoulli Masking       
& 60.3{\tiny$\pm$1.2} & 52.8{\tiny$\pm$1.3} & 0.6{\tiny$\pm$0.1}
& 37.7{\tiny$\pm$0.6} & 74.1{\tiny$\pm$0.7} & 0.4{\tiny$\pm$0.1}
& \textbf{32.1{\tiny$\pm$1.1}} & 68.1{\tiny$\pm$1.2} & 10.8{\tiny$\pm$0.3}
& \textbf{\meanstd{15.8}{0.4}} & \meanstd{86.0}{0.6} & \meanstd{3.7}{0.3}
\\
\midrule

\multirow{4}{*}{20\%}
& Random Bit Flip            
& 64.0{\tiny$\pm$1.6} & 48.8{\tiny$\pm$2.3} & 6.1{\tiny$\pm$0.0}
& 46.2{\tiny$\pm$0.3} & 66.0{\tiny$\pm$1.1} & 0.6{\tiny$\pm$0.0}
& 27.0{\tiny$\pm$2.1} & 79.0{\tiny$\pm$3.2} & 3.3{\tiny$\pm$0.9}
& \meanstd{16.1}{0.4} & \meanstd{87.4}{0.6} & \meanstd{2.6}{0.2}
\\

& Bernoulli Bit Flip 
& 64.3{\tiny$\pm$2.0} & 48.2{\tiny$\pm$2.6} & 0.3{\tiny$\pm$0.1}
& \textbf{57.4{\tiny$\pm$0.0}} & 53.5{\tiny$\pm$0.3} & 0.6{\tiny$\pm$0.1}
& \textbf{33.9{\tiny$\pm$0.2}} & 71.3{\tiny$\pm$1.8} & 4.7{\tiny$\pm$1.4}
& \meanstd{18.4}{0.3} & \meanstd{84.7}{0.5} & \meanstd{2.4}{0.2}
\\

& Random Flip + Masking   
& \textbf{64.4{\tiny$\pm$1.4}} & 47.0{\tiny$\pm$1.1} & 0.9{\tiny$\pm$0.1}
& 46.0{\tiny$\pm$0.4} & 66.1{\tiny$\pm$0.5} & 0.5{\tiny$\pm$0.1}
& 29.6{\tiny$\pm$3.2} & 77.2{\tiny$\pm$2.6} & 2.5{\tiny$\pm$0.8}
& \meanstd{17.9}{0.3} & \meanstd{85.2}{0.5} & \meanstd{2.5}{0.2}
\\

& Bernoulli Masking       
& 64.4{\tiny$\pm$1.3} & 47.0{\tiny$\pm$1.3} & 0.9{\tiny$\pm$0.1}
& 44.9{\tiny$\pm$0.6} & 66.2{\tiny$\pm$0.9} & 0.6{\tiny$\pm$0.1}
& 31.7{\tiny$\pm$0.3} & 75.2{\tiny$\pm$0.5} & 2.9{\tiny$\pm$0.2}
& \textbf{\meanstd{20.4}{0.3}} & \meanstd{81.2}{0.5} & \meanstd{3.4}{0.2}
\\

\midrule

\multirow{4}{*}{30\%}
& Random Bit Flip            & \textbf{67.7{\tiny $\pm$0.6}} & 43.6{\tiny $\pm$0.7} & 8.0{\tiny $\pm$0.2} & 49.0{\tiny $\pm$0.7} & 62.7{\tiny $\pm$0.5} & 0.8{\tiny $\pm$0.0} & 25.8{\tiny $\pm$1.1} & 80.5{\tiny $\pm$1.1} & 2.3{\tiny $\pm$0.2} 
& \meanstd{18.9}{0.3} & \meanstd{82.8}{0.5} & \meanstd{2.5}{0.1} \\

& Bernoulli Bit Flip & 66.1{\tiny $\pm$1.8} & 45.4{\tiny $\pm$3.4} & 7.7{\tiny $\pm$1.9} & \textbf{58.7{\tiny $\pm$1.0}} & 52.1{\tiny $\pm$1.2} & 0.6{\tiny $\pm$0.1} & \textbf{34.3{\tiny $\pm$4.0}} & 69.5{\tiny $\pm$4.5} & 4.8{\tiny $\pm$0.7} 
& \meanstd{21.2}{0.3} & \meanstd{79.5}{0.4} & \meanstd{2.3}{0.1} \\

& Random Flip + Masking   & 66.0{\tiny $\pm$1.3} & 46.5{\tiny $\pm$1.5} & 0.6{\tiny $\pm$0.1} & 47.7{\tiny $\pm$0.6} & 63.9{\tiny $\pm$0.7} & 0.7{\tiny $\pm$0.1} & 27.9{\tiny $\pm$1.0} & 78.9{\tiny $\pm$0.8} & 2.4{\tiny $\pm$0.2}
& \meanstd{20.7}{0.3} & \meanstd{80.0}{0.5} & \meanstd{2.4}{0.1} \\
& Bernoulli Masking       & 66.0{\tiny $\pm$1.4} & 46.5{\tiny $\pm$1.6} & 0.6{\tiny $\pm$0.1} & 47.7{\tiny $\pm$1.0} & 63.1{\tiny $\pm$0.9} & 0.7{\tiny $\pm$0.1} & 26.7{\tiny $\pm$3.8} & 78.7{\tiny $\pm$4.3} & 3.6{\tiny $\pm$1.3} 
& \textbf{\meanstd{23.6}{0.2}} & \meanstd{78.4}{0.4} & \meanstd{3.1}{0.2} \\
\midrule

\multirow{4}{*}{40\%}
& Random Bit Flip             & 69.2{\tiny $\pm$0.9} & 42.5{\tiny $\pm$2.0} & 6.7{\tiny $\pm$1.3} & 51.9{\tiny $\pm$0.5} & 58.9{\tiny $\pm$0.7} & 0.7{\tiny $\pm$0.0} & 25.6{\tiny $\pm$1.0} & 81.0{\tiny $\pm$1.0} & 1.8{\tiny $\pm$0.5} 
& \meanstd{20.4}{0.2} & \meanstd{79.2}{0.4} & \meanstd{2.4}{0.1} \\

& Bernoulli Bit Flip & 67.2{\tiny $\pm$2.9} & 43.6{\tiny $\pm$3.0} & 9.5{\tiny $\pm$1.1} & \textbf{59.4{\tiny $\pm$0.7}} & 51.3{\tiny $\pm$0.4} & 0.6{\tiny $\pm$0.1} & \textbf{30.9{\tiny $\pm$0.9}} & 74.8{\tiny $\pm$0.9} & 2.9{\tiny $\pm$0.4}
& \meanstd{22.5}{0.2} & \meanstd{76.1}{0.3} & \meanstd{2.2}{0.1} \\

& Random Flip + Masking   & 70.9{\tiny $\pm$1.8} & 38.5{\tiny $\pm$3.1} & 1.0{\tiny $\pm$0.3} & 51.9{\tiny $\pm$0.8} & 59.6{\tiny $\pm$0.9} & 0.5{\tiny $\pm$0.1} & 24.7{\tiny $\pm$1.6} & 81.7{\tiny $\pm$1.6} & 1.9{\tiny $\pm$0.6} 
& \meanstd{22.0}{0.2} & \meanstd{76.6}{0.3} & \meanstd{2.3}{0.1} \\

& Bernoulli Masking       & \textbf{72.2{\tiny $\pm$1.6}} & 36.2{\tiny $\pm$2.2} & 1.1{\tiny $\pm$0.2} & 49.8{\tiny $\pm$1.0} & 61.8{\tiny $\pm$0.6} & 0.6{\tiny $\pm$0.1} & 30.7{\tiny $\pm$3.1} & 75.3{\tiny $\pm$2.6} & 3.9{\tiny $\pm$0.5}
& \textbf{\meanstd{26.2}{0.1}} & \meanstd{75.6}{0.2} & \meanstd{2.9}{0.0} \\
\midrule

\multirow{4}{*}{50\%}
& Random Bit Flip             & 68.1{\tiny $\pm$0.4} & 41.1{\tiny $\pm$0.3} & 11.4{\tiny $\pm$1.5} & 56.0{\tiny $\pm$0.5} & 55.7{\tiny $\pm$0.8} & 0.6{\tiny $\pm$0.0} & 26.6{\tiny $\pm$0.0} & 80.1{\tiny $\pm$0.2} & 2.4{\tiny $\pm$0.2} 
& \meanstd{20.6}{0.2} & \meanstd{78.8}{0.4} & \meanstd{2.4}{0.1} \\

& Bernoulli Bit Flip & 68.3{\tiny $\pm$1.9} & 41.8{\tiny $\pm$2.4} & 8.2{\tiny $\pm$0.7} & \textbf{58.6{\tiny $\pm$0.7}} & 52.6{\tiny $\pm$0.6} & 0.5{\tiny $\pm$0.0} & 26.9{\tiny $\pm$0.6} & 80.0{\tiny $\pm$0.7} & 2.1{\tiny $\pm$0.3} 
& \meanstd{22.7}{0.2} & \meanstd{75.7}{0.3} & \meanstd{2.2}{0.1} \\

& Random Flip + Masking   & \textbf{68.5{\tiny $\pm$2.4}} & 42.3{\tiny $\pm$3.2} & 0.7{\tiny $\pm$0.1} & 52.1{\tiny $\pm$0.1} & 59.2{\tiny $\pm$0.0} & 0.6{\tiny $\pm$0.0} & \textbf{28.8{\tiny $\pm$2.4}} & 77.7{\tiny $\pm$2.6} & 3.1{\tiny $\pm$0.6}
& \meanstd{22.2}{0.2} & \meanstd{76.2}{0.3} & \meanstd{2.3}{0.1} \\

& Bernoulli Masking       & 67.2{\tiny $\pm$1.4} & 44.6{\tiny $\pm$2.1} & 0.7{\tiny $\pm$0.1} & 51.6{\tiny $\pm$0.7} & 59.2{\tiny $\pm$0.7} & 0.8{\tiny $\pm$0.2} & 25.7{\tiny $\pm$0.1} & 80.3{\tiny $\pm$0.1} & 3.3{\tiny $\pm$0.3} 
& \textbf{\meanstd{26.8}{0.2}} & \meanstd{74.8}{0.3} & \meanstd{2.8}{0.1} \\
\midrule

\multirow{4}{*}{60\%}
& Random Bit Flip             & \textbf{69.0{\tiny $\pm$1.9}} & 40.8{\tiny $\pm$0.8} & 10.1{\tiny $\pm$2.5} & 52.5{\tiny $\pm$0.4} & 58.8{\tiny $\pm$0.5} & 0.5{\tiny $\pm$0.0} & 25.6{\tiny $\pm$2.4} & 80.6{\tiny $\pm$3.1} & 2.5{\tiny $\pm$0.3} 
& \meanstd{20.9}{0.2} & \meanstd{78.3}{0.4} & \meanstd{2.3}{0.1} \\

& Bernoulli Bit Flip & 69.0{\tiny $\pm$1.6} & 41.0{\tiny $\pm$2.0} & 9.1{\tiny $\pm$0.0} & \textbf{57.5{\tiny $\pm$0.2}} & 53.4{\tiny $\pm$0.3} & 0.5{\tiny $\pm$0.0} & \textbf{32.2{\tiny $\pm$0.5}} & 75.6{\tiny $\pm$0.2} & 1.9{\tiny $\pm$0.1} 
& \meanstd{23.0}{0.2} & \meanstd{75.3}{0.3} & \meanstd{2.1}{0.1} \\
& Random Flip + Masking   & 66.0{\tiny $\pm$1.5} & 44.9{\tiny $\pm$2.2} & 1.0{\tiny $\pm$0.1} & 50.8{\tiny $\pm$2.2} & 60.4{\tiny $\pm$2.2} & 0.5{\tiny $\pm$0.1} & 21.4{\tiny $\pm$2.3} & 84.2{\tiny $\pm$2.0} & 1.8{\tiny $\pm$0.2}
& \meanstd{22.5}{0.2} & \meanstd{75.8}{0.3} & \meanstd{2.2}{0.1} \\
& Bernoulli Masking       & 66.0{\tiny $\pm$1.8} & 44.9{\tiny $\pm$2.4} & 1.1{\tiny $\pm$0.5} & 50.6{\tiny $\pm$0.9} & 60.5{\tiny $\pm$0.8} & 0.5{\tiny $\pm$0.1} & 23.1{\tiny $\pm$1.3} & 82.3{\tiny $\pm$1.2} & 3.1{\tiny $\pm$0.6} 
& \textbf{\meanstd{27.2}{0.2}} & \meanstd{74.0}{0.3} & \meanstd{2.7}{0.1} \\
\midrule

\multirow{4}{*}{70\%}
& Random Bit Flip             & 69.2{\tiny $\pm$1.0} & 39.7{\tiny $\pm$0.8} & 10.7{\tiny $\pm$3.1} & 49.3{\tiny $\pm$0.6} & 62.1{\tiny $\pm$1.0} & 0.7{\tiny $\pm$0.0} & 26.1{\tiny $\pm$0.6} & 80.8{\tiny $\pm$0.6} & 2.0{\tiny $\pm$0.3} 
& \meanstd{21.1}{0.2} & \meanstd{77.9}{0.3} & \meanstd{2.3}{0.1} \\

& Bernoulli Bit Flip & \textbf{70.6{\tiny $\pm$1.1}} & 40.9{\tiny $\pm$1.1} & 7.2{\tiny $\pm$0.6} & \textbf{61.5{\tiny $\pm$0.0}} & 48.8{\tiny $\pm$0.0} & 0.7{\tiny $\pm$0.0} & \textbf{29.1{\tiny $\pm$1.2}} & 77.7{\tiny $\pm$0.7} & 2.1{\tiny $\pm$0.1} 
& \meanstd{23.2}{0.2} & \meanstd{75.0}{0.3} & \meanstd{2.1}{0.1} \\

& Random Flip + Masking   & 67.9{\tiny $\pm$1.7} & 42.6{\tiny $\pm$2.5} & 1.0{\tiny $\pm$0.2} & 56.1{\tiny $\pm$0.4} & 53.9{\tiny $\pm$0.9} & 0.7{\tiny $\pm$0.0} & 22.8{\tiny $\pm$0.2} & 82.8{\tiny $\pm$0.8} & 2.5{\tiny $\pm$0.0} 
& \meanstd{22.7}{0.2} & \meanstd{75.5}{0.3} & \meanstd{2.2}{0.1} \\
& Bernoulli Masking       & 66.9{\tiny $\pm$1.3} & 44.6{\tiny $\pm$1.6} & 0.9{\tiny $\pm$0.3} & 49.0{\tiny $\pm$0.6} & 61.9{\tiny $\pm$0.7} & 0.6{\tiny $\pm$0.0} & 22.4{\tiny $\pm$0.2} & 82.5{\tiny $\pm$0.1} & 4.0{\tiny $\pm$0.3} 
& \textbf{\meanstd{27.6}{0.3}} & \meanstd{73.5}{0.4} & \meanstd{2.7}{0.1} \\
\midrule

\multirow{4}{*}{80\%}
& Random Bit Flip             & 68.3{\tiny $\pm$2.2} & 42.2{\tiny $\pm$2.7} & 8.0{\tiny $\pm$0.5} & 48.7{\tiny $\pm$0.6} & 62.7{\tiny $\pm$0.8} & 0.6{\tiny $\pm$0.0} & 28.0{\tiny $\pm$3.2} & 78.7{\tiny $\pm$3.1} & 2.1{\tiny $\pm$0.4} 
& \meanstd{21.3}{0.2} & \meanstd{77.6}{0.3} & \meanstd{2.3}{0.1} \\
& Bernoulli Bit Flip & \textbf{70.3{\tiny $\pm$1.6}} & 39.6{\tiny $\pm$2.5} & 8.4{\tiny $\pm$1.1} & \textbf{60.3{\tiny $\pm$1.0}} & 51.0{\tiny $\pm$1.0} & 0.4{\tiny $\pm$0.0} & \textbf{30.5{\tiny $\pm$1.0}} & 76.3{\tiny $\pm$0.2} & 2.0{\tiny $\pm$0.2} 
& \meanstd{23.4}{0.2} & \meanstd{74.8}{0.3} & \meanstd{2.1}{0.1} \\
& Random Flip + Masking   & 69.9{\tiny $\pm$1.9} & 37.0{\tiny $\pm$2.7} & 1.7{\tiny $\pm$0.3} & 49.4{\tiny $\pm$3.4} & 61.2{\tiny $\pm$3.5} & 0.7{\tiny $\pm$0.0} & 24.8{\tiny $\pm$0.3} & 81.4{\tiny $\pm$0.3} & 2.2{\tiny $\pm$0.1} 
& \meanstd{22.9}{0.2} & \meanstd{75.3}{0.3} & \meanstd{2.2}{0.1} \\
& Bernoulli Masking       & 66.1{\tiny $\pm$1.5} & 45.6{\tiny $\pm$2.1} & 0.6{\tiny $\pm$0.1} & 48.7{\tiny $\pm$0.7} & 63.3{\tiny $\pm$1.4} & 0.7{\tiny $\pm$0.1} & 25.0{\tiny $\pm$1.6} & 80.4{\tiny $\pm$1.5} & 3.5{\tiny $\pm$0.1} 
& \textbf{\meanstd{27.9}{0.3}} & \meanstd{73.1}{0.3} & \meanstd{2.6}{0.1} \\
\midrule

\multirow{4}{*}{90\%}
& Random Bit Flip             & 72.2{\tiny $\pm$1.5} & 34.5{\tiny $\pm$4.4} & 13.6{\tiny $\pm$4.4} & 49.0{\tiny $\pm$0.6} & 61.6{\tiny $\pm$1.1} & 0.6{\tiny $\pm$0.0} & 25.8{\tiny $\pm$3.3} & 80.3{\tiny $\pm$3.1} & 2.7{\tiny $\pm$0.2}
& \meanstd{21.5}{0.2} & \meanstd{77.3}{0.3} & \meanstd{2.3}{0.1} \\

& Bernoulli Bit Flip & \textbf{73.6{\tiny $\pm$2.4}} & 34.9{\tiny $\pm$4.2} & 10.4{\tiny $\pm$1.7} & \textbf{58.1{\tiny $\pm$0.7}} & 52.1{\tiny $\pm$1.2} & 0.8{\tiny $\pm$0.1} & \textbf{33.1{\tiny $\pm$3.4}} & 74.2{\tiny $\pm$3.5} & 2.6{\tiny $\pm$0.2} 
& \meanstd{23.5}{0.2} & \meanstd{74.6}{0.3} & \meanstd{2.1}{0.1} \\
& Random Flip + Masking   & 70.3{\tiny $\pm$1.7} & 38.0{\tiny $\pm$2.1} & 1.3{\tiny $\pm$0.2} & 50.7{\tiny $\pm$2.1} & 59.6{\tiny $\pm$3.4} & 0.8{\tiny $\pm$0.3} & 22.1{\tiny $\pm$0.9} & 83.3{\tiny $\pm$0.7} & 2.5{\tiny $\pm$0.1}
& \meanstd{23.0}{0.2} & \meanstd{75.0}{0.3} & \meanstd{2.2}{0.1} \\
& Bernoulli Masking       & 67.0{\tiny $\pm$1.8} & 41.3{\tiny $\pm$4.3} & 1.5{\tiny $\pm$0.9} & 47.2{\tiny $\pm$1.1} & 62.8{\tiny $\pm$1.1} & 1.0{\tiny $\pm$0.1} & 23.5{\tiny $\pm$0.6} & 79.8{\tiny $\pm$0.5} & 6.5{\tiny $\pm$0.3} 
& \textbf{\meanstd{28.2}{0.3}} & \meanstd{72.6}{0.3} & \meanstd{2.6}{0.1} \\
\bottomrule

\end{tabular}
 }
\end{table*}

%% file: tables/baseline_comparison_table.tex
\begin{table*}[t]
\centering
\footnotesize
\caption{Comparison of baseline without active learning. 
Results are in \% for multiple runs with mean{\tiny$\pm$std}.}
\label{tab:baseline_results}
\resizebox{\textwidth}{!}{
\begin{tabular}{l|ccc|ccc|ccc|ccc}
\toprule

\multirow{2}{*}{\textbf{Method}} 
& \multicolumn{3}{c|}{\textbf{APIGraph}} 
& \multicolumn{3}{c|}{\textbf{Chen-AZ}} 
& \multicolumn{3}{c|}{\textbf{LAMDA}} 
& \multicolumn{3}{c}{\textbf{MaMaDroid}} \\ \cline{2-13}
 & F1 & FNR & FPR 
 & F1 & FNR & FPR 
 & F1 & FNR & FPR 
 & F1 & FNR & FPR \\
\midrule

TRANSCENDENT 
& \meanstd{50.3}{0.8} & \meanstd{61.4}{1.6} & \meanstd{1.1}{0.0}
& \meanstd{29.3}{0.6} & \meanstd{78.2}{1.6} & \meanstd{1.3}{0.0} 
& \meanstd{30.4}{1.0} & \meanstd{75.8}{0.7} & \meanstd{16.4}{0.6}
& \meanstd{19.4}{0.4} & \meanstd{83.5}{0.3} & \meanstd{3.9}{0.1} \\

CADE        
& \meanstd{64.6}{1.2} & \meanstd{39.2}{1.2} & \meanstd{2.5}{0.3}
& \meanstd{56.0}{1.2} & \meanstd{50.5}{1.5} & \meanstd{2.1}{0.4}
& \meanstd{29.5}{0.7} & \meanstd{76.7}{1.6} & \meanstd{4.0}{1.2}
& \meanstd{20.3}{0.5} & \meanstd{82.4}{0.4} & \meanstd{3.6}{0.1} \\

Chen-AL     
& \meanstd{60.4}{1.0} & \meanstd{47.2}{1.3} & \meanstd{2.0}{0.1}
& \meanstd{47.5}{0.9} & \meanstd{61.3}{0.5} & \meanstd{1.6}{0.2}
& \meanstd{22.6}{0.1} & \meanstd{83.5}{0.1} & \meanstd{1.7}{0.0}
& \meanstd{20.1}{0.3} & \meanstd{81.7}{0.4} & \meanstd{3.2}{0.1} \\

FixMatch-Baseline 
& \meanstd{0.0}{0.0} & \meanstd{1.0}{0.0} & \meanstd{0.0}{0.0}
& \meanstd{0.0}{0.0} & \meanstd{1.0}{0.0} & \meanstd{0.0}{0.0}
& \meanstd{0.0}{0.0} & \meanstd{1.0}{0.0} & \meanstd{0.0}{0.0}
& \meanstd{0.0}{0.0} & \meanstd{1.0}{0.0} & \meanstd{0.0}{0.0} \\

FlexMatch-Baseline 
& \meanstd{0.0}{0.0} & \meanstd{1.0}{0.0} & \meanstd{0.0}{0.0}
& \meanstd{0.0}{0.0} & \meanstd{1.0}{0.0} & \meanstd{0.0}{0.0}
& \meanstd{0.0}{0.0} & \meanstd{1.0}{0.0} & \meanstd{0.0}{0.0}
& \meanstd{0.0}{0.0} & \meanstd{1.0}{0.0} & \meanstd{0.0}{0.0} \\

MORSE  
& \meanstd{64.5}{0.4} & \meanstd{49.6}{0.5} & \meanstd{0.2}{0.0} 
& \meanstd{55.5}{0.8} & \meanstd{54.9}{0.9} & \meanstd{0.6}{0.0} 
& \meanstd{27.1}{0.6} & \meanstd{79.6}{0.8} & \meanstd{1.8}{0.0}
& \meanstd{23.2}{0.2} & \meanstd{78.6}{0.3} & \meanstd{2.6}{0.1} \\

\textbf{CITADEL (Ours)} 
& \textbf{\meanstd{67.2}{2.9}} & \meanstd{43.6}{3.0} & \meanstd{9.5}{1.1}
& \textbf{\meanstd{59.4}{0.7}} & \meanstd{51.3}{0.4} & \meanstd{0.6}{0.1}
& \textbf{\meanstd{30.9}{0.9}} & \meanstd{74.8}{0.9} & \meanstd{2.9}{0.4}
& \textbf{\meanstd{26.2}{0.1}} & \meanstd{75.6}{0.2} & \meanstd{2.9}{0.0} \\

\bottomrule
\end{tabular}
}
\end{table*}

%% file: tables/active_learning_table.tex
\begin{table*}[t]
\scriptsize
\centering
\caption{
Comparison of active learning systems across four malware datasets (APIGraph, Chen-AZ, LAMDA, MaMaDroid). 
All values are reported in percentage for multiple runs with mean$\pm$std (std in tiny font). 
\textbf{SC:} Sample Selector. {\bf UNC:} Uncertainty, {\bf Cred:} Credibility, {\bf PS:} Pseudo Loss, {\bf MC:} Multi-Criteria (Ours), {\bf OOD} Out-of-Distribution.
}
\label{tab:main_results}

\resizebox{\textwidth}{!}{
\begin{tabular}{p{0.28cm}|p{1.9cm}|c|ccc|ccc|ccc|ccc}
\toprule

\rotatebox[origin=c]{90}{\multirow{2}{*}{\textbf{Budget}}}
 & \multirow{2}{*}{\textbf{Method}} & \multirow{2}{*}{\textbf{SC}} 
 & \multicolumn{3}{c|}{\textbf{APIGraph}} 
 & \multicolumn{3}{c|}{\textbf{Chen-AZ}} 
 & \multicolumn{3}{c|}{\textbf{LAMDA}} 
 & \multicolumn{3}{c}{\textbf{MaMaDroid}} \\ \cline{4-15}
 & & & F1 & FNR & FPR & F1 & FNR & FPR & F1 & FNR & FPR & F1 & FNR & FPR \\
\midrule

\multirow{8}{*}{50}
  & Binary SVM & UNC      
    & \meanstd{87.1}{0.2} & \meanstd{17.6}{0.4} & \meanstd{0.7}{0.0}
    & \meanstd{58.0}{0.2} & \meanstd{53.5}{0.4} & \meanstd{0.2}{0.0}
    & \meanstd{37.1}{0.2} & \meanstd{5.7}{0.3}  & \meanstd{87.9}{0.2}
    & \meanstd{18.0}{0.5} & \meanstd{77.0}{0.7} & \meanstd{12.4}{0.3} \\

  & Binary GBDT & UNC       
    & \meanstd{77.8}{0.2} & \meanstd{32.1}{0.4} & \meanstd{0.5}{0.0}
    & \meanstd{58.9}{0.2} & \meanstd{52.1}{0.2} & \meanstd{0.5}{0.0}
    & \meanstd{37.0}{0.3} & \meanstd{5.9}{0.2}  & \meanstd{87.5}{0.3}
    & \meanstd{18.3}{0.5} & \meanstd{76.5}{0.6} & \meanstd{12.6}{0.3} \\

  & TRANSCENDENT & Cred
    & \meanstd{75.0}{0.3} & \meanstd{32.0}{0.4} & \meanstd{0.0}{0.0}
    & \meanstd{38.4}{0.3} & \meanstd{69.4}{0.6} & \meanstd{1.6}{0.0}
    & \meanstd{32.0}{1.5} & \meanstd{37.8}{2.1} & \meanstd{63.7}{1.8}
    & \meanstd{19.0}{0.5} & \meanstd{75.5}{0.7} & \meanstd{12.0}{0.3} \\

  & CADE & OOD
    & \meanstd{83.9}{1.7} & \meanstd{19.7}{2.0} & \meanstd{1.0}{0.1}
    & \meanstd{49.7}{0.7} & \meanstd{62.5}{0.4} & \meanstd{0.3}{0.0}
    & \meanstd{34.2}{1.8} & \meanstd{44.0}{2.5} & \meanstd{56.8}{2.2}
    & \meanstd{20.5}{0.5} & \meanstd{73.5}{0.6} & \meanstd{11.4}{0.3} \\

  & Chen-AL & PS
    & \meanstd{89.2}{0.3} & \meanstd{15.1}{0.3} & \meanstd{0.5}{0.0}
    & \meanstd{69.5}{2.2} & \meanstd{39.6}{3.1} & \meanstd{0.5}{0.0}
    & \meanstd{37.4}{2.4} & \meanstd{3.5}{0.3} & \meanstd{93.7}{2.7}
    & \meanstd{22.0}{0.4} & \meanstd{72.0}{0.6} & \meanstd{10.8}{0.3} \\

  & MORSE & UNC & \meanstd{88.8}{0.3} & \meanstd{14.9}{0.4} & \meanstd{0.6}{0.1}
    & \meanstd{55.9}{1.2} & \meanstd{28.2}{0.9} & \meanstd{15.6}{0.4}
    & \meanstd{61.4}{0.4} & \meanstd{36.1}{0.5} & \meanstd{11.0}{0.3}
    & {\meanstd{25.3}{0.9}} & \meanstd{81.2}{1.3} & \meanstd{3.8}{0.5} \\
    

  & \textbf{CITADEL} & \textbf{MC}
    & \textbf{\meanstd{90.1}{0.1}} & \textbf{\meanstd{14.5}{0.2}} & \textbf{\meanstd{0.3}{0.0}}
    & \textbf{\meanstd{72.7}{1.0}} & \textbf{\meanstd{36.7}{1.6}} & \textbf{\meanstd{0.3}{0.0}}
    & \textbf{\meanstd{70.9}{1.2}} & \meanstd{33.6}{1.6} & \textbf{\meanstd{2.0}{0.1}}
    & \textbf{\meanstd{30.9}{0.9}} & \meanstd{74.8}{0.9} & \meanstd{2.9}{0.4}
    \\

\midrule
\multirow{8}{*}{100}
  & Binary SVM & UNC      
    & \meanstd{87.4}{0.3} & \meanstd{16.6}{0.4} & \meanstd{0.7}{0.0}
    & \meanstd{59.8}{0.2} & \meanstd{51.5}{0.2} & \meanstd{0.2}{0.0}
    & \meanstd{37.3}{0.3} & \meanstd{5.4}{0.3}  & \meanstd{88.5}{0.4}
    & \meanstd{20.2}{0.4} & \meanstd{72.7}{0.6} & \meanstd{11.1}{0.3} \\

  & Binary GBDT & UNC
    & \meanstd{80.1}{0.3} & \meanstd{27.9}{0.3} & \meanstd{0.7}{0.0}
    & \meanstd{59.4}{0.2} & \meanstd{50.3}{0.3} & \meanstd{0.7}{0.0}
    & \meanstd{37.2}{0.2} & \meanstd{5.6}{0.4}  & \meanstd{88.2}{0.5}
    & \meanstd{20.4}{0.4} & \meanstd{72.2}{0.6} & \meanstd{11.0}{0.3} \\

  & TRANSCENDENT & Cred
    & \meanstd{76.8}{0.2} & \meanstd{29.0}{0.3} & \meanstd{0.0}{0.0}
    & \meanstd{39.8}{0.4} & \meanstd{68.5}{0.7} & \meanstd{1.5}{0.0}
    & \meanstd{35.8}{1.3} & \meanstd{28.4}{1.9} & \meanstd{70.1}{1.6}
    & \meanstd{21.3}{0.4} & \meanstd{71.2}{0.6} & \meanstd{10.8}{0.3} \\

  & CADE & OOD
    & \meanstd{87.1}{0.5} & \meanstd{14.4}{1.1} & \meanstd{1.1}{0.2}
    & \meanstd{50.1}{0.1} & \meanstd{62.3}{0.1} & \meanstd{0.3}{0.0}
    & \meanstd{37.3}{1.6} & \meanstd{37.2}{2.1} & \meanstd{45.2}{2.0}
    & \meanstd{24.7}{0.4} & \meanstd{67.5}{0.5} & \meanstd{9.9}{0.2} \\

  & Chen-AL & PS
    & \meanstd{90.7}{0.2} & \meanstd{13.2}{0.3} & \meanstd{0.5}{0.0}
    & \meanstd{70.7}{0.7} & \meanstd{70.7}{0.7} & \meanstd{0.6}{0.0}
    & \meanstd{38.5}{2.2} & \meanstd{3.3}{0.3}  & \meanstd{93.5}{2.6}
    & \meanstd{26.3}{0.3} & \meanstd{65.9}{0.5} & \meanstd{9.4}{0.2} \\

  & MORSE & UNC & \meanstd{89.4}{0.3} & \meanstd{14.3}{0.4} & \meanstd{0.6}{0.1}
    & \meanstd{63.2}{1.0} & \meanstd{26.7}{0.8} & \meanstd{10.5}{0.3}
    & \meanstd{62.3}{0.4} & \meanstd{35.5}{0.5} & \meanstd{11.2}{0.3}
    & {\meanstd{28.4}{0.8}} & \meanstd{77.0}{1.1} & \meanstd{3.3}{0.4} \\
  

  & \textbf{CITADEL} & \textbf{MC}
    & \textbf{\meanstd{91.9}{0.1}} & \textbf{\meanstd{12.1}{0.2}} & \textbf{\meanstd{0.3}{0.0}}
    & \textbf{\meanstd{75.7}{2.1}} & \textbf{\meanstd{32.7}{2.9}} & \textbf{\meanstd{0.4}{0.1}}
    & \textbf{\meanstd{73.8}{0.5}} & \textbf{\meanstd{30.0}{0.0}} & \textbf{\meanstd{3.2}{0.4}}
    & \textbf{\meanstd{37.6}{2.0}} & \meanstd{47.6}{1.8} & \meanstd{18.0}{0.9}
    \\

\midrule
\multirow{8}{*}{200}
  & Binary SVM & UNC & \meanstd{87.9}{0.2} & \meanstd{14.6}{0.4} & \meanstd{0.5}{0.0} & \meanstd{63.5}{0.2} & \meanstd{46.4}{0.8} & \meanstd{0.3}{0.0} & \meanstd{37.7}{0.3} & \meanstd{4.8}{0.3} & \meanstd{89.7}{0.7} & \meanstd{21.6}{0.4} & \meanstd{69.1}{0.5} & \meanstd{9.6}{0.2} \\

  & Binary GBDT & UNC & \meanstd{82.6}{0.2} & \meanstd{24.7}{0.2} & \meanstd{0.6}{0.0} & \meanstd{63.9}{0.2} & \meanstd{45.0}{0.3} & \meanstd{0.8}{0.0} & \meanstd{37.2}{0.2} & \meanstd{5.6}{0.4} & \meanstd{88.2}{0.5} & \meanstd{21.8}{0.4} & \meanstd{68.7}{0.5} & \meanstd{9.5}{0.2} \\

  & TRANSCENDENT & Cred & \meanstd{78.2}{0.2} & \meanstd{27.0}{0.3} & \meanstd{0.0}{0.0} & \meanstd{41.2}{0.4} & \meanstd{67.4}{0.6} & \meanstd{1.3}{0.0} & \meanstd{39.0}{1.2} & \meanstd{21.3}{1.5} & \meanstd{76.3}{1.4} & \meanstd{22.8}{0.4} & \meanstd{67.4}{0.5} & \meanstd{9.2}{0.2} \\

  & CADE & OOD & \meanstd{88.9}{0.2} & \meanstd{14.0}{0.3} & \meanstd{0.7}{0.0} & \meanstd{50.2}{0.0} & \meanstd{62.2}{0.1} & \meanstd{0.3}{0.0} & \meanstd{38.5}{1.3} & \meanstd{31.8}{1.8} & \meanstd{38.5}{1.6} & \meanstd{26.3}{0.3} & \meanstd{63.0}{0.4} & \meanstd{8.4}{0.2} \\

  & Chen-AL & PS & \meanstd{91.7}{0.4} & \meanstd{11.4}{0.6} & \meanstd{0.4}{0.0} & \meanstd{73.6}{1.3} & \meanstd{33.7}{1.2} & \meanstd{0.5}{0.0} & \meanstd{41.0}{1.8} & \meanstd{2.8}{0.3} & \meanstd{91.2}{2.3} & \meanstd{28.0}{0.3} & \meanstd{61.5}{0.4} & \meanstd{7.8}{0.2} \\

  & MORSE & UNC & \meanstd{90.0}{0.2} & \meanstd{13.6}{0.3} & \meanstd{0.5}{0.1}
    & \meanstd{70.5}{0.9} & \meanstd{25.8}{0.7} & \meanstd{5.3}{0.2}
    & \meanstd{63.8}{0.3} & \meanstd{34.9}{0.4} & \meanstd{11.5}{0.3}
    & {\meanstd{32.6}{1.0}} & \meanstd{72.3}{1.0} & \meanstd{2.9}{0.3} \\
    

  & \textbf{CITADEL} & \textbf{MC}
    & \textbf{\meanstd{92.6}{0.2}} & \textbf{\meanstd{10.6}{0.2}} & \textbf{\meanstd{0.3}{0.0}}
    & \textbf{\meanstd{76.1}{0.4}} & \textbf{\meanstd{31.8}{0.6}} & \textbf{\meanstd{0.3}{0.0}}
    & \textbf{\meanstd{75.4}{0.4}} & \meanstd{26.5}{1.1} & \textbf{\meanstd{2.8}{0.2}}
    & \textbf{\meanstd{40.8}{1.5}} & \meanstd{37.6}{2.1} & \meanstd{20.4}{1.0}
    \\

\midrule
\multirow{8}{*}{400}
  & Binary SVM & UNC & \meanstd{88.9}{0.1} & \meanstd{12.5}{0.2} & \meanstd{0.8}{0.0} & \meanstd{70.7}{0.3} & \meanstd{38.2}{0.8} & \meanstd{0.4}{0.1} & \meanstd{38.4}{0.2} & \meanstd{3.5}{0.3} & \meanstd{92.0}{1.4} & \meanstd{22.7}{0.3} & \meanstd{65.5}{0.4} & \meanstd{8.1}{0.2} \\

  & Binary GBDT & UNC & \meanstd{86.4}{0.2} & \meanstd{20.0}{0.4} & \meanstd{0.5}{0.0} & \meanstd{70.0}{0.2} & \meanstd{38.0}{0.3} & \meanstd{0.4}{0.0} & \meanstd{38.3}{0.2} & \meanstd{3.6}{0.2} & \meanstd{91.8}{0.9} & \meanstd{22.9}{0.3} & \meanstd{65.1}{0.4} & \meanstd{8.0}{0.2} \\

  & TRANSCENDENT & Cred & \meanstd{78.9}{0.2} & \meanstd{25.1}{0.2} & \meanstd{0.0}{0.0} & \meanstd{43.9}{0.4} & \meanstd{65.4}{0.6} & \meanstd{0.9}{0.0} & \meanstd{40.6}{1.1} & \meanstd{15.6}{1.3} & \meanstd{82.8}{1.2} & \meanstd{24.0}{0.3} & \meanstd{63.8}{0.4} & \meanstd{7.8}{0.2} \\

  & CADE & OOD & \meanstd{89.1}{0.5} & \meanstd{13.2}{0.5} & \meanstd{0.8}{0.1} & \meanstd{60.1}{0.8} & \meanstd{51.4}{0.7} & \meanstd{0.4}{0.1} & \meanstd{45.4}{1.1} & \meanstd{59.2}{1.5} & \meanstd{10.1}{1.2} & \meanstd{28.0}{0.3} & \meanstd{59.0}{0.4} & \meanstd{6.9}{0.1} \\

  & Chen-AL & PS & \meanstd{92.4}{0.2} & \meanstd{10.2}{0.5} & \meanstd{0.4}{0.0} & \meanstd{75.5}{1.6} & \meanstd{31.4}{1.9} & \meanstd{0.5}{0.1} & \meanstd{43.0}{1.6} & \meanstd{2.5}{0.2} & \meanstd{89.8}{2.1} & \meanstd{30.0}{0.2} & \meanstd{57.0}{0.4} & \meanstd{6.2}{0.1} \\

  & MORSE & UNC & \meanstd{90.6}{0.2} & \meanstd{12.9}{0.3} & \meanstd{0.5}{0.0}
    & \meanstd{78.4}{1.1} & \meanstd{25.1}{0.6} & \meanstd{0.9}{0.1}
    & \meanstd{64.5}{0.3} & \meanstd{33.8}{0.4} & \meanstd{11.7}{0.3}
    & {\meanstd{36.0}{0.7}} & \meanstd{68.1}{0.9} & \meanstd{2.6}{0.3} \\
    

  & \textbf{CITADEL} & \textbf{MC}
    & \textbf{\meanstd{93.5}{0.3}} & \textbf{\meanstd{9.5}{0.7}} & \textbf{\meanstd{0.3}{0.0}}
    & \textbf{\meanstd{82.7}{2.6}} & \textbf{\meanstd{23.9}{3.0}} & \textbf{\meanstd{0.3}{0.1}}
    & \textbf{\meanstd{77.7}{0.1}} & \meanstd{24.0}{0.1} & \textbf{\meanstd{2.3}{0.4}}
    & \textbf{\meanstd{44.9}{0.0}} & \meanstd{52.7}{0.0} & \meanstd{5.0}{0.0}
    \\

\bottomrule
\end{tabular}
}
\end{table*}

%% file: tables/complexity_comparison_table.tex
\begin{table}[!t]
\centering
\scriptsize
\caption{Comparison of Runtime and \#of Operations.}
\label{tab:complexity_comparison}
\resizebox{\columnwidth}{!}{
\begin{tabular}{c|c|c|c|c|c|c}
\toprule
\textbf{n} & \textbf{Chen-AL (S)} & \textbf{\system (S)} & \textbf{Speedup} & \textbf{Chen-AL Op} & \textbf{\system Op} & \textbf{OpRed} \\
\midrule
$1\mathrm{e}{2}$   & 210.0  & 1.7   & \textbf{123.5$\times$} & 2.63e6  & 3.89e4  & \textbf{67.6$\times$} \\
$1\mathrm{e}{3}$   & 118.0  & 1.0   & \textbf{118.0$\times$} & 3.14e6  & 4.31e4  & \textbf{72.9$\times$} \\
$5\mathrm{e}{3}$   & 188.0  & 6.1   & \textbf{30.8$\times$}  & 5.35e6  & 2.13e5  & \textbf{25.1$\times$} \\
$1\mathrm{e}{4}$   & 273.0  & 10.2  & \textbf{26.8$\times$}  & 8.11e6  & 4.25e5  & \textbf{19.1$\times$} \\
$5\mathrm{e}{4}$   & 1030.0 & 45.4  & \textbf{22.7$\times$}  & 3.02e7  & 2.12e6  & \textbf{14.2$\times$} \\

 \textbf{$1\mathrm{e}{5}$}   & 2312.0 & 95.8  & \textbf{24.1$\times$}  & 5.79e7  & 4.24e6  & \textbf{13.6$\times$} \\
$5\mathrm{e}{5}$   & 9909.0 & 464.4 & \textbf{21.3$\times$}  & 2.58e8  & 2.12e7  & \textbf{12.2$\times$} \\
\bottomrule
\end{tabular}
}
\end{table}

%% file: tables/ablation-study-APIGraph-LAMDA.tex
\begin{table*}[!t]
\centering
\footnotesize
\caption{Ablation study of \system on APIGraph~\cite{api_graph_dataset}, MaMaDroid~\cite{mamadroid}, and LAMDA~\cite{haque2025lamda}. Results are reported in \%.}
\label{tab:ablation_apigraph_mamadroid_lamda}
\begin{tabular}{l|ccc|ccc|ccc}
\toprule
\multirow{2}{*}{\textbf{Configuration}} 
& \multicolumn{3}{c|}{\textbf{APIGraph}} 
& \multicolumn{3}{c|}{\textbf{MaMaDroid}} 
& \multicolumn{3}{c}{\textbf{LAMDA}} \\ 
\cline{2-10}
& \textbf{F1 Score} & \textbf{FNR} & \textbf{FPR} 
& \textbf{F1 Score} & \textbf{FNR} & \textbf{FPR} 
& \textbf{F1 Score} & \textbf{FNR} & \textbf{FPR} \\ 
\midrule

\multicolumn{10}{c}{\textbf{Baseline and Augmentation Only}} \\
\midrule
FixMatch-Baseline (No Aug, No AL) 
& 0.0{\tiny $\pm$0.0} & 1.0{\tiny $\pm$0.0} & 0.0{\tiny $\pm$0.0} 
& 0.0{\tiny $\pm$0.0} & 1.0{\tiny $\pm$0.0} & 0.0{\tiny $\pm$0.0}
& 0.0{\tiny $\pm$0.0} & 1.0{\tiny $\pm$0.0} & 0.0{\tiny $\pm$0.0} \\

+ Our augmentation only (no AL) 
& \textbf{67.2{\tiny $\pm$2.9}} & 43.6{\tiny $\pm$3.0} & 9.5{\tiny $\pm$1.1} 
& \textbf{\meanstd{26.2}{0.1}} & \meanstd{75.6}{0.2} & \meanstd{2.9}{0.0} 
& \textbf{30.9{\tiny $\pm$0.9}} & 74.8{\tiny $\pm$0.9} & 2.9{\tiny $\pm$0.4} \\

\midrule \midrule
\multicolumn{10}{c}{\textbf{Baseline + Augmentation + Active Learning}} \\
\midrule
+ AL (Random Sampling) 
& 79.9{\tiny $\pm$1.5} & 25.4{\tiny $\pm$2.0} & 1.2{\tiny $\pm$0.3} 
& \meanstd{27.8}{0.1} & \meanstd{76.6}{0.2} & \meanstd{2.9}{0.0} 
& 36.1{\tiny $\pm$3.8} & 67.2{\tiny $\pm$4.2} & 5.2{\tiny $\pm$0.6} \\

+ AL (Margin Only) 
& 92.3{\tiny $\pm$0.2} & 11.0{\tiny $\pm$0.3} & 0.4{\tiny $\pm$0.0} 
& \meanstd{42.7}{0.1} & \meanstd{54.1}{0.2} & \meanstd{5.3}{0.0}  
& 75.7{\tiny $\pm$1.4} & 25.7{\tiny $\pm$1.2} & 2.0{\tiny $\pm$0.2} \\

+ AL ($L_p$-Norm Only) 
& 92.1{\tiny $\pm$0.2} & 12.1{\tiny $\pm$0.2} & 0.3{\tiny $\pm$0.0} 
& \meanstd{43.1}{0.9} & \meanstd{53.7}{0.7} & \meanstd{5.2}{0.3}
& 75.3{\tiny $\pm$0.2} & 26.8{\tiny $\pm$0.3} & 2.4{\tiny $\pm$0.2} \\

+ AL (Low Confidence Only $<$75\%) 
& 85.1{\tiny $\pm$0.2} & 20.1{\tiny $\pm$0.3} & 0.5{\tiny $\pm$0.0} 
& \meanstd{28.4}{0.1} & \meanstd{73.9}{0.2} & \meanstd{2.6}{0.0} 
& 69.1{\tiny $\pm$0.2} & 35.9{\tiny $\pm$0.3} & 2.1{\tiny $\pm$0.2} \\

\midrule \midrule
\multicolumn{10}{c}{\textbf{Baseline + Augmentation + Active Learning With Multi-Criteria Selection}} \\
\midrule
+ Our Multi-Criteria (w/o Loss) 
& 93.0{\tiny $\pm$0.2} & 10.2{\tiny $\pm$0.2} & 0.3{\tiny $\pm$0.0} 
& \meanstd{44.6}{0.6} & \meanstd{53.0}{0.5} & \meanstd{5.2}{0.3} 
& 77.0{\tiny $\pm$0.2} & 24.8{\tiny $\pm$0.2} & 2.6{\tiny $\pm$0.3} \\

+ Our Multi-Criteria + Combined Loss 
& \textbf{93.5{\tiny $\pm$0.3}} & \textbf{9.5{\tiny $\pm$0.7}} & \textbf{0.3{\tiny $\pm$0.0}} 
&\textbf{\meanstd{44.9}{0.9}} & \meanstd{52.7}{0.8} & \meanstd{5.0}{0.4}
& \textbf{77.7{\tiny $\pm$0.1}} & \textbf{24.0{\tiny $\pm$0.1}} & \textbf{2.3{\tiny $\pm$0.4}} \\

\bottomrule
\end{tabular}
\end{table*}

%% file: 6_discussion.tex
\section{Discussion}

Our results suggest that semi-supervised active learning is a practical alternative to fully supervised malware detection under long-term distribution drift. Unlike prior methods that rely on semantic structures~\cite{chen2023continuous, yang2021cade}, \system remains effective even even such structures are absent, particularly under long-term evaluation scenarios. CITADEL consistently performs well with limited labeled data and achieves high computational efficiency. However, we observe diminishing gains beyond a 40\% label ratio, and classification remains challenging for singleton and ambiguous samples. 

\paragraph{Recommendation for Practitioners.} 
Based on our experience in developing CITADEL, we outline practical recommendations for adapting CITADEL. First, models should be initialized with reliable labeled data, as accurate early supervision supports stable downstream adaptation. 
Second, our results indicate that mild augmentation improves performance, whereas excessive augmentation can degrade performance (see Appendix~\ref{app:augSensitive} for sensitivity analysis). 
We further recommend using a high pseudo-label confidence threshold (\(\tau \geq 0.95\)), since reducing this threshold is found to negatively impact performance.

Third, class imbalance should be monitored carefully throughout training and retraining to avoid bias toward the majority class. 
Fourth, at scale, efficient training benefits from standard high-performance setups; including batch-wise computation with PyTorch tensors on CUDA-enabled GPUs; our replication artifacts follow this design. 
Finally, we recommend using \system's Multi-Criteria sample selection strategy since it prioritizes informative and ambiguous samples for labeling. In practice, integrating this selector with periodic human review supports continual adaptation to evolving drift while maintaining strong detection performance.


\paragraph{ML Attacks against CITADEL.}

As with other adaptive ML-based malware detectors, CITADEL operates in adversarial settings where evasion and poisoning attacks are well known challenges.
In particular, the use of pseudo-labeling and selective querying can, in principle, be targeted by clean-label poisoning attempts that aim to gradually influence the decision boundaries.
Our multi-criteria sample selection is designed to mitigate such risks by prioritizing ambiguous and informative samples rather than relying on a single heuristic. However, further strengthening against adversarial manipulation remains an active direction of future work.

%% file: 7_conclusion.tex
\section{Conclusion}

In this work, we present \system, a semi-supervised active learning framework for Android malware detection under long-term concept drift. \system combines domain-specific feature augmentations for binary malware features vectors with a multi-criteria active sample selection mechanism. Extensive evaluations on four real-world Android malware datasets show that \system consistently outperforms state-of-the-art baselines. On the most temporally drifted LAMDA dataset, \system achieves up to a 14\% F2 improvement while using only 40\% of the labeled data. It also provides substantial efficiency gains, with up to $24\times$ faster training and $13\times$ fewer operations. These results suggest that semi-supervised active learning-based approaches such as \system hold strong potential for real-world deployment due to their robustness, efficiency, and adaptability in the face of evolving malware threats.

%% file: appendix.tex
\section{Family and Singleton Statistics in the Datasets}
\label{app:singleton}

\input{tables/datasets-singleton}
Table~\ref{tab:families-singletons} presents the year-wise counts of singleton malware samples—those appearing only once—for the same datasets. APIGraph reports singletons from 2012–2018, peaking at 68 in 2013, with no data afterward. Chen-AZ includes singleton counts from 2019–2021, peaking at 42 in 2019. LAMDA shows extensive singleton data from 2013–2025, with a major surge between 2018 and 2022, peaking at 24,791 in 2020. Dashes indicate years without data. These singleton trends reflect temporal variations and dataset-specific dynamics, potentially signaling malware evolution or labeling noise. Additionally, Table \ref{tab:families-singletons} also summarizes the year-wise counts of malware families across the four datasets. APIGraph lists family totals from 2012–2018, increasing from 104 families in 2012 to a peak of 199 in 2016, followed by a gradual decline. Chen-AZ provides family counts only for 2019–2021, ranging from 51 to 121 during this period. LAMDA's family counts span from 2013 to 2025, starting with 213 families in 2013 and increasing over the years, reaching as high as 651 families in 2022. MaMaDroid reports family data for the same years as LAMDA, but its family count is relatively less. It reached a maximum of 134 families in a single year, and its lowest is one family in 2013–2014. Among all datasets, LAMDA exhibits the broadest temporal coverage and the highest family diversity.

\section{Visual Analysis of CITADEL vs Chen-AL Performance}
\label{app:CITADEL_Chen_visual_comp}

\begin{figure*}[t]
\centering
\setlength{\tabcolsep}{2pt}

\begin{tabular}{c c c}

\includegraphics[width=0.31\textwidth]{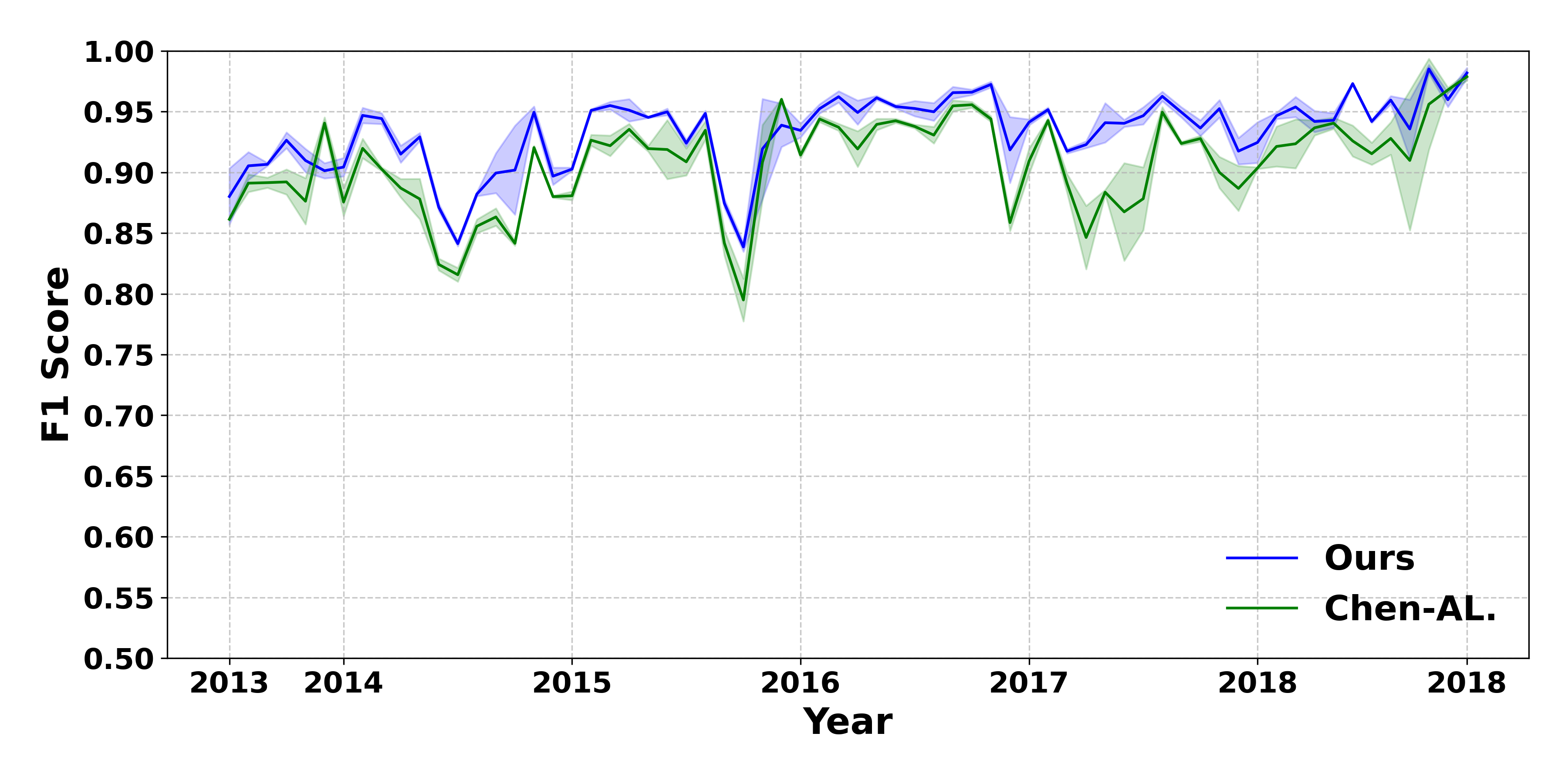} &
\includegraphics[width=0.31\textwidth]{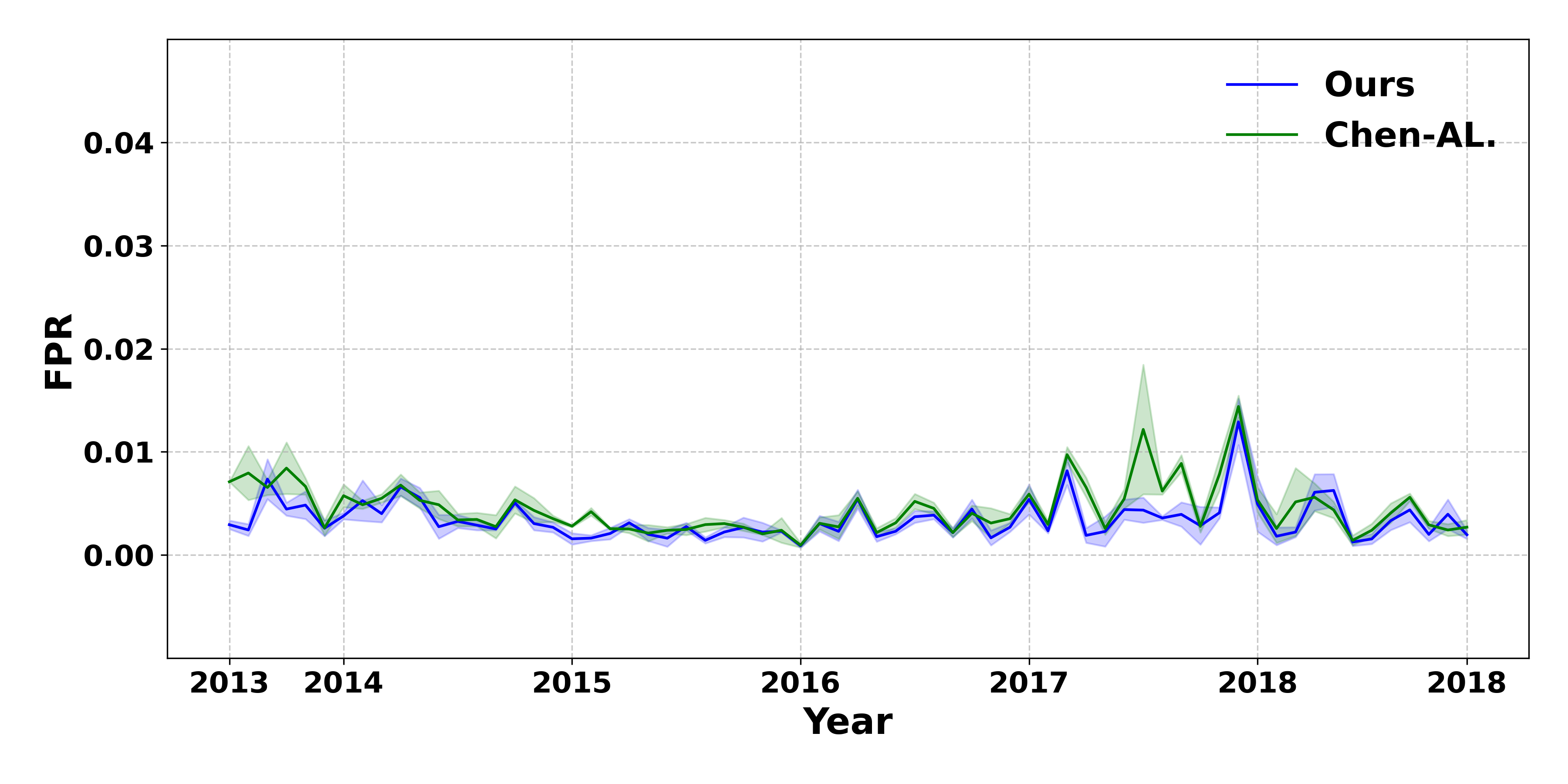} &
\includegraphics[width=0.31\textwidth]{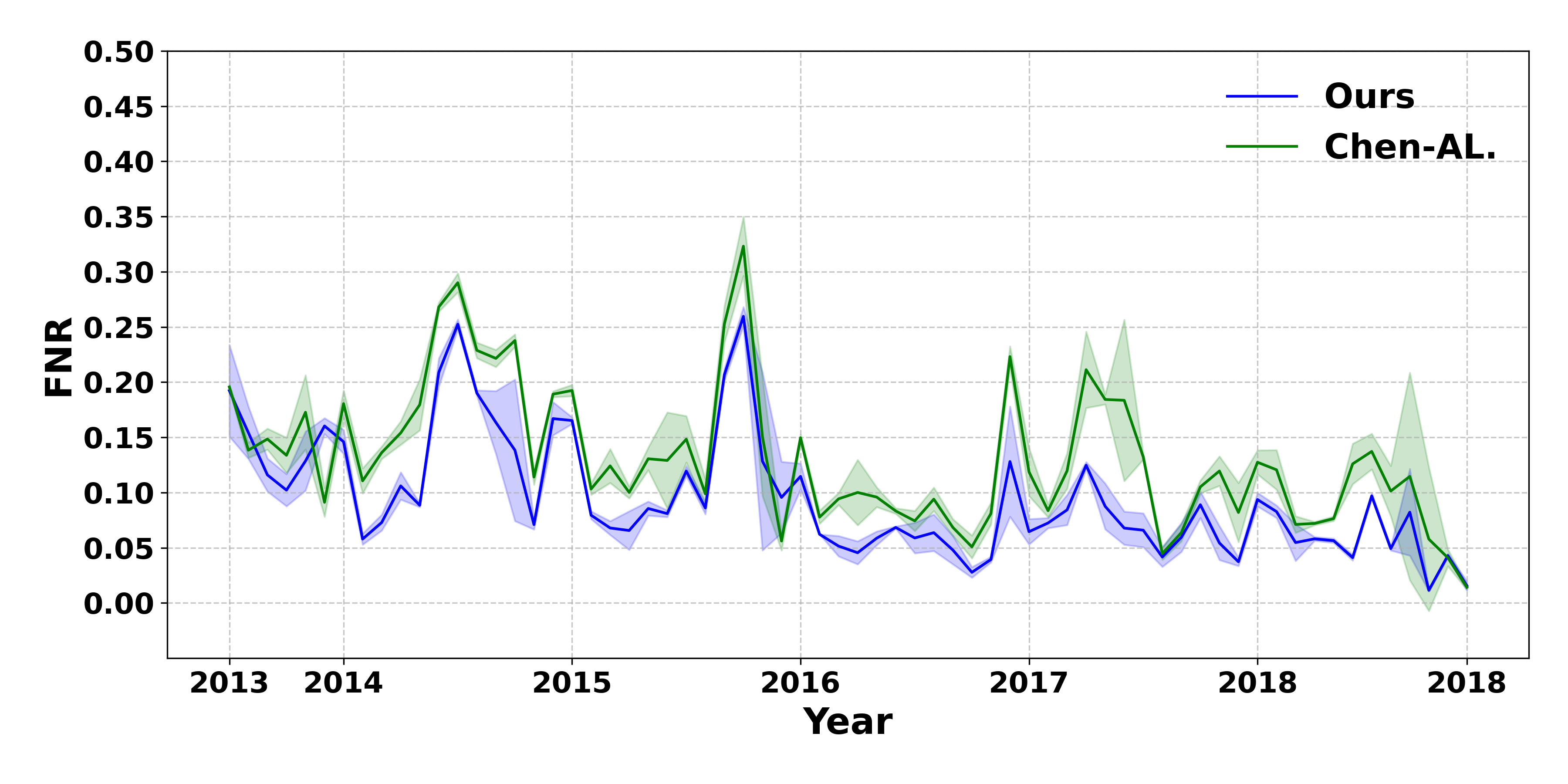} \\
\multicolumn{3}{c}{(a) APIGraph} \\[7pt]

\includegraphics[width=0.31\textwidth]{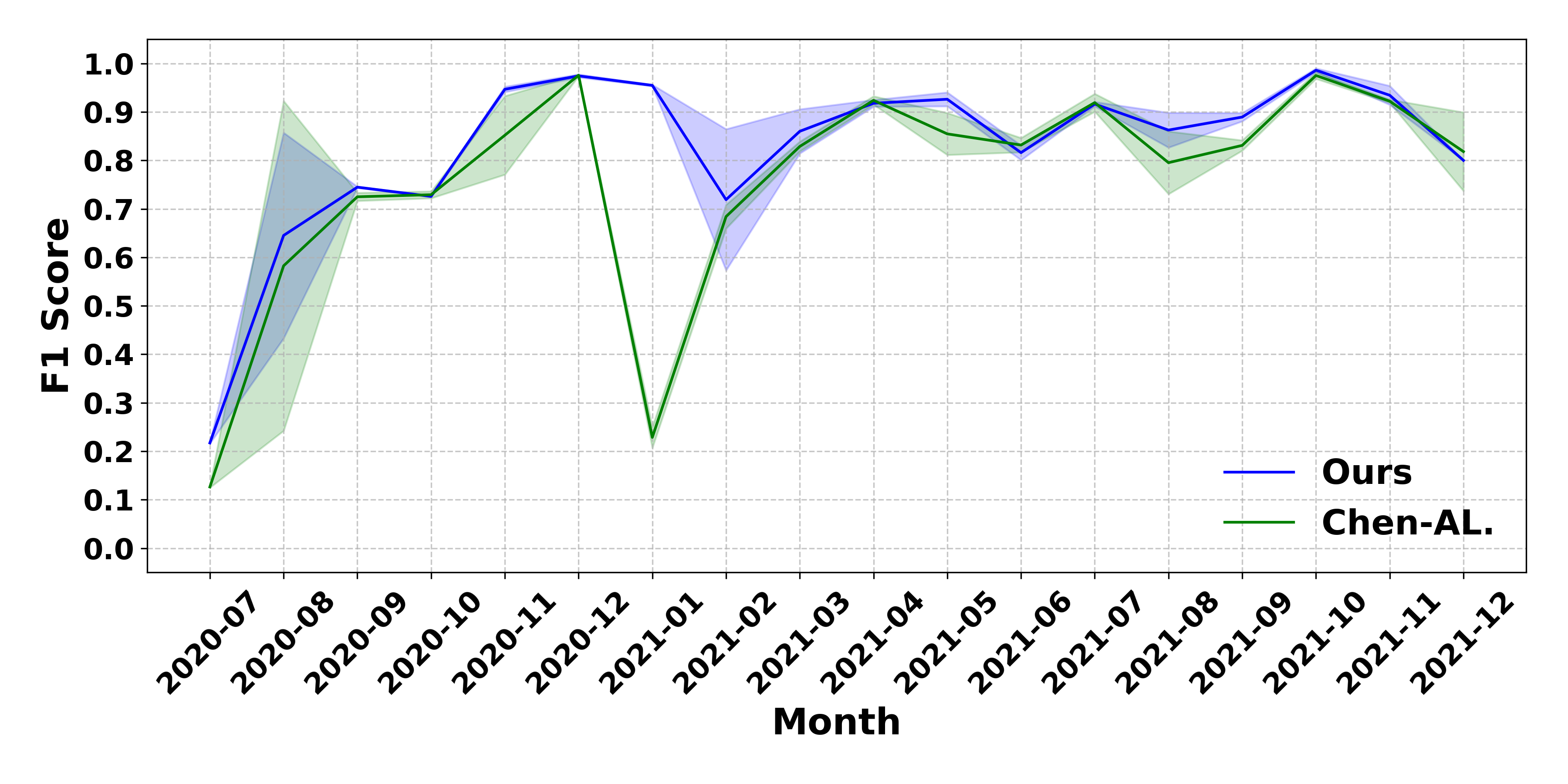} &
\includegraphics[width=0.31\textwidth]{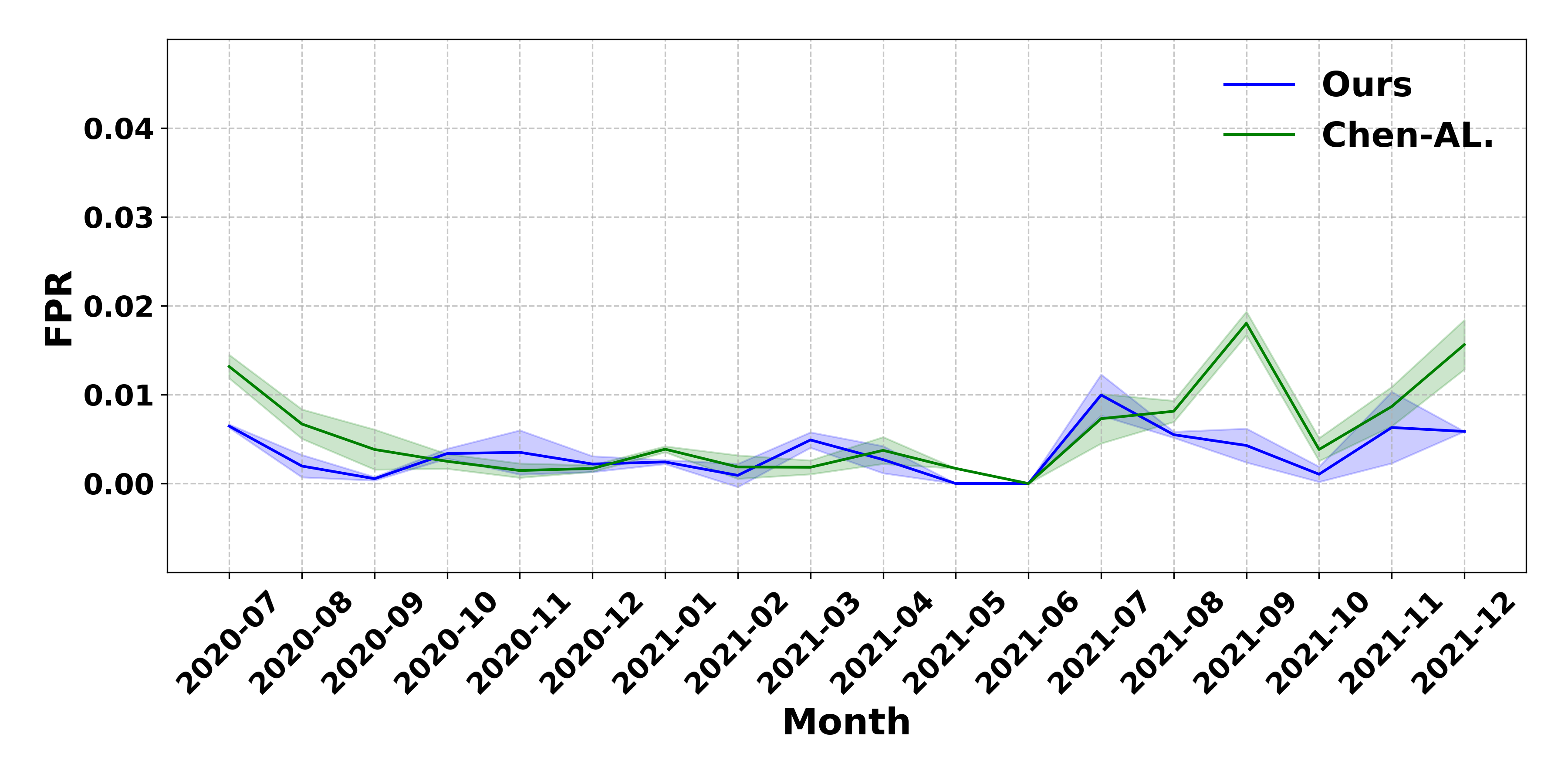} &
\includegraphics[width=0.31\textwidth]{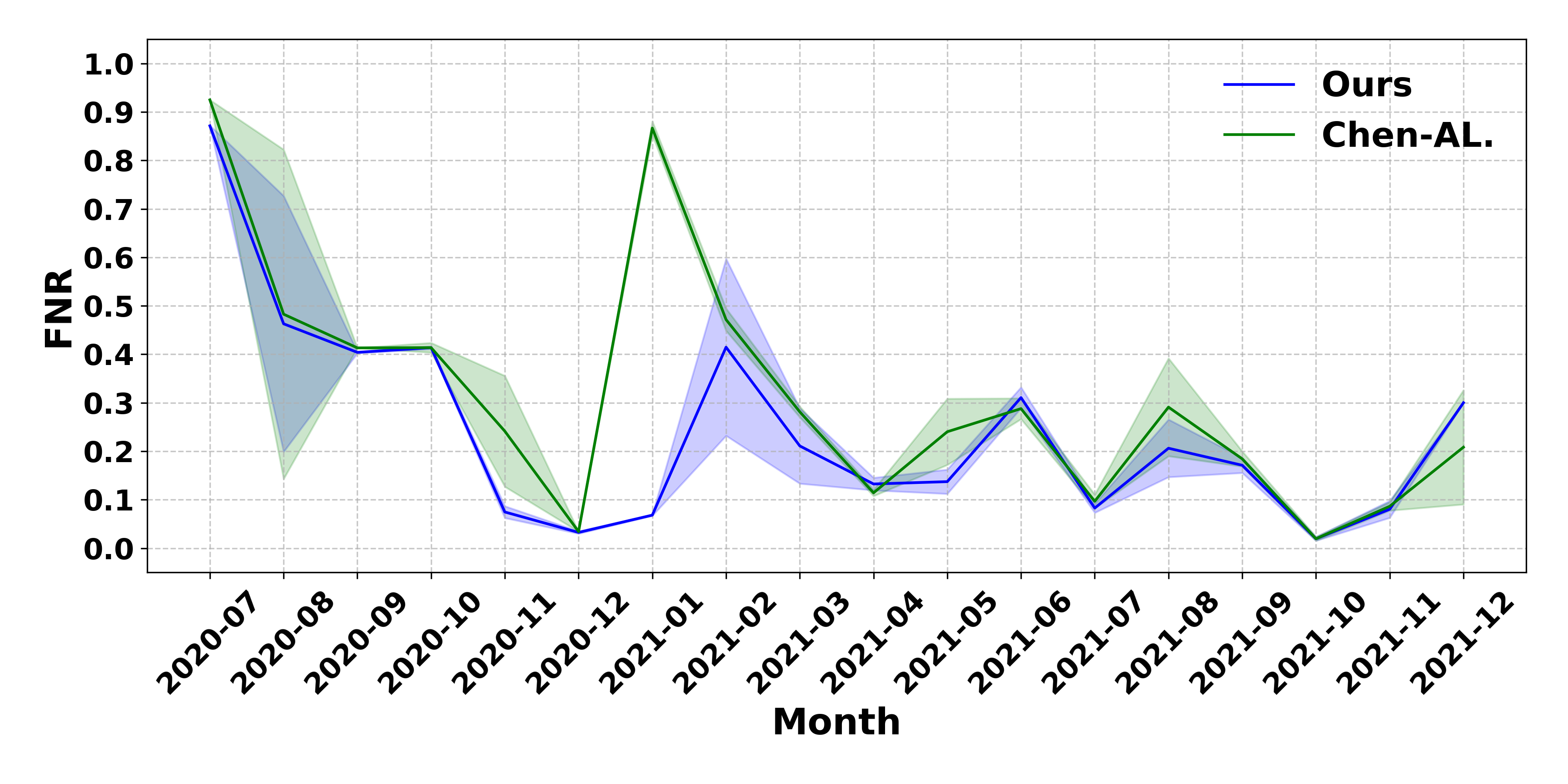} \\
\multicolumn{3}{c}{(b) Chen-AZ} \\[7pt]

\includegraphics[width=0.31\textwidth]{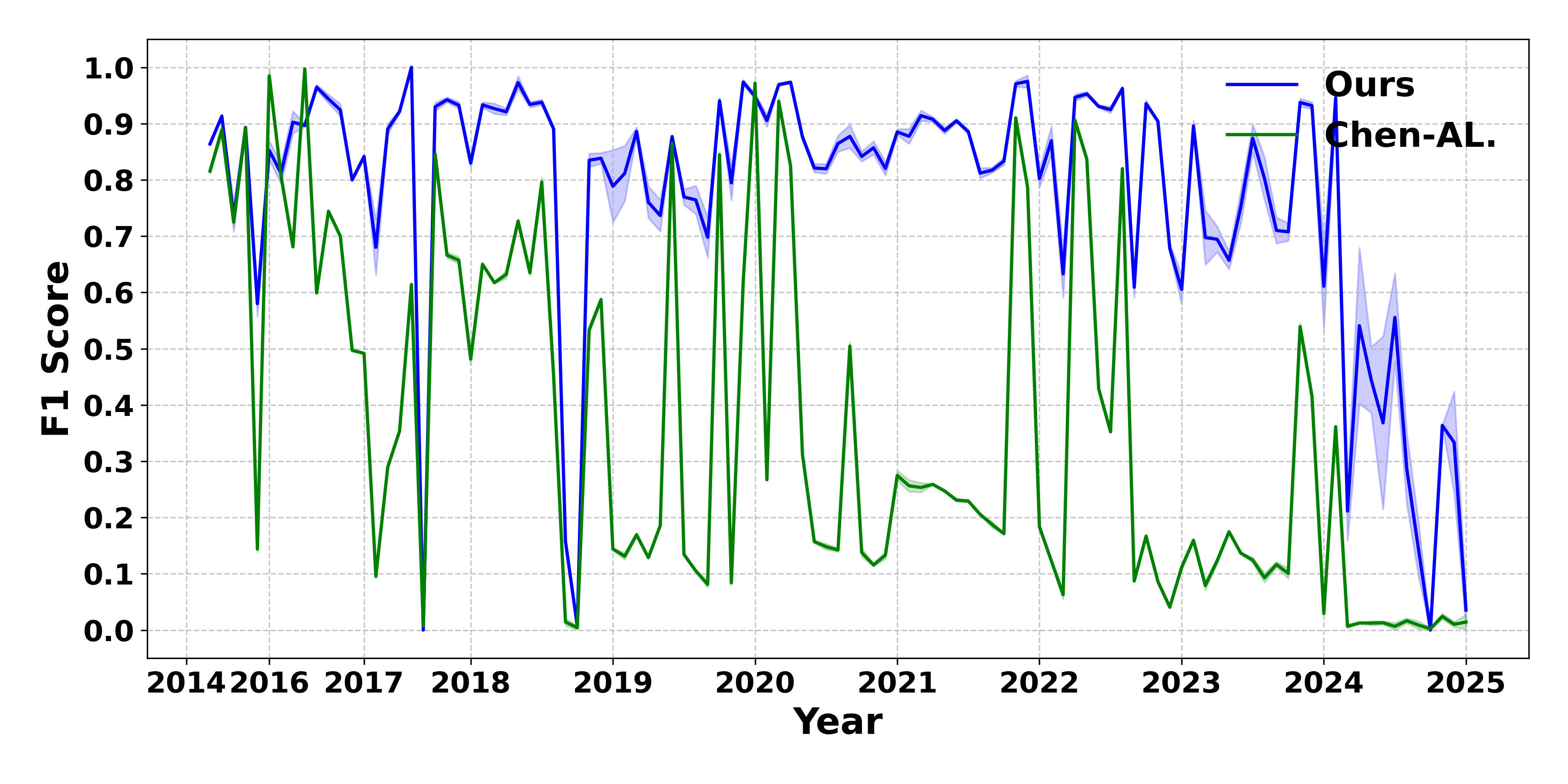} &
\includegraphics[width=0.31\textwidth]{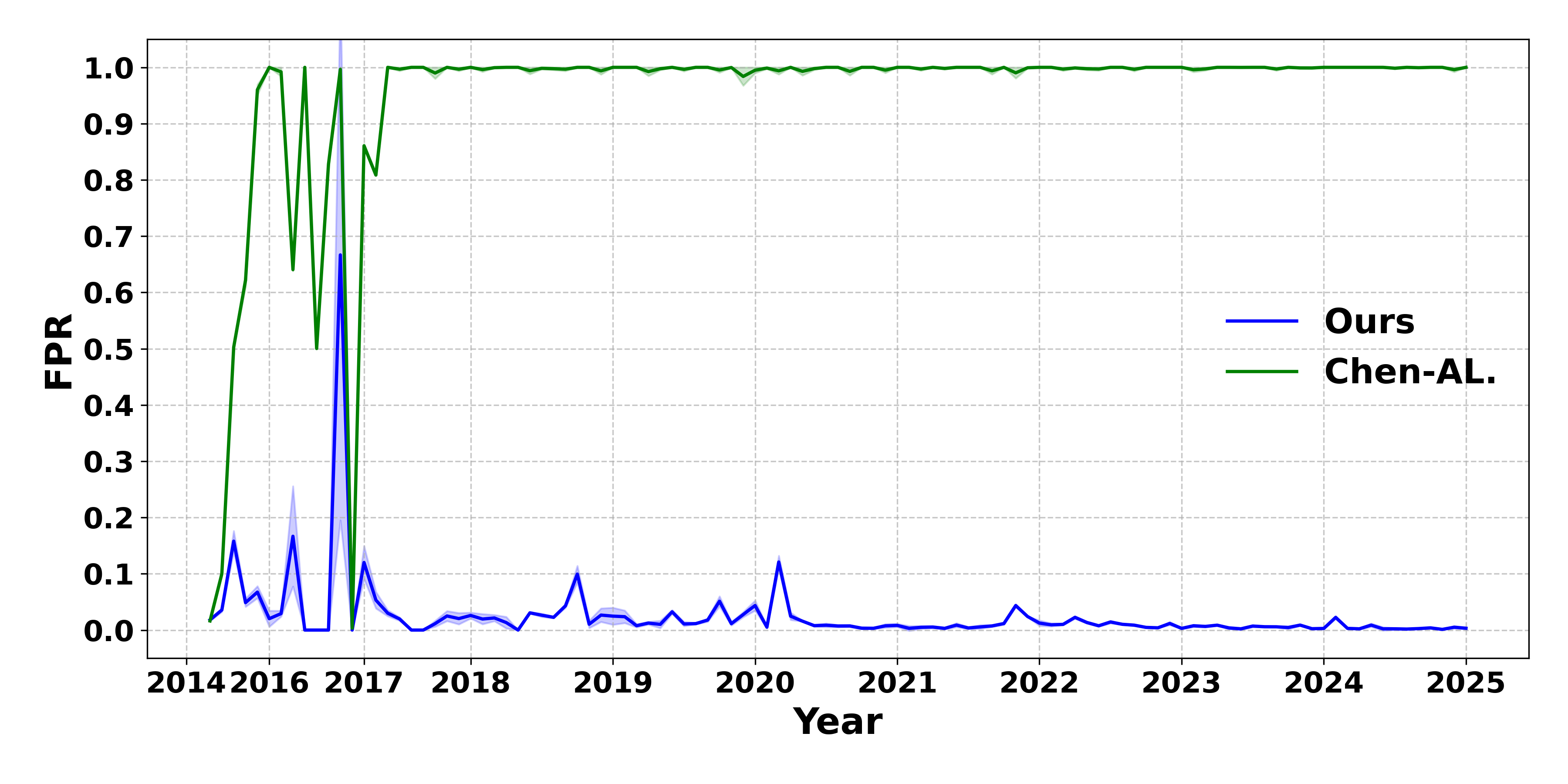} &
\includegraphics[width=0.31\textwidth]{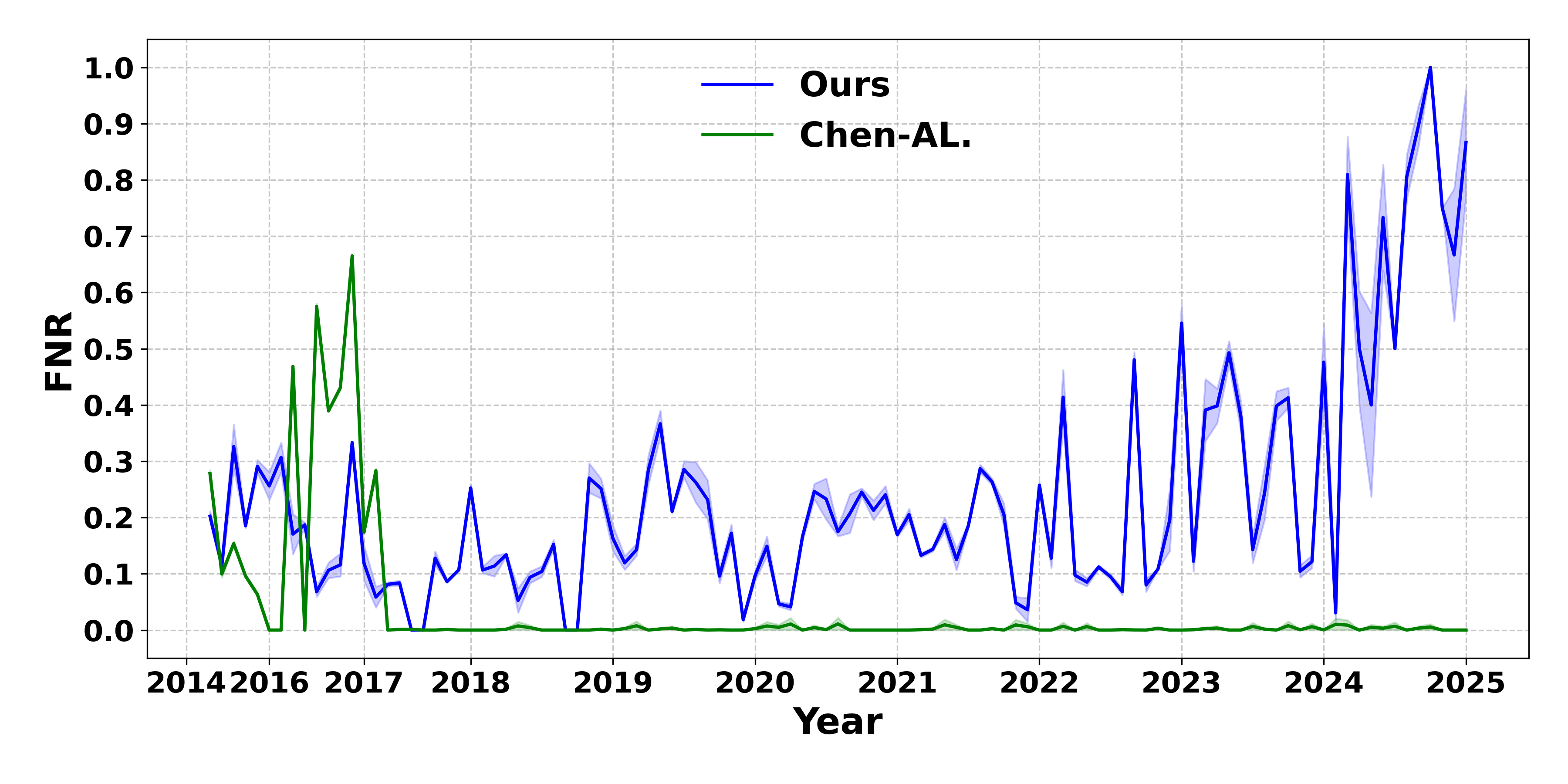} \\
\multicolumn{3}{c}{(c) LAMDA} \\[7pt]

\includegraphics[width=0.31\textwidth]{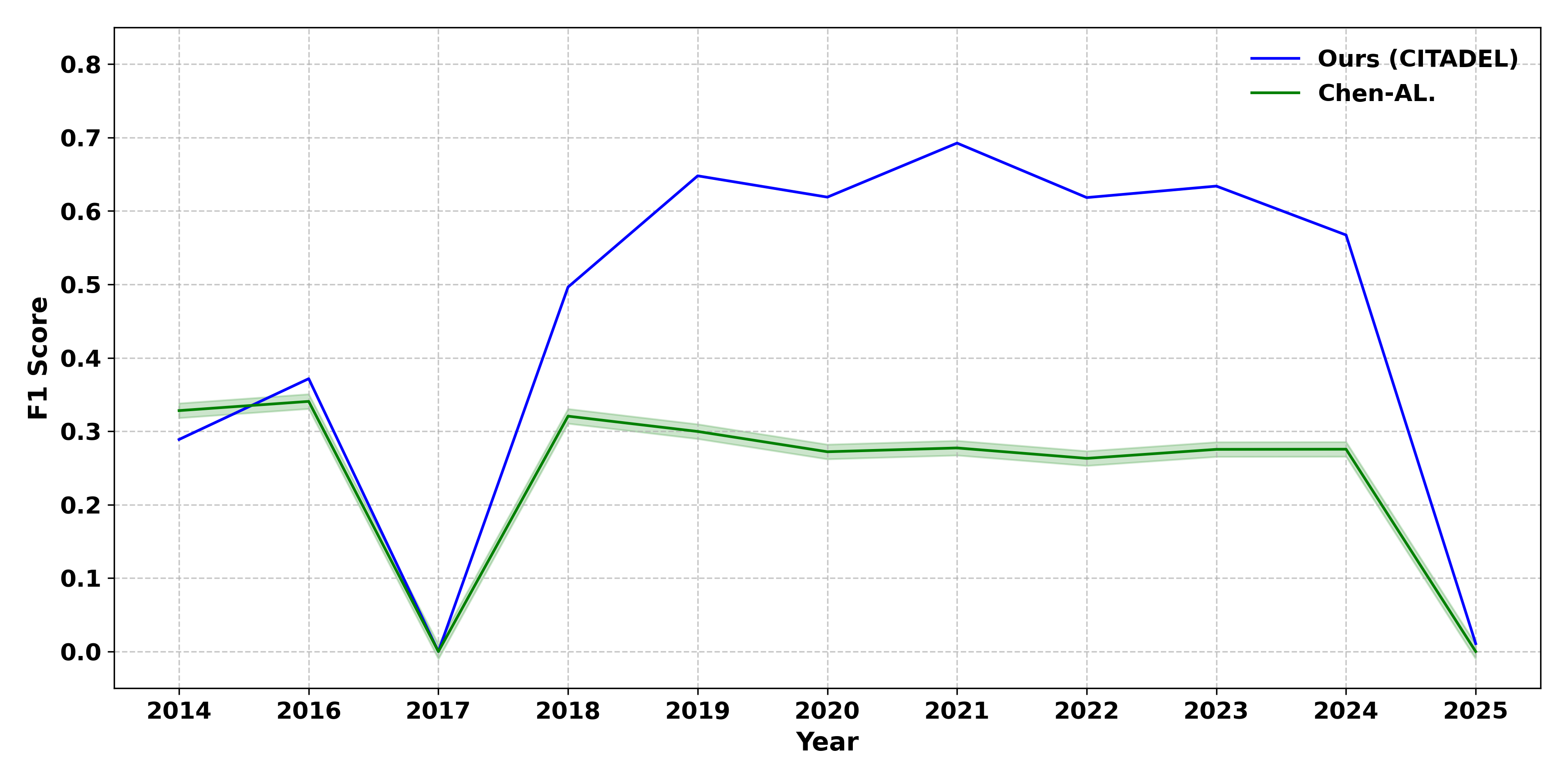} &
\includegraphics[width=0.31\textwidth]{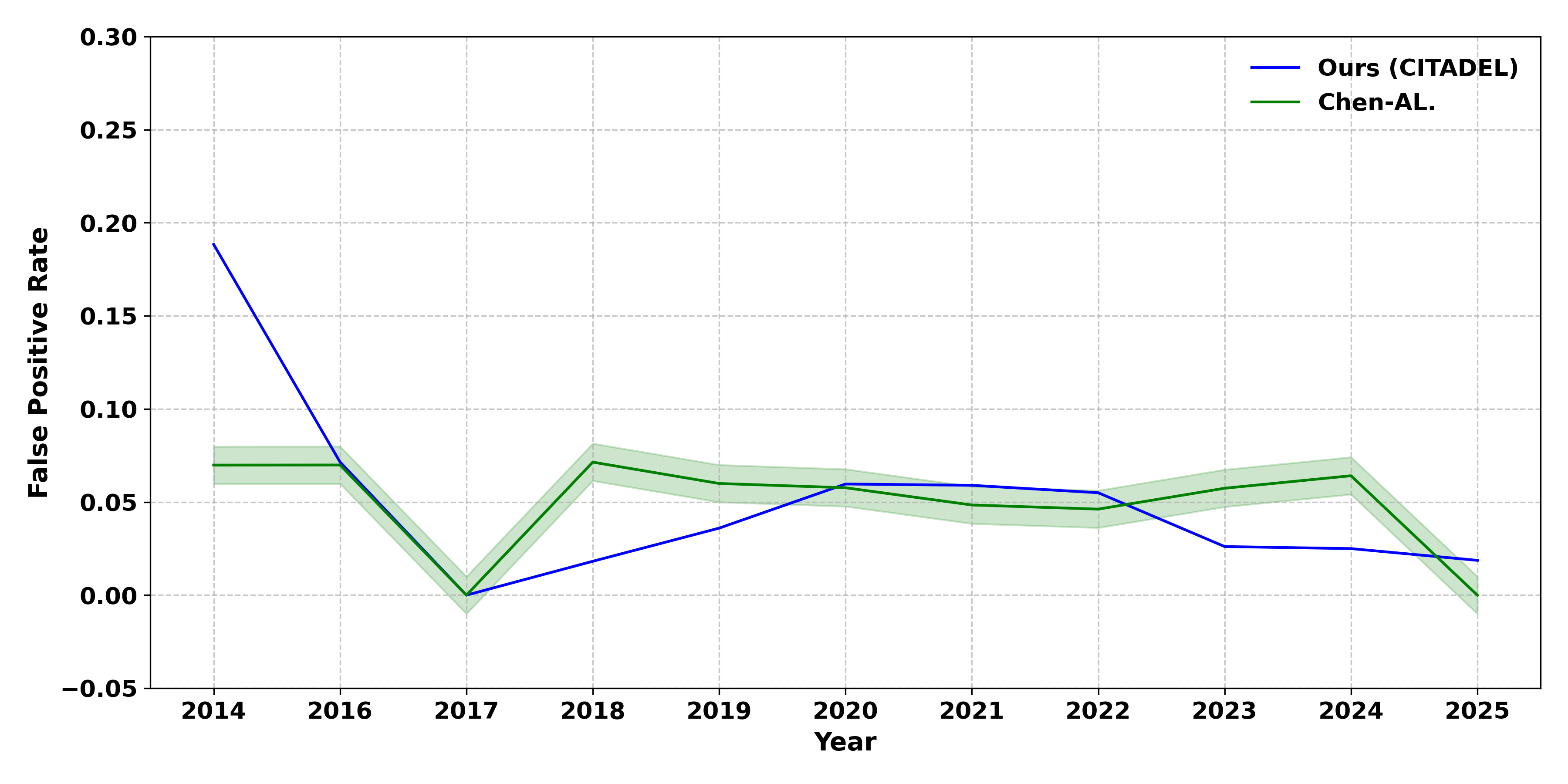} &
\includegraphics[width=0.31\textwidth]{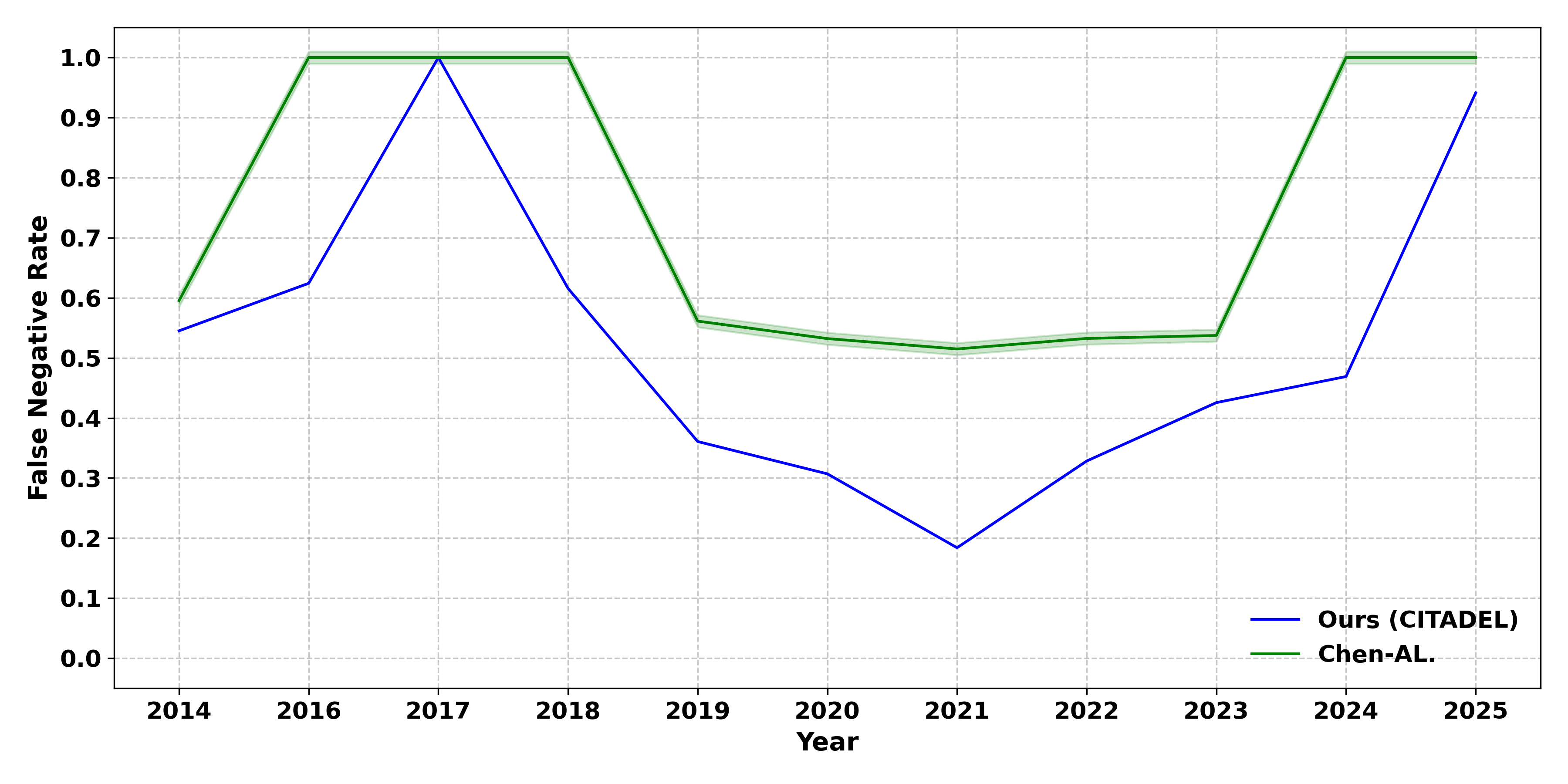} \\
\multicolumn{3}{c}{(d) MaMaDroid}

\end{tabular}

\vspace{0.2cm}
\caption{
\textbf{Chen-AL vs. \system}. Longitudinal performance comparison across datasets under concept drift.
Each subfigure corresponds to a dataset, and columns report F1 Score, False Positive Rate (FPR), and False Negative Rate (FNR).
}
\label{fig:longitudinal-comparison}
\end{figure*}

\begin{figure*}[!t]
    \centering
    \includegraphics[width=\textwidth]{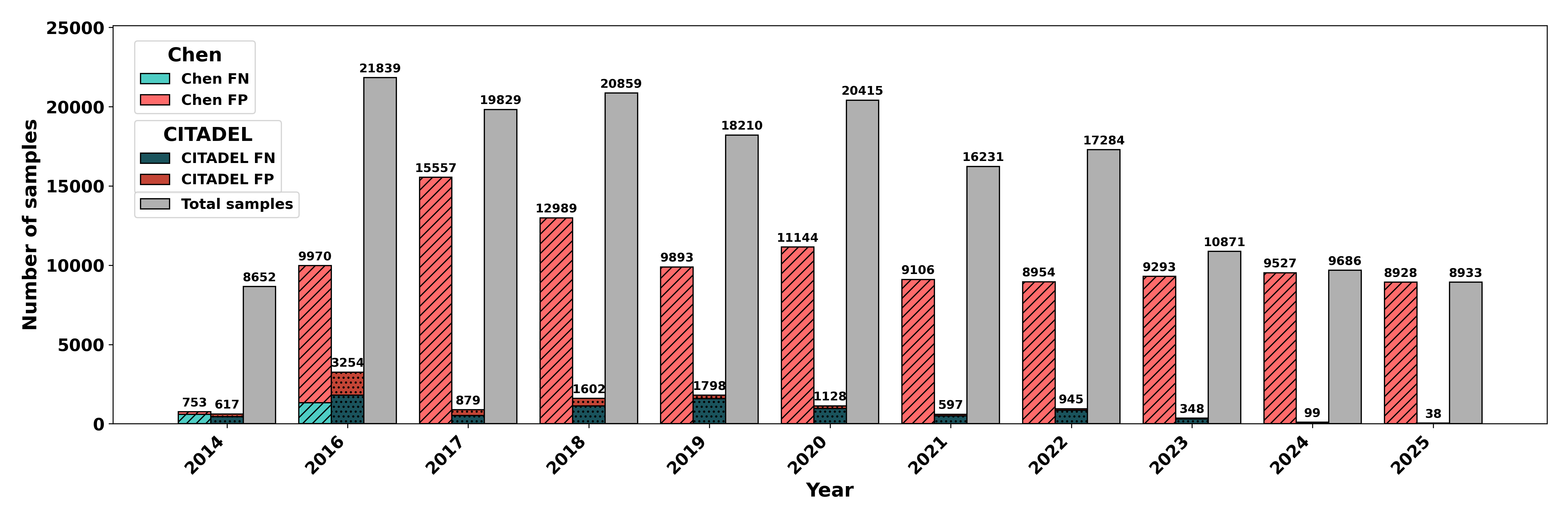}
    \caption{Yearwise comparison of False Negative (FN) and False Positive (FP) for Chen-AL vs CITADEL.}
    \label{fig:chen_vs_citadel_fn_fp}
\end{figure*}

Figures~\ref{fig:longitudinal-comparison} visualize monthly trends in F1-score, false positive rate (FPR), and false negative rate (FNR) for \system versus Chen-AL~\cite{chen2023continuous} on the APIGraph, Chen-AZ (AndroZoo), LAMDA, and MaMaDroid datasets, respectively. In addition, we provide an in-depth comparison between Chen-AL~\cite{chen2023continuous} and \system in terms of the total count of False Negative (FN) and False Positive (FP) over the year on LAMDA dataset~\cite{haque2025lamda} in Table~\ref{tab:lamda_fn_fp_onecol} and Figure~\ref{fig:chen_vs_citadel_fn_fp}. Clearly, after 2017, Chen-AL~\cite{chen2023continuous} fails to generalize between malware and benign samples. The low false-negative (FN) count and high false-positive (FP) count relative to the total number of samples indicate that the model increasingly classifies benign samples as malware, resulting in an effectively zero FN rate. Consequently, Chen-AL does not cope with dynamic decision boundary shifts induced by longitudinal concept drift, particularly on the LAMDA dataset~\cite{haque2025lamda}. Furthermore, substantial label drift causes a significant performance degradation in prior techniques, whereas \system consistently retains its performance even under these challenging conditions.

\input{tables/CITADEL_vs_Chen}

\section{CITADEL Hyperparameter tuning}
\label{app:hyperparameter}

We conducted the hyperparameter tuning experiments. First, we conducted the different label ratio experiment to show the reason for selecting 40\% labeled ratio for active learning scenario. We saw the performance plateau after 40\% labeled ratio. 


\begin{table}[t]
\centering
\caption{Active learning results of \system{} (CITADEL) under different labeled ratios on the APIGraph dataset. Results are shown as mean{\tiny$\pm$std} across three runs. Performance plateaus after 40\% labeled data, indicating label efficiency.}
\renewcommand{\arraystretch}{1.1}
\setlength{\tabcolsep}{3.5pt}
\begin{tabular}{@{}cccc@{}}
\toprule
\textbf{Labeled Ratio (\%)} & \textbf{Mean F1} & \textbf{Mean FNR} & \textbf{Mean FPR} \\ 
\midrule
10 & 92.99{\tiny$\pm$0.15} & 9.68{\tiny$\pm$0.12} & 0.39{\tiny$\pm$0.05} \\
20 & 93.09{\tiny$\pm$0.12} & 9.49{\tiny$\pm$0.10} & 0.41{\tiny$\pm$0.04} \\
30 & 93.39{\tiny$\pm$0.14} & 9.15{\tiny$\pm$0.09} & 0.38{\tiny$\pm$0.05} \\
40 & \textbf{93.57{\tiny$\pm$0.10}} & \textbf{8.64{\tiny$\pm$0.08}} & \textbf{0.40{\tiny$\pm$0.04}} \\
50 & 93.60{\tiny$\pm$0.12} & 8.79{\tiny$\pm$0.09} & 0.37{\tiny$\pm$0.05} \\
60 & 93.58{\tiny$\pm$0.11} & 8.81{\tiny$\pm$0.08} & 0.36{\tiny$\pm$0.04} \\
70 & 93.45{\tiny$\pm$0.13} & 9.16{\tiny$\pm$0.09} & 0.36{\tiny$\pm$0.05} \\
80 & 93.47{\tiny$\pm$0.10} & 9.08{\tiny$\pm$0.09} & 0.35{\tiny$\pm$0.05} \\
90 & 93.51{\tiny$\pm$0.12} & 8.95{\tiny$\pm$0.10} & 0.35{\tiny$\pm$0.04} \\
\bottomrule
\end{tabular}
\label{tab:citadel_label_ratio}
\end{table}

\subsection{Augmentation Sensitivity Analysis}
\label{app:augSensitive}

\paragraph{Experimental Setup.} 
We investigate the impact of weak ($p_w$) and strong ($p_s$) perturbation probabilities in CITADEL’s Bernoulli distribution--based augmentation on the APIGraph dataset under an active learning setting with a fixed labeling budget of 400 samples. We first conduct a coarse sweep in which the weak and strong perturbation probabilities are set to the same value ($p_w = p_s$) and varied from 0.0 to 0.9 at intervals of 0.1 to capture overall performance trends. Based on the observed behavior, we then perform a fine-grained sweep over a narrower range ($p_w = p_s \in [0.01, 0.09]$) to more precisely identify favorable augmentation strengths.

\begin{figure*}[t]
    \centering
    \includegraphics[width=0.32\textwidth]{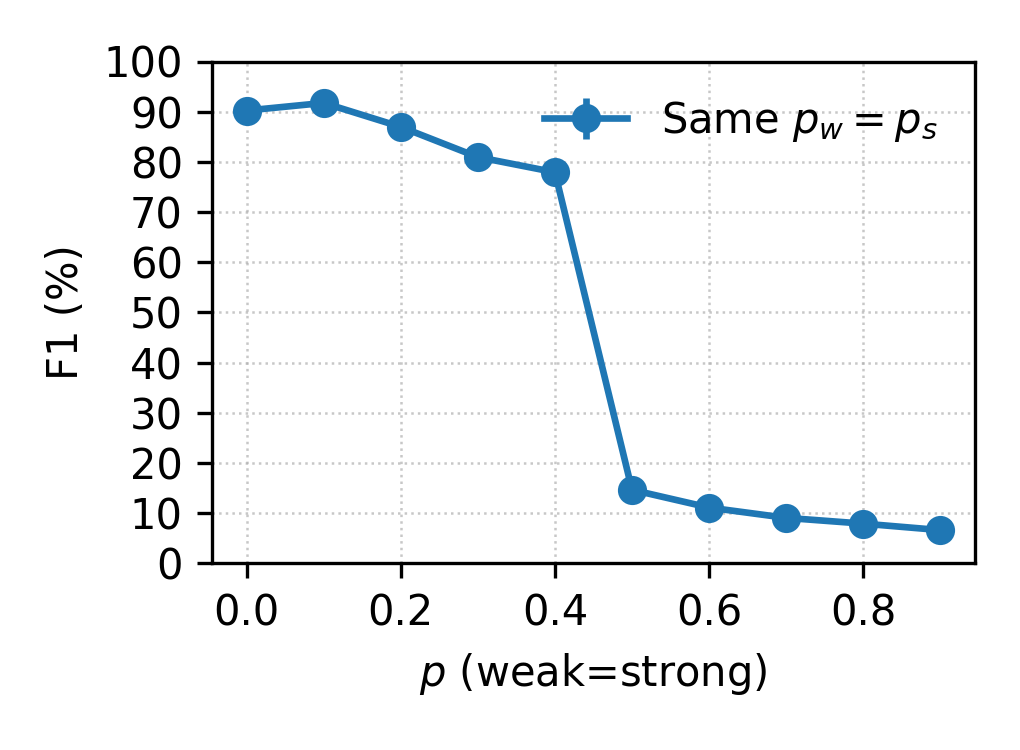}
    \hfill
    \includegraphics[width=0.32\textwidth]{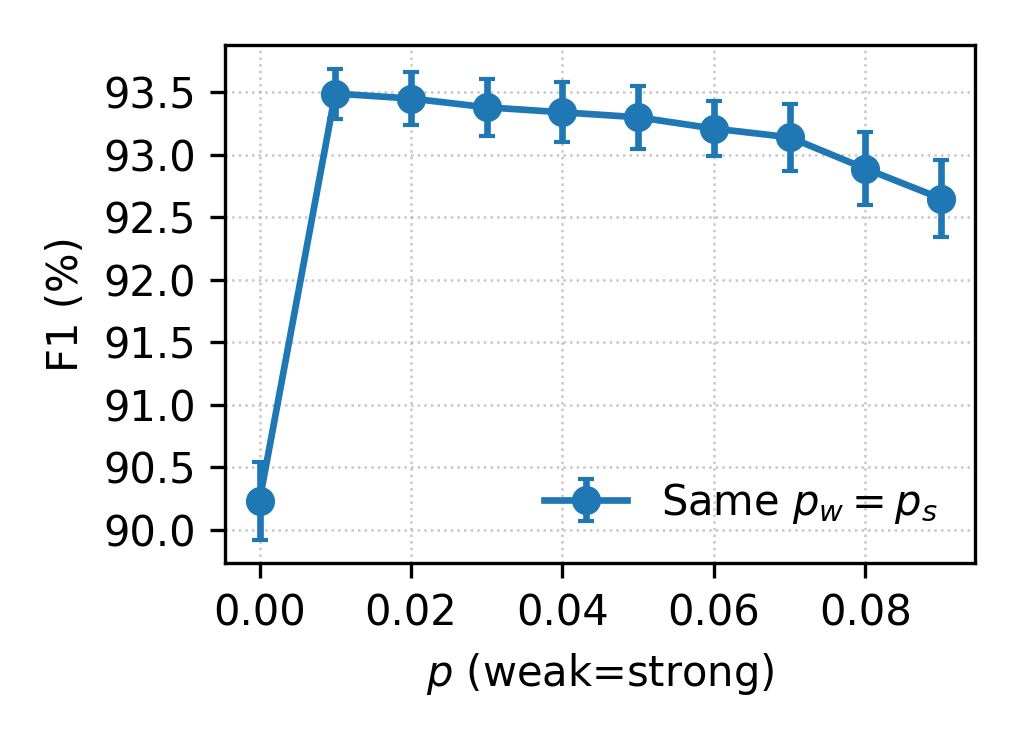}
    \hfill
    \includegraphics[width=0.32\textwidth]{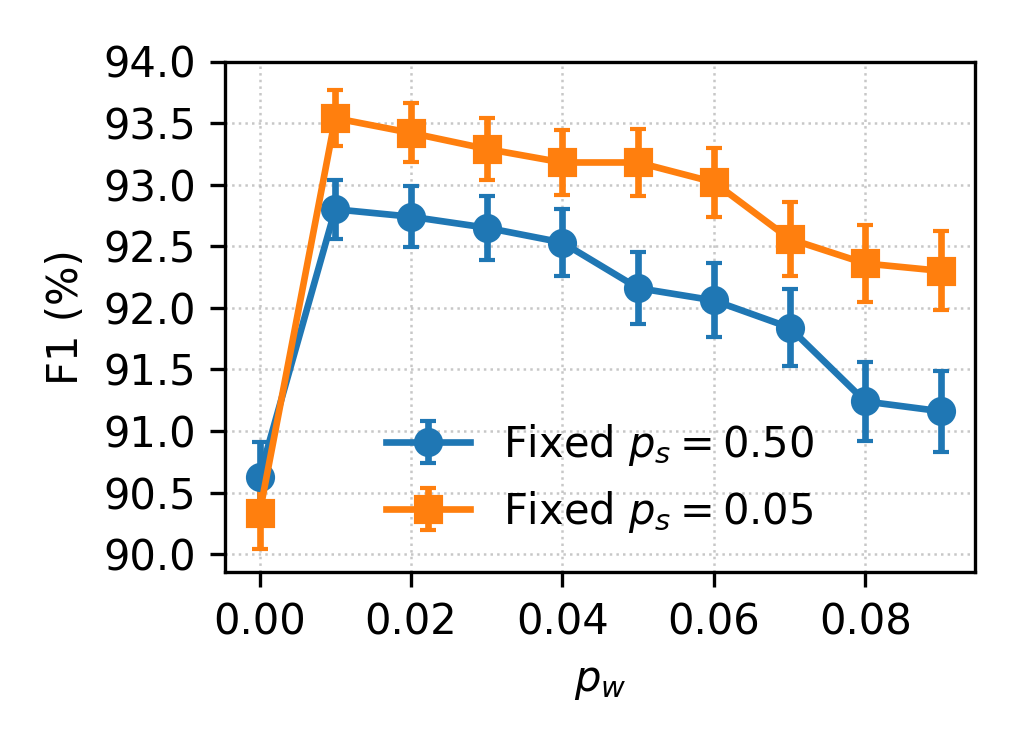}
    \caption{Sensitivity of CITADEL to augmentation probabilities on APIGraph.
(\textbf{Left}) Same weak and strong perturbation ($p_w=p_s$), coarse sweep.
(\textbf{Middle}) Same weak and strong perturbation ($p_w=p_s$), fine sweep.
(\textbf{Right}) Varying $p_w$ with fixed $p_s$.}

    \label{fig:aug-prob-apigraph}
\end{figure*}

To further examine the individual effects of weak and strong augmentations, we next control their probabilities independently. We fix the strong perturbation probability at a high value ($p_s = 0.5$) while varying the weak perturbation probability $p_w$ from 0.00 to 0.09. We then repeat this experiment with a low strong perturbation probability ($p_s = 0.05$), again varying $p_w$ over the same range. The results are reported in Table~\ref{tab:aug-prob-appendix-transposed} and illustrated in Figure~\ref{fig:aug-prob-apigraph}.

\paragraph{Analysis.} Across all configurations, introducing small amounts of perturbation consistently improves performance, with F1 scores increasing from 90.2\% to a maximum of 93.5\%. As the weak perturbation probability increases further, performance gradually declines. This degradation occurs because excessive weak augmentation introduces substantial noise into both labeled and unlabeled samples, which can distort class semantics and destabilize learning. In contrast, mild weak perturbations improve generalization by promoting robustness while preserving discriminative structure.

Strong augmentation exhibits a different trend. Even at higher strong perturbation probabilities, performance degradation is less severe. This behavior arises because strong augmentation is applied exclusively to the unsupervised branch, and CITADEL employs a confidence-based filtering mechanism that prevents heavily distorted samples from contributing to the consistency objective. Overall, these results highlight the importance of separately controlling weak and strong perturbations. The best performance is achieved with a small weak perturbation probability and a low strong perturbation probability—specifically, $p_w = 0.01$ and $p_s = 0.05$—which yields an effective balance between robustness and discriminative learning.


\begin{table*}[t]
\centering
\small
\setlength{\tabcolsep}{2.5pt}
\caption{Finding the best weak, $p_w$ and strong, $p_s$ augmentation probability for CITADEL on APIGraph.}
\label{tab:aug-prob-appendix-transposed}
\begin{tabular}{lcccccccccc}
\toprule

\multicolumn{11}{c}{\textbf{Same Weak and Strong Perturbation ($p_w = p_s$)}} \\
\midrule
$\boldsymbol{p}$ 
& 0.00 & 0.10 & 0.20 & 0.30 & 0.40 & 0.50 & 0.60 & 0.70 & 0.80 & 0.90 \\
\midrule
F1 & 90.23$\pm$\tiny{0.31} & 91.74$\pm$\tiny{0.27} & 86.96$\pm$\tiny{0.39} & 80.94$\pm$\tiny{0.48} & 77.92$\pm$\tiny{0.52} & 14.62$\pm$\tiny{0.22} & 11.08$\pm$\tiny{0.19} & 
9.03$\pm$\tiny{0.45} & 7.88$\pm$\tiny{0.39} & 6.59$\pm$\tiny{0.32} \\

\midrule
\midrule
$\boldsymbol{p}$ 
& 0.00 & 0.01 & 0.02 & 0.03 & 0.04 & 0.05 & 0.06 & 0.07 & 0.08 & 0.09 \\
\midrule
F1  &
90.23$\pm$\tiny{0.31} &
93.49$\pm$\tiny{0.20} &
93.45$\pm$\tiny{0.21} &
93.38$\pm$\tiny{0.23} &
93.34$\pm$\tiny{0.24} &
93.30$\pm$\tiny{0.25} &
93.21$\pm$\tiny{0.22} &
93.14$\pm$\tiny{0.27} &
92.89$\pm$\tiny{0.29} &
92.65$\pm$\tiny{0.31} \\

\midrule
\bottomrule
\multicolumn{11}{c}{\textbf{Varying Weak, Strong Fixed}} \\
\midrule
$\boldsymbol{p_w}$ 
& 0.00 & 0.01 & 0.02 & 0.03 & 0.04 & 0.05 & 0.06 & 0.07 & 0.08 & 0.09 \\
\midrule
F1 ($p_s=0.50$)  &
90.63$\pm$\tiny{0.28} &
92.80$\pm$\tiny{0.24} &
92.74$\pm$\tiny{0.25} &
92.65$\pm$\tiny{0.26} &
92.53$\pm$\tiny{0.27} &
92.16$\pm$\tiny{0.29} &
92.06$\pm$\tiny{0.30} &
91.84$\pm$\tiny{0.31} &
91.24$\pm$\tiny{0.32} &
91.16$\pm$\tiny{0.33} \\

\midrule
\midrule

F1 ($p_s=0.05$)  &
90.33$\pm$\tiny{0.29} &
\textbf{93.54$\pm$\tiny{0.23}} &
93.42$\pm$\tiny{0.24} &
93.29$\pm$\tiny{0.25} &
93.18$\pm$\tiny{0.26} &
93.18$\pm$\tiny{0.27} &
93.02$\pm$\tiny{0.28} &
92.56$\pm$\tiny{0.30} &
92.36$\pm$\tiny{0.31} &
92.30$\pm$\tiny{0.32} \\

\bottomrule
\end{tabular}%
\end{table*}

\subsection{Threshold and Coefficients} 
While \textsc{CITADEL} adopts several hyperparameters from FixMatch (e.g., $\tau = 0.95$ and $\lambda_{u} = 1$) as initial references, we further fine-tuned these and introduced a new hyperparameter, $\lambda_{\text{con}} = 0.5$. Our experiments show that both $\tau$ and $\lambda_{u}$ are sensitive to model stability—decreasing $\tau$ or increasing $\lambda_{u}$ introduces additional noise from unreliable pseudo-labels, thereby degrading performance when learning from unlabeled data. The selected configuration ($\tau = 0.95$, $\lambda_{u} = 1$) achieves the best trade-off between leveraging unlabeled data and maintaining detector reliability, confirming that our final settings are empirically justified for this domain. In addition, \textsc{CITADEL} introduces two additional hyperparameters, $\lambda_{\text{con}}=0.5$ and $p$, whose optimal values were determined through extensive experimental validation for the malware specific-domain.

\begin{figure}[t]
    \centering
    \includegraphics[width=0.47\textwidth]{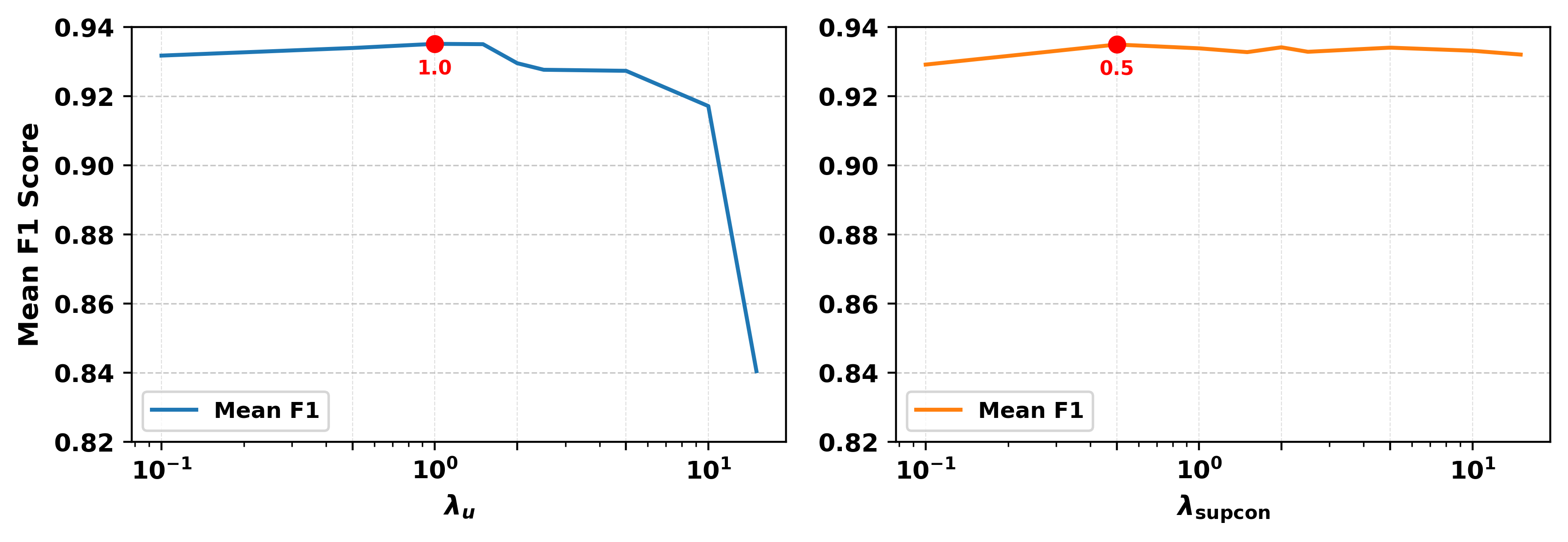}
    \caption{Hyperparameter tuning for $\lambda_u$ and $\lambda_{con}$.}
    \label{fig:hyperparameter_lambda}
\end{figure}

\begin{figure}[t]
    \centering
    \includegraphics[width=0.45\textwidth]{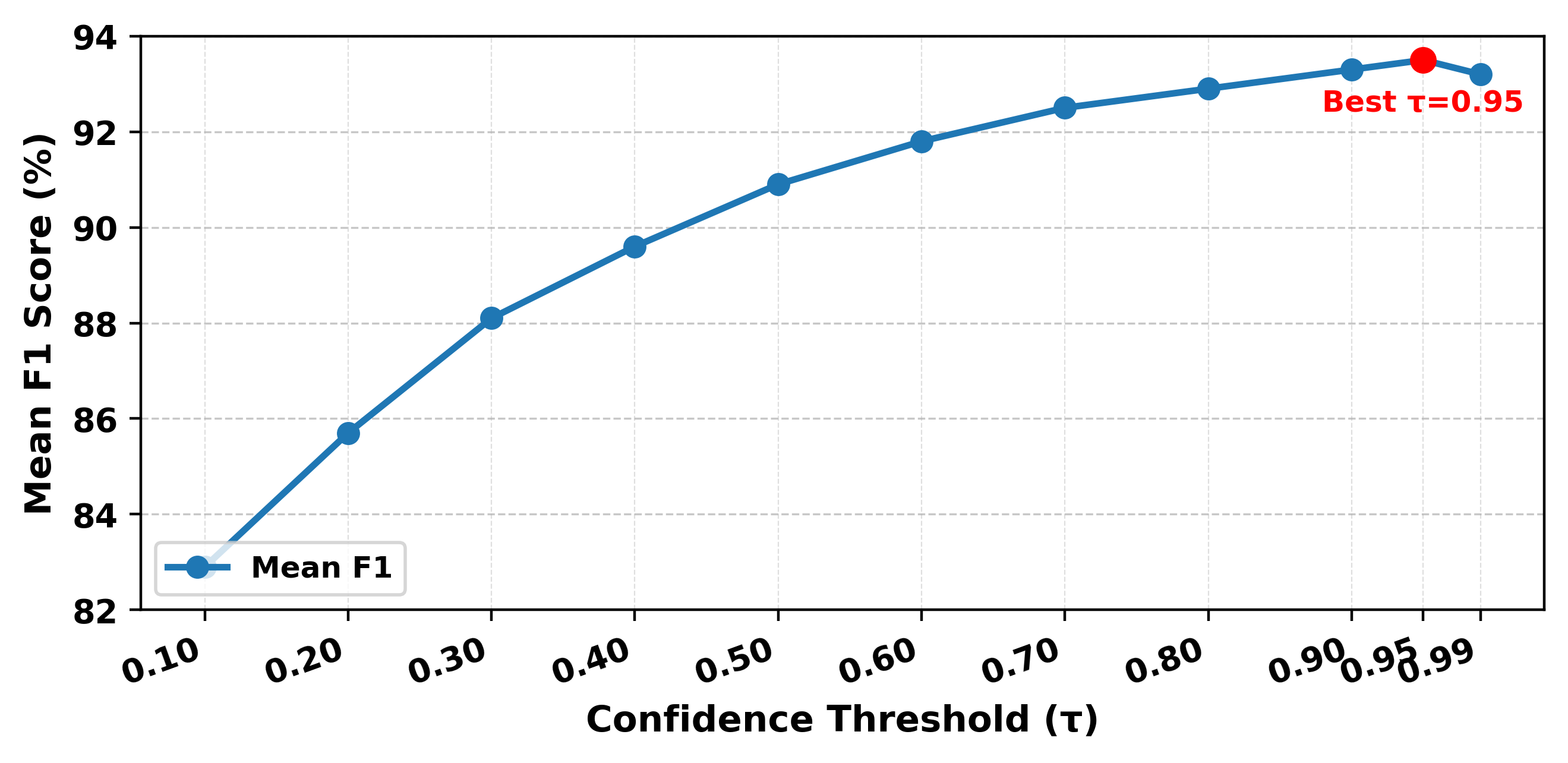}
    \caption{Hyperparameter tuning for threshold ($\tau$).}
    \label{fig:hyperparameter_threshold}
\end{figure}

\section{CITADEL Performance under Label Noise}
\label{app:label_noise}

\paragraph{Experiments.} 
To evaluate the performance of CITADEL against noisy supervision, we conduct a controlled experiment using the APIGraph dataset. We start with cleanly labeled Android malware samples and progressively inject synthetic label noise by randomly flipping a portion of the ground-truth labels to incorrect ones. The label noise rate varies from 0\% to 90\% in increments of 10\%. The training data may contain misclassified or mislabeled samples, which is common in large-scale malware datasets due to incorrect labels provided by automated anti-virus engines.


\paragraph{Results.} CITADEL demonstrates a strong tolerance to label corruption. As shown in Table~\ref{tab:label_noise_citadel}, CITADEL maintains an F1-score above 88\% for up to 20\% label noise, and even with extreme corruption levels of 90\% noisy labels, it achieves an F1-score of approximately 70\%. This corresponds to only a 23\% relative drop compared to the noise-free case. These results highlight that CITADEL retains stable, even under heavily corrupted supervision.

\begin{table}[!t]
\centering
\caption{CITADEL performance under varying label noise rates. 
All values are in percentage (\meanstd{mean}{std}) over multiple runs.}
\label{tab:label_noise_citadel}

\begin{tabular}{c|ccc}
\toprule
\textbf{Noise (\%)} & \textbf{F1} & \textbf{FNR} & \textbf{FPR} \\
\midrule
0  & \meanstd{93.5}{0.3} & \meanstd{8.4}{0.3} & \meanstd{0.4}{0.1} \\
10 & \meanstd{91.3}{0.3} & \meanstd{12.4}{0.4} & \meanstd{0.4}{0.1} \\
20 & \meanstd{87.8}{0.4} & \meanstd{14.0}{0.5} & \meanstd{0.9}{0.1} \\
30 & \meanstd{84.8}{0.4} & \meanstd{15.5}{0.5} & \meanstd{1.5}{0.2} \\
40 & \meanstd{81.5}{0.5} & \meanstd{17.0}{0.6} & \meanstd{2.5}{0.3} \\
50 & \meanstd{78.1}{0.6} & \meanstd{18.4}{0.6} & \meanstd{4.3}{0.4} \\
60 & \meanstd{76.2}{0.6} & \meanstd{19.5}{0.6} & \meanstd{6.9}{0.5} \\
70 & \meanstd{74.3}{0.7} & \meanstd{21.5}{0.7} & \meanstd{9.8}{0.6} \\
80 & \meanstd{73.6}{0.7} & \meanstd{14.4}{0.8} & \meanstd{11.7}{0.7} \\
90 & \meanstd{70.0}{0.8} & \meanstd{16.6}{0.8} & \meanstd{13.1}{0.8} \\
\bottomrule
\end{tabular}
\end{table}

\section{CITADEL with Curriculum Learning}
\label{app:curriculum_learning}
\paragraph{Experiments.} To evaluate whether Curriculum Learning (CL) can mitigate malware drift, we designed a baseline experiment where training samples were ordered from easy to hard following the CL principle proposed by~\cite{zhou2020curriculum}. Unlike Active Learning (AL), which adaptively selects new informative samples during training, CL assumes a smooth progression in task difficulty. This assumption rarely holds in malware detection, where new variants appear abruptly. We used the 2012 training subset of the APIGraph dataset and ranked all instances by \textit{instance hardness} following~\cite{zhou2020curriculum}. To simulate limited human annotation effort, we additionally labeled 400 hard samples per month as in~\cite{kong2021adaptive}. Both settings used identical hyperparameters and data splits as our main experiments for fair comparison with active learning.

\paragraph{Results.}Table~\ref{tab:curriculum_learning} summarizes the comparison. Without CL, the model achieved an F1 score of 71.2\%, with relatively high FNR and FPR. Incorporating CL and limited hard-sample feedback improved the F1 score to 83.0\%, showing moderate gains in generalization and convergence stability.
However, our proposed Active Learning (CITADEL) approach—jointly considering margin, distance, and confidence (Table~IV)—achieved a substantially higher 93.5\% F1, with a much lower FNR of 8.4\% and FPR of 0.4\%. These results demonstrate that while CL helps on static datasets, AL is far more effective under evolving malware conditions due to its adaptability and targeted sample selection.

\begin{table}[!t]
\centering
\caption{Comparison of Curriculum Learning (CurL) and Active Learning (AL) for CITADEL under identical labeling budgets (400 hard samples/month).}
\label{tab:curriculum_learning}
\small
\begin{tabular}{lccc}
\toprule
\textbf{Method} & \textbf{F1} & \textbf{FNR} & \textbf{FPR} \\
\midrule
w/o CurL & 71.2\tiny{$\pm$0.6} & 36.2\tiny{$\pm$0.7} & 1.20\tiny{$\pm$0.0} \\
CurL + 400 samples & 83.0\tiny{$\pm$0.5} & 24.0\tiny{$\pm$0.6} & 0.90\tiny{$\pm$0.0} \\
\textbf{AL (CITADEL)} & \textbf{93.5\tiny{$\pm$0.3}} & \textbf{8.4\tiny{$\pm$0.3}} & \textbf{0.4\tiny{$\pm$0.1}} \\
\bottomrule
\end{tabular}
\end{table}

\section{Experiments on EMBER Feature}
\label{app:exp_other_feature}
\paragraph{Experiments.} 
On the EMBER dataset~\cite{anderson2018ember}, we use samples from 2018-01 to 2018-02 for initial training, and begin active learning from 2018-03 to 2018-12 using different methods—Chen-AL~\cite{chen2023continuous}, CADE~\cite{yang2021cade}, and \system. For this feature space, since Bernoulli bit-flip is not applicable, we employ Bernoulli feature masking as the augmentation strategy in \system. This approach regularizes the model by randomly masking input features, thereby improving generalization under limited labeling budgets.

\paragraph{Results.} 
Table~\ref{tab:ember_results} presents the performance of CADE, Chen-AL, and \system under different labeling budgets on the EMBER dataset. \system consistently achieves the best results across all budgets. At a budget of 50, \system attains an F1-score of 87.9\%, outperforming CADE (85.4\%) and Chen-AL (83.4\%). For budget 400, \system further improves to an F1-score of 89.4\%, while CADE reaches 86.0\%.

\begin{table}[!t]
\centering
\caption{Performance comparison of CADE, Chen-AL, and CITADEL under different labeling budgets on the EMBER dataset.}
\renewcommand{\arraystretch}{1.05}
\setlength{\tabcolsep}{4pt}
\small
\begin{tabular}{@{}clccc@{}}
\toprule
\textbf{Budget} & \textbf{Method} & \textbf{F1} & \textbf{FNR} & \textbf{FPR} \\ 
\midrule
\multirow{3}{*}{50} 
 & CADE     & 85.4\tiny{$\pm$0.5} & 22.5\tiny{$\pm$0.4} & 4.0\tiny{$\pm$0.2} \\ 
 & Chen-AL  & 83.4\tiny{$\pm$0.8} & 17.3\tiny{$\pm$0.5} & 18.5\tiny{$\pm$0.6} \\ 
 & CITADEL  & 87.9\tiny{$\pm$0.5} & 16.8\tiny{$\pm$0.4} & 6.6\tiny{$\pm$0.3} \\ 
\midrule
\multirow{3}{*}{100} 
 & CADE     & 85.7\tiny{$\pm$0.5} & 22.1\tiny{$\pm$0.4} & 4.0\tiny{$\pm$0.2} \\ 
 & Chen-AL  & 83.7\tiny{$\pm$0.8} & 17.0\tiny{$\pm$0.5} & 18.1\tiny{$\pm$0.5} \\ 
 & CITADEL  & 88.4\tiny{$\pm$0.5} & 16.1\tiny{$\pm$0.4} & 6.4\tiny{$\pm$0.3} \\ 
\midrule
\multirow{3}{*}{200} 
 & CADE     & 85.9\tiny{$\pm$0.4} & 21.8\tiny{$\pm$0.3} & 4.1\tiny{$\pm$0.2} \\ 
 & Chen-AL  & 84.0\tiny{$\pm$0.7} & 16.7\tiny{$\pm$0.4} & 17.6\tiny{$\pm$0.5} \\ 
 & CITADEL  & 88.9\tiny{$\pm$0.4} & 15.3\tiny{$\pm$0.3} & 6.3\tiny{$\pm$0.2} \\ 
\midrule
\multirow{3}{*}{400} 
 & CADE     & 86.0\tiny{$\pm$0.4} & 21.5\tiny{$\pm$0.3} & 4.2\tiny{$\pm$0.2} \\ 
 & Chen-AL  & 84.3\tiny{$\pm$0.4} & 16.5\tiny{$\pm$0.3} & 16.9\tiny{$\pm$0.3} \\ 
 & CITADEL  & 89.4\tiny{$\pm$0.3} & 14.5\tiny{$\pm$0.3} & 6.3\tiny{$\pm$0.2} \\ 
\bottomrule
\end{tabular}
\label{tab:ember_results}
\end{table}


\section{Computational Resources}
\label{app:computational_resources}

All experiments are conducted on a server with dual-socket Intel Xeon Gold 6430 CPUs, providing 64 physical cores (128 logical cores) and 1TB of main memory. The server is equipped with 4$\times$ NVIDIA H100 NVL GPUs, each with 94GB of GPU memory. The software environment uses CUDA 12.8 with NVIDIA driver version 570.124.X. Each run of CITADEL uses approximately 3--4GB of main memory and 2--3~GB of GPU memory. \system can also be executed on cloud-based platforms such as Google Colab, where an NVIDIA A100 or H100 GPU is recommended for acceleration.



%% file: tables/datasets-singleton.tex

\begin{table*}[t]
\scriptsize
\setlength{\tabcolsep}{2.0pt}
\centering
\caption{Year-wise comparison of total malware families (F) and singleton samples (S) across four datasets (2012–2025).}
\begin{tabular}{l|*{14}{cc}}
\toprule
\textbf{Dataset} &
\multicolumn{2}{c}{\textbf{2012}} &
\multicolumn{2}{c}{\textbf{2013}} &
\multicolumn{2}{c}{\textbf{2014}} &
\multicolumn{2}{c}{\textbf{2015}} &
\multicolumn{2}{c}{\textbf{2016}} &
\multicolumn{2}{c}{\textbf{2017}} &
\multicolumn{2}{c}{\textbf{2018}} &
\multicolumn{2}{c}{\textbf{2019}} &
\multicolumn{2}{c}{\textbf{2020}} &
\multicolumn{2}{c}{\textbf{2021}} &
\multicolumn{2}{c}{\textbf{2022}} &
\multicolumn{2}{c}{\textbf{2023}} &
\multicolumn{2}{c}{\textbf{2024}} &
\multicolumn{2}{c}{\textbf{2025}} \\

\cmidrule(lr){2-29}
 & \textbf{F} & \textbf{S}
 & \textbf{F} & \textbf{S}
 & \textbf{F} & \textbf{S}
 & \textbf{F} & \textbf{S}
 & \textbf{F} & \textbf{S}
 & \textbf{F} & \textbf{S}
 & \textbf{F} & \textbf{S}
 & \textbf{F} & \textbf{S}
 & \textbf{F} & \textbf{S}
 & \textbf{F} & \textbf{S}
 & \textbf{F} & \textbf{S}
 & \textbf{F} & \textbf{S}
 & \textbf{F} & \textbf{S}
 & \textbf{F} & \textbf{S} \\
\midrule

\textbf{APIGraph}~\cite{api_graph_dataset} 
& 104 & 36 & 172 & 68 & 175 & 55 & 193 & 53 & 199 & 68 & 147 & 48 & 128 & 45 & - & - & - & - & - & - & - & - & - & - & - & - & - & - \\

\textbf{Chen-AZ}~\cite{chen2023continuous} 
& - & - & - & - & - & - & - & - & - & - & - & - & - & - & 121 & 42 & 82 & 31 & 51 & 17 & - & - & - & - & - & - & - & - \\

\textbf{LAMDA}~\cite{haque2025lamda} 
& - & - & 213 & 1320 & 231 & 2046 & - & - & 375 & 4819 & 207 & 7324 & 373 & 16531 & 635 & 15396 & 588 & 24791 & 295 & 24083 & 651 & 20147 & 224 & 4839 & 64 & 538 & 8 & 18 \\

\textbf{MaMaDroid~\cite{mamadroid}} 
& - & - & 1 & 32 & 1 & 47 & - & - & 96 & 132 & 48 & 391 & 80 & 521 & 134 & 291 & 117 & 455 & 62 & 805 & 130 & 515 & 93 & 717 & 57 & 526 & 7 & 11 \\

\bottomrule
\end{tabular}
\label{tab:families-singletons}
\end{table*}

%% file: tables/CITADEL_vs_Chen.tex
\begin{table}[t]
\centering
\setlength{\tabcolsep}{3.5pt}
\caption{Year-wise comparison on the LAMDA dataset. For each year, we report the total number of evaluated samples and the number of misclassified samples, decomposed into false negatives (FN) and false positives (FP), comparing Chen-AL~\cite{chen2023continuous}. with CITADEL.}
\label{tab:lamda_fn_fp_onecol}
\begin{tabular}{r r r r r r}
\toprule
\multirow{2}{*}{Year} & \multirow{2}{*}{Total} &
\multicolumn{2}{c}{Chen-AL~\cite{chen2023continuous}} & \multicolumn{2}{c}{CITADEL} \\
\cmidrule(lr){3-4}\cmidrule(lr){5-6}
& & FN & FP & FN & FP \\
\midrule
2014 &  8652 &  586 &  167 &  439 &  178 \\
2016 & 21839 & 1324 & 8646 & 1796 & 1458 \\
2017 & 19829 &    0 & 15557 &  525 &  354 \\
2018 & 20859 &    0 & 12989 & 1123 &  479 \\
2019 & 18210 &    0 &  9893 & 1573 &  225 \\
2020 & 20415 &    0 & 11144 &  967 &  161 \\
2021 & 16231 &    0 &  9106 &  496 &  101 \\
2022 & 17284 &    0 &  8954 &  856 &   89 \\
2023 & 10871 &    0 &  9293 &  302 &   46 \\
2024 &  9686 &    0 &  9527 &   57 &   42 \\
2025 &  8933 &    0 &  8928 &    4 &   34 \\
\bottomrule
\end{tabular}
\end{table}